\documentclass[a4paper,11pt]{article}
\pdfoutput=1
\usepackage{jheppub}
\usepackage[T1]{fontenc}
\usepackage{amsmath,amssymb,amsthm}
\usepackage{dsfont}
\usepackage{graphicx}
\usepackage{booktabs,array}
\usepackage[dvipsnames]{xcolor}
\usepackage{tikz}
\usepackage{ytableau}
\ytableausetup{boxsize=0.55em,aligntableaux=center}
\newcommand{\yd}[1]{\ydiagram{#1}}                     
\newcommand{\irr}[2]{[\,#1\,;\,#2\,]}                   
\usetikzlibrary{arrows.meta,decorations.pathmorphing}

\newcommand{\Nc}{N_c}
\newcommand{\Tb}{\mathbf T}
\newcommand{\ket}[1]{|#1\rangle}
\newcommand{\bra}[1]{\langle#1|}
\newcommand{\braket}[2]{\langle#1|#2\rangle}
\newcommand{\orthobase}{\texttt{OrthoBase}}

\theoremstyle{plain}
\newtheorem{theorem}{Theorem}[section]
\newtheorem{lemma}[theorem]{Lemma}

\title{A billion colour states: exact soft-gluon evolution beyond leading colour for many-jet QCD}
\author{Bakar Chargeishvili}
\affiliation{Institute for Theoretical Physics, Karlsruhe Institute of Technology (KIT),\\Wolfgang-Gaede-Stra\ss e 1, 76131 Karlsruhe, Germany}
\emailAdd{bakar.chargeishvili@kit.edu}
\preprint{KA-TP-33-2026}
\abstract{We present a new method, and its implementation, for constructing orthogonal colour bases and computing colour factors
for any process with quarks, antiquarks and gluons, and use it to compute one-loop soft-gluon evolution exactly in full
colour, keeping all colour interference. Each basis vector is labelled, independently of $\Nc$, by the sequence of colour
representations that appear when the partons are coupled one at a time. We derive a closed formula for the norms of the
basis vectors, and colour operators act only on the part of a label between the partons involved, through small tables.
Basis vectors can therefore be generated and used one at a time, without listing the complete basis: norms and matrix
elements of single basis vectors are obtained for $10^4$ gluons in seconds. The same locality lets the soft anomalous
dimension act on colour states without being stored. This allows optimised implementations on CPUs and GPUs, with which
we evolve colour states exactly for up to fourteen gluons on a single GPU and for fifteen gluons, a colour space of
$827\,888\,726$ dimensions, on a single compute node. We validate the basis and the evolution against independent codes and published results for processes with few partons, the only ones for which such results are available. We then measure the accuracy of leading colour as a function of the number of partons. For
exclusive jet rates with a veto on soft radiation outside the jets, at a ratio of about 14 between the hard and the veto
scale, leading colour overestimates the full-colour result by a median factor of 1.3 for eight partons, 1.8 for ten and
6.6 for fourteen. Pairs of hard jets are in a colour singlet with probabilities of several per cent, which vanish at
leading colour. The formal $1/\Nc^2$ suppression of subleading colour therefore does not make it small for many-jet final
states.
}

\begin{document}
\maketitle
\flushbottom

\clearpage
\section{Introduction}\label{sec:intro}

Events with many jets are common at the Large Hadron Collider (LHC), and their description relies on the resummation of
soft-gluon effects, either analytically or through parton showers. Soft gluons are sensitive to the colour state of all
hard partons. When radiation is restricted, for instance by a jet veto or a rapidity gap, the colour state of the hard
partons evolves between the hard scale and the veto scale. This evolution is governed by the soft anomalous dimension,
a matrix that describes how the exchange of soft gluons mixes the colour structures of the amplitude. It acts in a
space whose dimension grows exponentially with the number of partons, by a factor of about six per gluon.

Soft anomalous dimension matrices were first computed for four-parton processes
\cite{Botts:1989kf,Sotiropoulos:1993rd,Kidonakis:1998nf} and have since been obtained for many processes and to higher
orders \cite{Kidonakis:2009ev,Becher:2009cu,Gardi:2009qi,Almelid:2015jia,Becher:2019avh}. Full-colour calculations have been
carried out for processes with few partons, in particular for gaps between jets and for non-global and super-leading
logarithms \cite{Forshaw:2005sx,Sjodahl:2008fz,DuranDelgado:2011tp,Forshaw:2006fk,Becher:2021zkk,
Becher:2023mtx,Becher:2024nqc}. For more partons the colour treatment is almost always approximated. Parton showers
evolve the colour state at leading colour \cite{Gustafson:1987rq}. In this approximation the colour state is described
by colour flows, that is, by ways of connecting the colour lines of the partons, and every pair of connected partons, a
colour dipole, radiates independently; the neglected terms are formally suppressed by $1/\Nc^2$, where $\Nc$ is the
number of colours.
Considerable effort has gone into going beyond this approximation, through subleading-colour corrections in showers
\cite{Nagy:2007ty,Nagy:2012bt,Platzer:2012np,Platzer:2013fha,Isaacson:2018zdi,Platzer:2018pmd,Nagy:2019pjp,
Hoche:2020pxj,Hamilton:2020rcu,Herren:2022jej,vanBeekveld:2024wws}, evolution at the amplitude level
\cite{AngelesMartinez:2018cfz,Forshaw:2019ver,DeAngelis:2020rvq,Platzer:2022jny,Forshaw:2025bmo}, and the treatment of
non-global logarithms at finite $\Nc$ \cite{Weigert:2003mm,Hatta:2013iba,Hagiwara:2015bia,Khelifa-Kerfa:2015mma}. The soft
function of multi-jet final states was evolved in full colour at next-to-leading logarithmic accuracy, in
non-orthogonal colour bases, for all QCD processes with up to seven partons in ref.~\cite{Gerwick:2014gya}, where
finite-$\Nc$ corrections were found to be significant for processes with many gluons, and amplitude-level evolution in
full colour has been applied to jet vetoes in processes with up to four hard coloured partons
\cite{Forshaw:2025bmo}. The common
limitation is the size of the colour space: for fifteen gluons it has $8.3\times10^8$ dimensions, and the number of
colour flows used at leading colour exceeds $10^{10}$.

As a consequence, it is not known how accurate leading colour is for processes with many partons. The corrections are
formally suppressed, but the number of parton pairs whose exchanges leading colour neglects grows with the number of
partons, and there has been no exact
calculation against which the approximation could be tested beyond a handful of partons. In this paper we provide such
calculations. We construct an orthogonal basis of colour space for any process with quarks, antiquarks and gluons, in
which the colour operators act locally, and use it to evolve colour states exactly with the one-loop soft anomalous
dimension for up to fifteen gluons. We then compare full and leading colour for several observables as a function of
the number of partons.

Before listing our main findings we fix some notation, defined in detail in section~\ref{sec:colourspace}. The number
of partons $n$ counts incoming and outgoing partons. The evolution variable $t$, eq.~\eqref{eq:cs:evolution}, measures
the logarithmic distance between the hard and the veto scale; for $\alpha_s=0.118$, $t=0.1$ corresponds to a ratio of
about 14 between the two scales. We write $R(t)$ for the colour-summed squared amplitude after evolution divided by its
value before; for a veto it is the colour factor of the probability that no soft gluon above the veto scale is emitted
into the vetoed region. Its values in full and in leading colour are $R_{\rm FC}$ and $R_{\rm LC}$. Our main findings
are the following.
\begin{enumerate}
\item Orthogonal multiplet bases for any process can be constructed by coupling the partons one at a time, with every
  gluon treated as a quark and an antiquark index, and fixing each step by the eigenvalue of a Casimir operator. The
  norms of the basis vectors are given by a closed formula. A colour operator acts only on the part of a label between
  the partons it involves, through small tables that depend only on neighbouring representations. Norms and matrix
  elements of single basis vectors can therefore be computed for $10^4$ gluons.
\item The soft evolution $e^{-t\Gamma}$ can be computed exactly without storing $\Gamma$. We evolve tree-level colour states
  of 13 and 14 gluons on a GPU for 24 and 19 kinematic points, and of 15 gluons, with $827\,888\,726$ colour states, on
  CPU nodes. The method itself has no fixed limit in $n$; beyond fifteen partons the limit is memory.
\item The accuracy of leading colour decreases steadily with the number of partons. For evolution with the anomalous
  dimension of the hard process (its one-loop virtual corrections), the median of $|\ln(R_{\rm LC}/R_{\rm FC})|$ at $t=0.05$ grows from 0.02--0.03 for four to six partons
  to 0.48 and 0.67 for twelve and thirteen. For exclusive jet
  rates with a soft veto, leading colour overestimates the full-colour result at $t=0.1$ by a median factor of 1.27 for
  $n=8$ partons, 1.83 for $n=10$ and 6.6 for $n=14$. For the gap fraction, the fraction of events without radiation
  above the veto scale in a rapidity interval between the jets, the factor reaches 2.2 at $n=14$.
\item For real weights, the smallest eigenvalue of the soft anomalous dimension lies below the leading-colour eigenvalue
  of every colour flow by an amount of order one, which does not decrease with $n$. At late times full colour therefore
  decays more slowly than leading colour; within $t\le0.2$ this is reached only for up to seven partons. Pairs of the three hardest jets are in a colour singlet with a probability of
  3--9\%, which vanishes at leading colour.
\item With real emissions included for up to five hard partons, the imaginary parts ($i\pi$ terms, Coulomb phases) of the
  anomalous dimension increase the gap fraction of $gg\to ggg$ by up to 72\% at $t=0.2$, mostly through the virtual
  evolution. For $q\bar q\to ggg$ full colour without Coulomb phases exceeds leading colour by 7--17\%.
\item Matrix product states write each coefficient of a colour state as a product of matrices, one per parton, and
  compress the state by limiting the size of these matrices. Starting from a single basis vector, they reproduce the
  exact evolution to within 2\% only for $t\lesssim0.05$ at generic kinematics. At the same matrix size they are 30 to
  more than 1000 times more accurate when the partons are ordered in angle, as in an angular-ordered shower. For such
  simple initial states the coefficients of $t$, $t^2$ and $t^3$ in $R(t)$ can be computed exactly for 500 gluons.
\end{enumerate}

The paper is organised as follows. Section~\ref{sec:colourspace} introduces colour space, the soft anomalous dimension
and leading colour. Section~\ref{sec:bases} reviews earlier colour bases and describes our construction.
Section~\ref{sec:evolution} explains the exact evolution, and section~\ref{sec:validation} summarises its validation.
Section~\ref{sec:lcvsfc} contains the comparison of full and leading colour, and section~\ref{sec:beyond} discusses
approximate methods for more partons. We conclude in section~\ref{sec:conclusions}. The appendices collect our
conventions, the proofs, a worked example, numerical details, the availability of the code and the projection of tree
amplitudes.
\section{Colour space and soft evolution}\label{sec:colourspace}

In this section we fix the notation for colour space and colour operators, define the soft evolution
studied in this paper and the observable $R(t)$ that we compute, and state what we mean by leading colour.
We close with the reason why an orthogonal basis is the natural setting for an exact calculation.
Conventions not needed in the main text are collected in appendix~\ref{app:conventions}.

\subsection{Colour space and colour operators}\label{sec:cs:space}

We consider a process with $n$ partons and treat all of them as outgoing; an incoming quark is an outgoing
antiquark and vice versa. The colour part of the amplitude is a vector $\ket{\mathcal M}$ in the space of
colour-singlet tensors
\begin{equation}
  \mathcal V_n=\mathrm{Inv}\bigl(V_1\otimes V_2\otimes\cdots\otimes V_n\bigr),
  \label{eq:cs:space}
\end{equation}
where $V_i$ is the fundamental representation $\mathbf N_c$ for a quark, the antifundamental
$\bar{\mathbf N}_c$ for an antiquark and the adjoint representation for a gluon. Following
refs.~\cite{Catani:1996vz,Catani:1998bh} we write $\Tb_i$ for the colour charge of parton $i$: it acts as
the generator $t^a$ on a quark index, as $-(t^a)^T$ on an antiquark index and as $(T^a)_{bc}=-if^{abc}$ on a
gluon index, and as the identity on all other partons. We normalise the generators as
$\mathrm{Tr}\,t^at^b=\tfrac12\delta^{ab}$, so that $\Tb_i^2=C_i$ with $C_F=(\Nc^2-1)/(2\Nc)$ for quarks and
$C_A=\Nc$ for gluons. Colour conservation reads
\begin{equation}
  \sum_{i=1}^n\Tb_i\ket{\mathcal M}=0 .
  \label{eq:cs:conservation}
\end{equation}
The colour-summed squared amplitude is the norm $\braket{\mathcal M}{\mathcal M}$ in the scalar product that
contracts all colour indices.

The space $\mathcal V_n$ grows quickly with $n$. For $n$ gluons and $\Nc=3$ its dimension is $8$, $145$,
$19\,280$ and $3\,609\,760$ for $n=4,6,9,12$, and it reaches $827\,888\,726$ for $n=15$. These numbers are
dimensions, so they do not depend on the choice of basis; any exact treatment of colour for fifteen gluons has
to handle vectors of this length.

\paragraph{Example.}
For $q\bar q\to q\bar q$, written as four outgoing partons $q_1\bar q_2q_3\bar q_4$, the space
$\mathcal V_4$ is two-dimensional. A common choice of basis is the pair of colour flows
\begin{equation}
  \ket{c_1}=\delta_{i_1\bar\imath_2}\,\delta_{i_3\bar\imath_4},\qquad
  \ket{c_2}=\delta_{i_1\bar\imath_4}\,\delta_{i_3\bar\imath_2},
  \label{eq:cs:qqflows}
\end{equation}
in which each Kronecker delta connects a quark to an antiquark. The two vectors are not orthogonal,
\begin{equation}
  \braket{c_1}{c_1}=\braket{c_2}{c_2}=\Nc^2,\qquad \braket{c_1}{c_2}=\Nc ,
\end{equation}
and their overlap is suppressed by only one power of $\Nc$ relative to the norms. The combinations
$\ket{c_1}$ and $\ket{c_2}-\ket{c_1}/\Nc$ are orthogonal: they project the pair $(q_1,\bar q_2)$ onto a
colour singlet and a colour octet, respectively. This is the simplest instance of the multiplet bases
constructed in section~\ref{sec:bases}.

\subsection{Soft anomalous dimension and evolution}\label{sec:cs:gamma}

A veto accepts an event with $n-2$ hard jets only if no additional gluon with transverse momentum above $Q_0$ enters a specified region $V$. The region can be a rapidity gap between jets or the part of the detector outside the jet cones. For primary radiation from the hard partons, real and virtual contributions cancel below $Q_0$ and outside $V$. Inside $V$ above $Q_0$, the virtual corrections remain and produce logarithms of $Q/Q_0$, where $Q$ is the hard scale. We resum these corrections by evolving the hard colour state with the soft anomalous dimension \cite{Kidonakis:1998nf,Catani:1998bh,Becher:2009cu,Becher:2009qa,Gardi:2009qi},
\begin{equation}
  \Gamma=\sum_{1\le i<j\le n}\omega_{ij}\,\Tb_i\cdot\Tb_j .
  \label{eq:cs:gamma}
\end{equation}

The weight $\omega_{ij}$ specifies the contribution from a soft gluon exchanged between hard partons of momenta $p_i$ and $p_j$. For a veto we use the \emph{veto weights}
\begin{equation}
  \omega_{ij}=-\Omega_{ij},\qquad
  \Omega_{ij}=\frac12\int_V\mathrm dy\,\frac{\mathrm d\phi}{2\pi}\,
  \frac{p_i\cdot p_j}{(p_i\cdot\hat k)(p_j\cdot\hat k)},\qquad
  \hat k=(\cosh y,\cos\phi,\sin\phi,\sinh y),
  \label{eq:cs:vetoweights}
\end{equation}
where $y$ and $\phi$ are the rapidity and azimuth of the soft gluon, whose direction is $\hat k$. We call these \emph{gap weights} when $V$ is a rapidity gap. Section~\ref{sec:lcvsfc} specifies the veto regions used below.

We also study the virtual corrections to the hard process itself. At renormalisation scale $\mu$ the corresponding \emph{hard weights} are
\begin{equation}
  \omega_{ij}=-\ln\frac{-s_{ij}-i0}{\mu^2},\qquad s_{ij}=2p_i\cdot p_j .
  \label{eq:cs:hardweights}
\end{equation}
Evolution with these weights describes the renormalisation-group running of the hard function between two scales \cite{Becher:2009cu}. A pair of incoming or a pair of outgoing partons contributes an imaginary term $+i\pi$, which we call a Coulomb phase. With partons 1 and 2 incoming, colour conservation gives
\begin{equation}
  \mathrm{Im}\,\Gamma=\pi\Bigl[(\Tb_1+\Tb_2)^2-\frac12\sum_{i=1}^nC_i\Bigr].
  \label{eq:cs:coulomb}
\end{equation}
Up to a multiple of the identity, this is $\pi$ times the squared total colour charge of the incoming pair \cite{Catani:1998bh,Forshaw:2006fk}. For massless partons the same incoming-pair operator describes Coulomb mixing for any number of final-state partons (massive coloured final-state particles give an additional Coulomb phase \cite{Banerjee:2025kkq}), while the number of terms in the real part grows like $n^2$. Coupling the incoming partons first makes the Coulomb operator diagonal (section~\ref{sec:bratteli}).

At fixed coupling the colour state evolves from $Q$ down to $Q_0$ as
\begin{equation}
  \ket{\mathcal M(t)}=e^{-t\Gamma}\ket{\mathcal M},\qquad t=\frac{\alpha_s}{2\pi}\ln\frac{Q^2}{Q_0^2} .
  \label{eq:cs:evolution}
\end{equation}
With one-loop running, $t=(2/\beta_0)\ln[\alpha_s(Q_0)/\alpha_s(Q)]$, where $\beta_0$ is the one-loop coefficient of the QCD beta function. For either choice of weights we compute the ratio
\begin{equation}
  R(t)=\frac{\braket{\mathcal M(t)}{\mathcal M(t)}}{\braket{\mathcal M}{\mathcal M}} .
  \label{eq:cs:R}
\end{equation}
For veto weights, $R(t)$ is the colour factor for no primary emission above $Q_0$ into $V$. It contains the complete colour dependence of this global contribution at next-to-leading logarithmic accuracy \cite{Dasgupta:2001sh}. We omit the colour-diagonal factors from double logarithms and collinear anomalous dimensions. For hard weights, the same ratio measures the change in the squared hard amplitude between the two scales.

The evolution variable $t$ measures how far apart the two scales are. For $\alpha_s=0.118$ the scale ratio
$Q/Q_0=e^{\pi t/\alpha_s}$ is about $4$ for $t=0.05$, $14$ for $t=0.1$ and $205$ for $t=0.2$. Dijet
measurements at the LHC with a veto on additional jets between the two leading ones use a veto scale of
$Q_0=20$~GeV and jets with mean transverse momenta of up to several hundred GeV
\cite{ATLAS:2011yyh,ATLAS:2014lzu}, which corresponds to $t\approx0.05$--$0.12$. The value $t=0.2$ is an extreme scale
ratio and serves as an upper end. We therefore show results for $0\le t\le0.2$ and quote numbers at the three values
above.

Two properties of $\Gamma$ are used throughout. First, by eq.~\eqref{eq:cs:conservation},
$\sum_{j\neq i}\Tb_i\cdot\Tb_j=-C_i$ on $\mathcal V_n$. A shift $\omega_{ij}\to\omega_{ij}+a_i+a_j+c$ therefore
changes $\Gamma$ by a multiple of the identity, which multiplies $R(t)$ by a number. For gluons, and for LC+ of
section~\ref{sec:cs:lc}, this number is the same in full and in leading colour, so that ratios such as
$R_{\rm LC}/R_{\rm FC}$ depend only on the part of $\omega_{ij}$ that is not of this form. For a quark or antiquark,
leading colour replaces $C_F$ by $\Nc/2$ in this factor, and the ratio changes at relative order $1/\Nc^2$; in comparisons
that involve quarks we therefore always use weights with their mean subtracted (section~\ref{sec:lcvsfc}). Second, for $n\le5$ the matrix $\Gamma$ is small (at most $9\times9$ for $n=4$ and $44\times44$ for $n=5$ at generic $\Nc$, and $8\times8$ and $32\times32$ at $\Nc=3$)
and is known in closed form for many processes \cite{Kidonakis:1998nf,Sjodahl:2009wx}. The difficulty
addressed here is the size of $\mathcal V_n$ at large $n$, not the form of $\Gamma$.

\subsection{Colour flows and leading colour}\label{sec:cs:lc}

Leading colour is most easily explained with colour lines \cite{tHooft:1973alw}. A quark carries one colour
line, an antiquark one anticolour line, and, up to terms suppressed by $1/\Nc^2$, a gluon carries one of each.
A \emph{colour flow} is one way of joining all these lines into closed loops or open strings. For $q\bar q\to
q\bar q$ the two flows are the vectors $\ket{c_1}$ and $\ket{c_2}$ of eq.~\eqref{eq:cs:qqflows}: in $\ket{c_1}$
the line of quark 1 ends on antiquark 2 and the line of quark 3 on antiquark 4, in $\ket{c_2}$ they are
crossed. For $n$ gluons a flow is a cyclic ordering of the gluons, $\mathrm{Tr}(t^{a_{\sigma_1}}\cdots
t^{a_{\sigma_n}})$: each gluon's colour line continues as the anticolour line of the next. Two partons that
are joined by a line in a flow are called \emph{colour connected} in that flow. A flow with $n$ gluons
therefore consists of $n$ colour-connected pairs, or colour dipoles.

\begin{figure}[t]
  \centering
  \resizebox{\textwidth}{!}{
\begin{tikzpicture}[
    line/.style={very thick,-{Stealth[length=2.2mm]}},
    gl/.style={thick,decorate,decoration={coil,aspect=0.6,segment length=1.6mm,amplitude=1.1mm}},
    lab/.style={font=\small},
    ptext/.style={font=\small,align=center}]
  \newcommand{\partons}[1]{%
    \node[lab,left]  at (#1+0.0,1.2) {$q_1$};
    \node[lab,left]  at (#1+0.0,0.0) {$q_3$};
    \node[lab,right] at (#1+1.8,1.2) {$\bar q_2$};
    \node[lab,right] at (#1+1.8,0.0) {$\bar q_4$};}
  \partons{0}
  \draw[line] (0,1.2) -- (1.8,1.2);
  \draw[line] (0,0.0) -- (1.8,0.0);
  \node[ptext,anchor=north] at (0.9,-0.4) {(a)\\[2pt] $\ket{c_1}$};
  \partons{3.6}
  \draw[line] (3.6,1.2) -- (5.4,0.0);
  \draw[line] (3.6,0.0) -- (5.4,1.2);
  \node[ptext,anchor=north] at (4.5,-0.4) {(b)\\[2pt] $\ket{c_2}$};
  \partons{7.2}
  \draw[line] (7.2,1.2) -- (9.0,1.2);
  \draw[line] (7.2,0.0) -- (9.0,0.0);
  \draw[gl] (7.55,1.2) to[out=70,in=110] (8.65,1.2);
  \node[ptext,anchor=north] at (8.1,-0.4) {(c)\\[2pt] $\Tb_1\!\cdot\Tb_2\ket{c_1}$\\[1pt] $=-C_F\ket{c_1}$};
  \partons{11.8}
  \draw[line] (11.8,1.2) -- (13.6,1.2);
  \draw[line] (11.8,0.0) -- (13.6,0.0);
  \draw[gl] (12.4,1.2) -- (12.4,0.0);
  \node[ptext,anchor=north] at (12.7,-0.4) {(d)\\[2pt] $\Tb_1\!\cdot\Tb_3\ket{c_1}$\\[1pt] $=\ket{c_2}/2-\ket{c_1}/(2\Nc)$};
\end{tikzpicture}}
  \caption{Colour lines for $q_1\bar q_2q_3\bar q_4$. (a), (b): the two colour flows of
  eq.~\eqref{eq:cs:qqflows}; each arrow is a colour line from a quark to an antiquark. (c): a soft gluon
  exchanged between colour-connected partons leaves the flow unchanged, with a coefficient of order $\Nc$.
  (d): an exchange between partons that are not connected reconnects the lines and produces the other flow,
  with a coefficient of order one. Leading colour keeps only processes of type (c).}
  \label{fig:colourlines}
\end{figure}
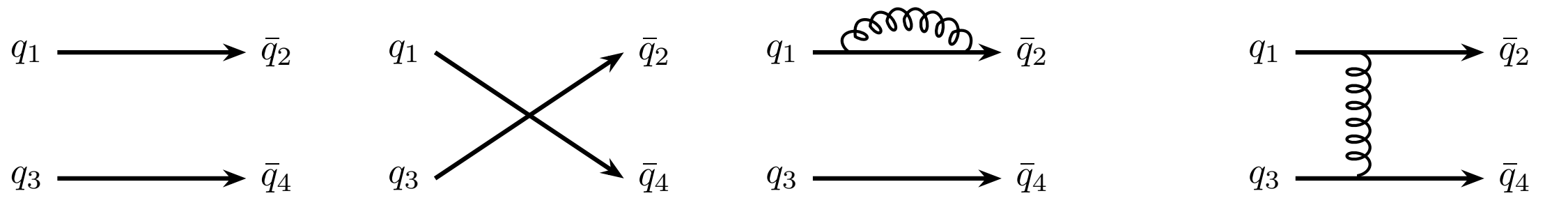

The effect of one soft-gluon exchange depends on whether the two partons are colour connected
(figure~\ref{fig:colourlines}). For the
running example, the Fierz identity $t^a_{ij}t^a_{kl}=\frac12(\delta_{il}\delta_{kj}-\delta_{ij}\delta_{kl}/\Nc)$
gives
\begin{equation}
  \Tb_1\cdot\Tb_2\ket{c_1}=-C_F\ket{c_1},\qquad
  \Tb_1\cdot\Tb_3\ket{c_1}=\frac12\ket{c_2}-\frac1{2\Nc}\ket{c_1}.
  \label{eq:cs:qqaction}
\end{equation}
An exchange between the connected partons 1 and 2 leaves the flow unchanged and comes with a factor
$C_F=\Nc/2-1/(2\Nc)$, of order $\Nc$. An exchange between the unconnected partons 1 and 3 reconnects the colour
lines and turns $\ket{c_1}$ into $\ket{c_2}$, with a coefficient of order one. The same pattern holds for any
number of partons: connected pairs give a diagonal term of order $\Nc$, unconnected pairs reconnect the lines.
In addition, different flows overlap only at relative order $1/\Nc$ or smaller (see the example above).

Leading colour (LC) keeps only the order-$\Nc$ terms. Each flow $\ket\sigma$ then evolves on its own, with
\begin{equation}
  \Gamma\ket\sigma\simeq\lambda^{\rm LC}_\sigma\ket\sigma,\qquad
  \lambda^{\rm LC}_\sigma=-\frac{\Nc}{2}\sum_{(ij)\ \text{connected in}\ \sigma}\omega_{ij},
  \label{eq:cs:lambdaLC}
\end{equation}
and different flows are treated as orthogonal. Writing the hard amplitude as $\ket{\mathcal M}=\sum_\sigma
A_\sigma\ket\sigma$, the evolution factor becomes an incoherent sum over flows,
\begin{equation}
  R_{\rm LC}(t)=\frac{\sum_\sigma|A_\sigma|^2\,e^{-2t\,\mathrm{Re}\,\lambda^{\rm LC}_\sigma}}{\sum_\sigma|A_\sigma|^2} .
  \label{eq:cs:RLC}
\end{equation}
Physically, every colour dipole radiates independently. This is the colour treatment of standard parton showers
and of many resummations \cite{Platzer:2012np,Nagy:2012bt,Hamilton:2020rcu}. The Coulomb phase drops out of
eq.~\eqref{eq:cs:RLC}. Reconnections of colour lines, the corresponding interference between flows, and the
$1/\Nc$ corrections to the Casimirs are all neglected; for $q\bar q\to q\bar q$ they are suppressed by
$1/\Nc^2\approx0.1$ relative to the leading terms, but their number grows with $n$.

We also consider the improvement of ref.~\cite{Nagy:2012bt}, which we call LC+. It keeps the flows and their
independent evolution, but shares the exact Casimir of each parton among its colour-connected neighbours: a quark or
antiquark, which has one neighbour, contributes $s_q=C_F$, and a gluon, which has two, contributes $s_g=C_A/2$ to each of
its pairs. The flow eigenvalue becomes
\begin{equation}
  \lambda^{+}_\sigma=-\sum_{(ij)\ \text{connected in}\ \sigma}\tfrac12\,(s_i+s_j)\,\omega_{ij},
  \label{eq:cs:lambdaLCp}
\end{equation}
so that a directly connected $q\bar q$ pair evolves with $C_F$ instead of $\Nc/2$, and the evolved flows are combined with
their exact overlaps,
\begin{equation}
  R_{\rm LC+}(t)=\frac{\bigl\|\sum_\sigma A_\sigma\,e^{-t\lambda^+_\sigma}\ket\sigma\bigr\|^2}{\bigl\|\sum_\sigma A_\sigma\ket\sigma\bigr\|^2},
  \label{eq:cs:RLCp}
\end{equation}
with real weights (no Coulomb phases). In full colour (FC) we evolve with eq.~\eqref{eq:cs:evolution} without approximation.

\subsection{Why an orthogonal basis}\label{sec:cs:why}

Colour flows are intuitive, but they are an inconvenient basis for an exact calculation. Their scalar products are non-zero at
every order in $1/\Nc$, so computing $R(t)$ requires the full matrix of overlaps. For $n\ge\Nc+1$ gluons they
are also linearly dependent at fixed $\Nc$, so the coefficients
$A_\sigma$ are not unique. Finally, the number of single-trace flows of section~\ref{sec:cs:lc}, $(n-1)!$, exceeds the dimension of
$\mathcal V_n$ by a large factor: for $n=15$ it is $8.7\times10^{10}$ against $8.3\times10^8$.

In an orthonormal basis $\{\ket{u_a}\}$ of $\mathcal V_n$ all three problems disappear. A colour state is a
vector $x$ of length $\dim\mathcal V_n$ with $x_a=\braket{u_a}{\mathcal M}$, the colour-summed square is
$\sum_a|x_a|^2$, and the evolution is a matrix exponential,
\begin{equation}
  x(t)=e^{-t\Gamma}x(0),\qquad R(t)=\frac{\sum_a|x_a(t)|^2}{\sum_a|x_a(0)|^2},\qquad
  \Gamma_{ab}=\bra{u_a}\Gamma\ket{u_b} .
  \label{eq:cs:orthoR}
\end{equation}
For real weights $\Gamma_{ab}$ is a real symmetric matrix, because the basis vectors constructed below have real
coefficients; in general it is complex symmetric.
The remaining task is to construct such a basis for any process and to apply $\Gamma$ to vectors with up to
$10^9$ components. Section~\ref{sec:bases} does the first, section~\ref{sec:evolution} the second.
\section{Orthogonal multiplet bases}\label{sec:bases}

This section constructs an orthogonal basis of $\mathcal V_n$ for any process with quarks, antiquarks and
gluons. After a short review of earlier colour bases in section~\ref{sec:history}, we explain the construction
in three steps: the labels (section~\ref{sec:bratteli}), the operators that tell the basis vectors apart
(section~\ref{sec:casimirs}), and the treatment of multiplets that occur more than once
(section~\ref{sec:copies}). Section~\ref{sec:norms} states the orthogonality theorem and gives the norms in
closed form, and section~\ref{sec:local} shows that colour operators act locally on the labels, which is what
makes the evolution of section~\ref{sec:evolution} possible. The example $q\bar q\to q\bar q$ of
section~\ref{sec:colourspace} is continued throughout.

\subsection{Colour bases before this work}\label{sec:history}

The decomposition of amplitudes into traces of generators goes back to the Chan--Paton factors and the
large-$\Nc$ expansion \cite{Paton:1969je,tHooft:1973alw}. Trace bases and colour-ordered amplitudes became the
standard tool for multi-parton tree amplitudes \cite{Parke:1986gb,Mangano:1987xk,Mangano:1988kk,Kleiss:1988ne,
Berends:1987me,Mangano:1990by}, together with the adjoint basis of refs.~\cite{DelDuca:1999iql,DelDuca:1999rs},
the colour-flow decomposition \cite{Maltoni:2002mq,Kilian:2012pz} and their extensions to quarks
\cite{Melia:2013bta,Melia:2013epa,Melia:2015ika,Johansson:2015oia,Ochirov:2016ewn,Ochirov:2019mtf}. These bases
are well suited to amplitudes, but they are not orthogonal and, for many partons, not linearly independent.

Soft-gluon resummation needs the soft anomalous dimension as a matrix in colour space. Such matrices were first
computed in refs.~\cite{Botts:1989kf,Sotiropoulos:1993rd}. The programme of
refs.~\cite{Kidonakis:1996aq,Kidonakis:1997gm,Contopanagos:1996nh,Kidonakis:1998bk,Kidonakis:1998nf} worked
them out for all $2\to2$ processes in $s$-channel bases (in which the two incoming partons are projected onto
irreducible representations), which are orthogonal for four partons, and applied them
to heavy-quark and jet production and to rapidity gaps
\cite{Bonciani:1998vc,Laenen:1998qw,Oderda:1998en,Berger:2001ns}. Over the following two decades this line of
work was extended to many further processes and to two and three loops
\cite{Kidonakis:1999ze,Kidonakis:2001nj,Kidonakis:2003tx,Kidonakis:2006bu,Kidonakis:2009ev,Kidonakis:2010ux,
Kidonakis:2010dk,Kidonakis:2019nqa,Kidonakis:2020gxo,Kidonakis:2021yrh,Kidonakis:2023juy}. Orthogonal
$s$-channel bases for the threshold resummation of heavy coloured pair production, including the all-order
colour structure, were constructed in refs.~\cite{Kulesza:2008jb,Kulesza:2009kq,Beneke:2009rj,Beneke:2010da,
Ahrens:2010zv}, and the two-loop structure
of the massive case in refs.~\cite{Mitov:2009sv,Ferroglia:2009ep,Ferroglia:2009ii}. For rapidity gaps and
large-angle soft radiation, the matrices for $2\to2$ and $2\to3$ processes, their symmetries and a general
recipe were given in refs.~\cite{Forshaw:2005sx,Seymour:2005ze,Kyrieleis:2005dt,Dokshitzer:2005ek,
Dokshitzer:2005ig,Sjodahl:2008fz,Seymour:2008xr,Sjodahl:2009wx,DuranDelgado:2011tp}. The matrices for
$t\bar t+\text{jet}$ were obtained in ref.~\cite{Chargeishvili:2026tqa}.

For processes with more partons, orthogonal bases are no longer obtained by inspection. Birdtrack diagrams
\cite{Cvitanovic:1976am,Cvitanovic:1980bu,Cvitanovic:2008zz} provide a graphical calculus for group-theory
projectors, and Hermitian Young projectors \cite{Keppeler:2013yla,Alcock-Zeilinger:2016sxc,
Alcock-Zeilinger:2016cva,Alcock-Zeilinger:2018fha,Keppeler:2017kwt} make it possible to project onto irreducible
representations of products of quarks and antiquarks. On this basis, ref.~\cite{Keppeler:2012ih} constructed
orthogonal multiplet bases for any number of partons, with labels that do not depend on $\Nc$. The construction
was developed further in refs.~\cite{Du:2015apa,Sjodahl:2015qoa,Sjodahl:2018cca,Keppeler:2023msu,
Sjodahl:2024fqn,Keppeler:2025aax}. An independent implementation that generates such bases automatically for
general $2\to n$ processes was presented in ref.~\cite{Chargeishvili:2024pnq}; it uses the Z3
satisfiability-modulo-theories solver \cite{deMoura:2008} for the decomposition into irreducible representations
and FORM \cite{Vermaseren:2000nd} for the tensor algebra. General-purpose packages for colour algebra are also available \cite{Sjodahl:2012nk,
Sjodahl:2014opa}. Projectors built from Casimir operators
were studied recently in refs.~\cite{Cougoulic:2025wpl,Peigne:2026xtz}. Multiplet bases are used for colour
matrix-element corrections in parton showers \cite{Platzer:2018pmd} and for colour sums of multi-gluon
amplitudes \cite{Bolinder:2025gbj,Frederix:2026cay}.

The construction of this section is independent of these works. It does not use birdtracks or Young
projectors. Instead, every basis vector is fixed by the eigenvalues of a chain of Casimir operators along the
parton ordering. In mathematics these operators are the Jucys--Murphy elements of the symmetric group and of the
walled Brauer algebra \cite{Jucys:1974,Murphy:1981,Okounkov:1996,Brundan:2012,Jung:2020,Grinko:2022mpd,
Grinko:2023cdq}; to our knowledge they have not been used to construct colour bases. Section~\ref{sec:casimirs} explains the connection in physics terms; no prior knowledge of these
objects is needed. Where the two
constructions can be compared, the resulting vectors agree with those of ref.~\cite{Keppeler:2012ih} up to
conventions (section~\ref{sec:ks}).

\subsection{Adding one parton at a time}\label{sec:bratteli}

The colour part of every amplitude is invariant under colour rotations: by colour conservation,
eq.~\eqref{eq:cs:conservation}, the $n$ partons together form a colour singlet, and $\mathcal V_n$ consists of exactly
these singlet tensors. A basis of $\mathcal V_n$ is therefore a list of the independent ways in which $n$ partons can
combine into a singlet. We enumerate these ways by coupling the partons one at a time, in a fixed order, and recording
after each parton the colour multiplet, or irreducible representation of $\mathrm{SU}(\Nc)$, carried by the partons
coupled so far. The intermediate multiplets need not be singlets; only the last one must be. They are useful as
labels because two vectors in which the first $k$ partons are in different multiplets are automatically orthogonal,
just as states of several spins with different intermediate total spins are orthogonal when angular momenta are
coupled one at a time. For $q_1\bar q_2q_3\bar q_4$ at $\Nc=3$, the steps are
\begin{equation}
  \mathbf 3\;\xrightarrow{\ \otimes\bar{\mathbf 3}\ }\;\mathbf 1\oplus\mathbf 8\;\xrightarrow{\ \otimes\mathbf 3\ }\;
  \mathbf 3\;(\text{from }\mathbf 1\text{ and from }\mathbf 8)\oplus\cdots\;\xrightarrow{\ \otimes\bar{\mathbf 3}\ }\;\mathbf 1 .
\end{equation}
Here $\cdots$ stands for the multiplets $\bar{\mathbf 6}$ and $\mathbf{15}$ of $\mathbf 8\otimes\mathbf 3$, from which the
last antiquark cannot reach a singlet, so they do not contribute to $\mathcal V_4$. Only two routes reach the final
singlet. They pass through $(\mathbf 3,\mathbf 1,\mathbf 3,\mathbf 1)$
and $(\mathbf 3,\mathbf 8,\mathbf 3,\mathbf 1)$. The first gives $\ket{c_1}$ of section~\ref{sec:cs:space}.
In the second, the pair $q_1\bar q_2$ is an octet.

\begin{figure}[t]
  \centering
\begin{tikzpicture}[
    irr/.style={draw,rounded corners=2pt,inner sep=2.5pt,minimum width=6mm,font=\small,fill=white},
    e/.style={thick,draw=MidnightBlue!80},
    ee/.style={draw=BurntOrange!80!black,line width=0.7pt,double=white,double distance=1.2pt},
    st/.style={font=\small,text=black}]
  \begin{scope}
    \node[irr] (a0) at (0,0)   {$\mathbf 1$};
    \node[irr] (a1) at (1.4,0) {$\mathbf 3$};
    \node[irr] (a2s) at (2.8,0.7)  {$\mathbf 1$};
    \node[irr] (a2o) at (2.8,-0.7) {$\mathbf 8$};
    \node[irr] (a3) at (4.2,0) {$\mathbf 3$};
    \node[irr] (a4) at (5.6,0) {$\mathbf 1$};
    \draw[e] (a0)--(a1) (a1)--(a2s) (a1)--(a2o) (a2s)--(a3) (a2o)--(a3) (a3)--(a4);
    \foreach \x/\l in {0/{},1.4/q_1,2.8/\bar q_2,4.2/q_3,5.6/\bar q_4} \node[st] at (\x,-2.15) {$\l$};
    \node[font=\small] at (2.8,-2.7) {(a)};
  \end{scope}
  \begin{scope}[xshift=7.6cm]
    \node[irr] (b0) at (0,0)   {$\mathbf 1$};
    \node[irr] (b1) at (1.5,0) {$\mathbf 8$};
    \node[irr] (b27) at (3,1.5)  {$\mathbf{27}$};
    \node[irr] (b10) at (3,0.75) {$\mathbf{10}$};
    \node[irr] (b10b) at (3,0)   {$\overline{\mathbf{10}}$};
    \node[irr] (b8) at (3,-0.75) {$\mathbf 8$};
    \node[irr] (b1m) at (3,-1.5) {$\mathbf 1$};
    \node[irr] (b3) at (4.5,0) {$\mathbf 8$};
    \node[irr] (b4) at (6,0)   {$\mathbf 1$};
    \draw[e] (b0)--(b1) (b1)--(b27) (b1)--(b10) (b1)--(b10b) (b1)--(b1m)
             (b27)--(b3) (b10)--(b3) (b10b)--(b3) (b1m)--(b3) (b3)--(b4);
    \draw[ee] (b1)--(b8);
    \draw[ee] (b8)--(b3);
    \node[irr] at (b8) {$\mathbf 8$};
    \foreach \x/\l in {0/{},1.5/g_1,3/g_2,4.5/g_3,6/g_4} \node[st] at (\x,-2.15) {$\l$};
    \node[font=\small] at (3,-2.7) {(b)};
  \end{scope}
\end{tikzpicture}
  \caption{Bratteli diagrams at $\Nc=3$. The leftmost node represents the empty set of partons.
  Each subsequent node gives the multiplet $X_k$ after the parton shown below it. (a) For
  $q_1\bar q_2q_3\bar q_4$ the two paths to the final singlet pass through a singlet or an octet
  after parton 2. (b) For four gluons the numbers give the multiplet dimensions. Each orange double
  line stands for two $\mathbf 8\to\mathbf 8$ couplings, antisymmetric ($f$) and symmetric ($d$),
  as defined in section~\ref{sec:copies}. The two choices on each double line give four routes through
  the middle octet. The four other paths bring the total to eight basis vectors.}
  \label{fig:bratteli}
\end{figure}
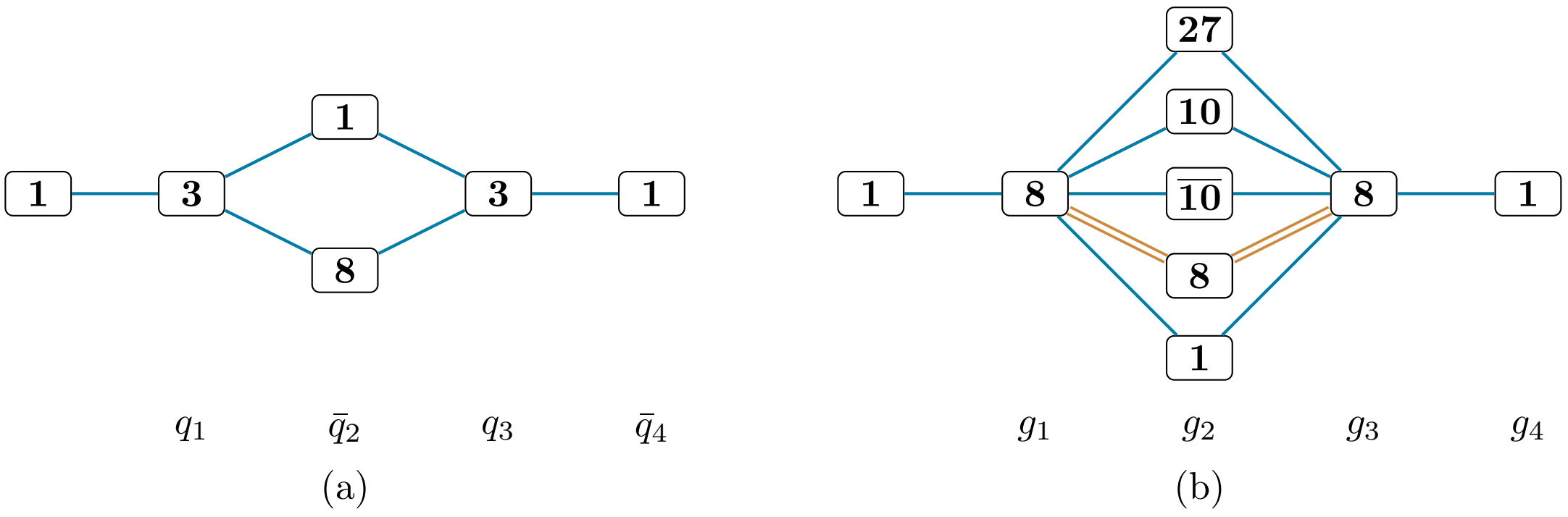

In general we write $X_k$ for the multiplet of the first $k$ partons. Its successive values define a
\emph{path},
\begin{equation}
  (X_1,X_2,\dots,X_n=\mathbf 1).
  \label{eq:b:path}
\end{equation}
A Bratteli diagram draws the allowed couplings, with one layer per parton and one edge per coupling
\cite{Bratteli:1972}. Figure~\ref{fig:bratteli} shows the four-parton example and the corresponding
diagram for four gluons. Counting routes to the singlet gives the dimensions in section~\ref{sec:cs:space}.
When two couplings join the same multiplets, we count each separately (section~\ref{sec:copies}).
We use this count to locate individual basis vectors in section~\ref{sec:local}.

To label multiplets independently of $\Nc$, we use a pair of Young diagrams $\irr{\alpha}{\beta}$
\cite{King:1971rs,Stembridge:1987}. A diagram consists of left-aligned rows of boxes, each row at most
as long as the one above it. Each box represents one uncontracted index. The shape records the symmetry
under exchange of indices of the same type. We need two shapes because $\alpha$ describes antiquark
indices and $\beta$ quark indices. Writing $\varnothing$ for an empty diagram, the fundamental is
$\irr{\varnothing}{\yd1}$, the antifundamental $\irr{\yd1}{\varnothing}$ and the singlet
$\irr{\varnothing}{\varnothing}$. The adjoint, the octet at $\Nc=3$, is $\irr{\yd1}{\yd1}$ for every
$\Nc$. We add the numbers of non-empty rows in $\alpha$ and $\beta$. If their sum exceeds $\Nc$, the
vector vanishes and we drop the label. For pure-gluon singlet paths at $\Nc=3$, this first happens
with four gluons.

For a gluon we write the adjoint index $a$ as two indices $(i,\bar\jmath)$ using the traceless matrix
$(t^a)_{i\bar\jmath}$. The decomposition $\mathbf N_c\otimes\bar{\mathbf N}_c=\mathbf 1\oplus\text{adjoint}$
makes this replacement exact once we remove the singlet trace proportional to $\delta_{i\bar\jmath}$.
We can therefore couple a gluon in two steps, first its quark leg and then its antiquark leg.
Each leg changes one box. A quark adds a box to $\beta$ or removes one from $\alpha$.
An antiquark adds a box to $\alpha$ or removes one from $\beta$.

As an example we couple a second gluon to an octet $\irr{\yd1}{\yd1}$, i.e.\ we decompose
$\mathbf 8\otimes\mathbf 8$, as in the middle of figure~\ref{fig:bratteli}(b). The quark leg either adds a box to $\beta$,
in the first row or in a new second row, or removes the box of $\alpha$. At $\Nc=3$ this is the decomposition
$\mathbf 8\otimes\mathbf 3=\mathbf{15}\oplus\bar{\mathbf 6}\oplus\mathbf 3$, with
$\mathbf{15}=\irr{\yd1}{\yd2}$, $\bar{\mathbf 6}=\irr{\yd1}{\yd{1,1}}$ and $\mathbf 3=\irr{\varnothing}{\yd1}$.
The antiquark leg then adds a box to $\alpha$ or removes one from $\beta$:
\begin{equation}
  \begin{aligned}
  \mathbf{15}\otimes\bar{\mathbf 3}&=\irr{\yd2}{\yd2}\oplus\irr{\yd{1,1}}{\yd2}\oplus\irr{\yd1}{\yd1}
    =\mathbf{27}\oplus\mathbf{10}\oplus\mathbf 8,\\
  \bar{\mathbf 6}\otimes\bar{\mathbf 3}&=\irr{\yd2}{\yd{1,1}}\oplus\irr{\yd1}{\yd1}
    =\overline{\mathbf{10}}\oplus\mathbf 8,\\
  \mathbf 3\otimes\bar{\mathbf 3}&=\irr{\yd1}{\yd1}\oplus\irr{\varnothing}{\varnothing}=\mathbf 8\oplus\mathbf 1 .
  \end{aligned}
  \label{eq:b:88}
\end{equation}
The dimensions add up to $72=8\times9$, the product of the octet with the $\Nc^2=9$ states of a quark--antiquark
pair. Three features of this example recur throughout. First, the label $\irr{\yd{1,1}}{\yd{1,1}}$, which would also
follow from $\bar{\mathbf 6}$, has four rows and exists only for $\Nc\ge4$; it is the ninth basis vector of four gluons
in table~\ref{tab:dims}, which is absent at $\Nc=3$. Second, three of the routes in eq.~\eqref{eq:b:88} end in an octet, but a gluon
produces only two. The reason is that the two legs together carry $\Nc^2=9$ states,
$\mathbf 3\otimes\bar{\mathbf 3}=\mathbf 8\oplus\mathbf 1$, of which only the octet is a gluon; the singlet, the trace
$\delta_{i\bar\jmath}$, is not. Coupled to the first octet, this singlet gives one more octet,
$\mathbf 8\otimes\mathbf 1=\mathbf 8$, which we remove. This leaves the two octets of
$\mathbf 8\otimes\mathbf 8=\mathbf 1\oplus\mathbf 8\oplus\mathbf 8\oplus\mathbf{10}\oplus\overline{\mathbf{10}}\oplus\mathbf{27}$,
the antisymmetric and symmetric couplings $f$ and $d$ of section~\ref{sec:copies}. Third, $\mathbf{10}$ and
$\overline{\mathbf{10}}$ are reached through different multiplets after the quark leg, $\mathbf{15}$ and $\bar{\mathbf 6}$.
We use the multiplet after the quark leg to construct the vectors, and this is what separates the two
(section~\ref{sec:casimirs}). The path of eq.~\eqref{eq:b:path} itself records only $X_k$ after both legs of gluon~$k$.

\subsection{Telling multiplets apart}\label{sec:casimirs}

A path is only a label. To turn it into a colour tensor we need an operator whose eigenvalue tells us which multiplet
the first $k$ partons are in, in the same way as the total spin $\mathbf J^2$ of the first $k$ spins identifies the
intermediate couplings of angular momenta. The natural choice is the quadratic Casimir of the first $k$ partons, the
square of their total charge $\Tb_{\le k}=\Tb_1+\dots+\Tb_k$. It is more convenient to use the difference between
consecutive Casimirs. With $\Tb_{<k}$ the total charge of the partons before $k$, this is the colour correlation between
parton $k$ and all earlier partons,
\begin{equation}
  D_k=\Tb_k\cdot\Tb_{<k}=\frac12\Bigl(\Tb_{\le k}^2-\Tb_{<k}^2-\Tb_k^2\Bigr).
  \label{eq:b:D}
\end{equation}
The operators $D_1,\dots,D_n$ are Hermitian and commute with each other. Write $C(X)$ for the quadratic Casimir of
multiplet $X$, $X_0=\mathbf 1$ for the multiplet before the first parton and $C_k=\Tb_k^2$. On a vector with path
$(X_1,\dots,X_n)$ the eigenvalue at step $k$ is
\begin{equation}
  D_k=\frac12\bigl(C(X_k)-C(X_{k-1})-C_k\bigr).
  \label{eq:b:Deig}
\end{equation}
For the pair $q_1\bar q_2$ of the running example, $X_1=\mathbf 3$ and $C_2=C_F$ is the Casimir of the antiquark,
so eq.~\eqref{eq:b:Deig} gives $D_2=\Tb_1\cdot\Tb_2=\frac12(0-C_F-C_F)=-C_F$ for $X_2=\mathbf 1$ and
$D_2=\frac12(C_A-C_F-C_F)=1/(2\Nc)$ for $X_2=\mathbf 8$. The two values differ, so $D_2$ tells the two paths apart.

For gluons this is not yet enough. After a complete gluon the Casimir can take the same value for different
multiplets: the $\mathbf{10}$ and the $\overline{\mathbf{10}}$ in $\mathbf 8\otimes\mathbf 8$ both have $C=6$ at
$\Nc=3$. The remedy is the splitting of section~\ref{sec:bratteli}: each gluon is coupled as a quark leg followed by an
antiquark leg, and we evaluate eq.~\eqref{eq:b:Deig} after every leg, with $k$ now counting legs and $C_k=C_F$ for
every leg.

The example of eq.~\eqref{eq:b:88} shows how this works. Let the first gluon be in the octet, $C(\mathbf 8)=3$, and
couple the quark leg of the second gluon. At $\Nc=3$ eq.~\eqref{eq:b:Deig} gives
\begin{equation}
  D=\tfrac12\bigl(C(Y)-3-\tfrac43\bigr)=+\tfrac12,\;-\tfrac12,\;-\tfrac32
  \qquad\text{for}\qquad Y=\mathbf{15},\;\bar{\mathbf 6},\;\mathbf 3 ,
\end{equation}
with $C(\mathbf{15})=\frac{16}3$, $C(\bar{\mathbf 6})=\frac{10}3$ and $C(\mathbf 3)=\frac43$. The three outcomes are
distinguished. Since the $\mathbf{10}$ is reached through the $\mathbf{15}$ and the $\overline{\mathbf{10}}$ through the
$\bar{\mathbf 6}$, the measurement after the quark leg also separates these two, although their Casimirs after the whole
gluon coincide.

This is true in general because a single leg changes the multiplet by one box, and the change of the Casimir depends
only on which box it is. We characterise a box by its \emph{content} $c$, its column number minus its row number,
counting from the top left. Let $s_k$ be the number of quark legs minus the number of antiquark legs before leg $k$,
and define the shifted operator $\ell_k=2D_k+s_k/\Nc$ for a quark leg and $\ell_k=2D_k-s_k/\Nc$ for an antiquark leg.
Its eigenvalue is
\begin{equation}
  \ell_k=\begin{cases} c & \text{if leg $k$ adds a box of content $c$},\\
  -(\Nc+c) & \text{if leg $k$ removes a box of content $c$},\end{cases}\;,
  \label{eq:b:content}
\end{equation}
where a quark leg adds boxes to $\beta$ and removes them from $\alpha$, and an antiquark leg does the reverse
(lemma~\ref{lem:p:spec}). In the example above, $s_k=0$ and the three outcomes add a box of content $+1$, add a box of
content $-1$, and remove a box of content $0$, giving $\ell_k=+1$, $-1$ and $-3$, i.e.\ $D=\ell_k/2$ as found. For the
pair $q_1\bar q_2$ one has $s_2=1$, and the singlet and the octet give $\ell_2=-\Nc$ and $0$, in agreement with the values
of $D_2$ above. The boxes that can be added to a diagram all have different contents, and so have the boxes that can
be removed. The possible outcomes of a leg therefore always have different eigenvalues (as functions of $\Nc$), and two vectors that differ
anywhere along their sequence of leg multiplets differ in at least one eigenvalue of a Hermitian operator and are
orthogonal.

The operator $\ell_k$ also has a simple form. By the Fierz identity, $2\Tb_k\cdot\Tb_l$ is, up to a constant, the
exchange of the two indices if both are quark (or both antiquark) indices, and minus their contraction if one is a
quark and the other an antiquark index. The shift in $\ell_k$ removes these constants, so that $\ell_k$ is the sum of
the exchanges of index $k$ with all earlier indices of the same type minus the sum of its contractions with all earlier
indices of the opposite type (appendix~\ref{app:p:setting}). For quarks alone this is the Jucys--Murphy element of the
symmetric group \cite{Jucys:1974,Murphy:1981,Okounkov:1996}, and with antiquarks the Jucys--Murphy element of the walled
Brauer algebra \cite{Brundan:2012,Jung:2020,Grinko:2023cdq}; eq.~\eqref{eq:b:content} is their classical spectrum.

The basis vectors are constructed by projection. We start from a \emph{seed}, a product of Kronecker deltas that has a
non-zero component along the chosen leg path: each delta joins the leg that adds a box to the leg that later removes
it (appendix~\ref{app:p:vectors}). At each leg $k$ we keep the component with the chosen eigenvalue $\lambda_k$ and
remove the components with the other possible eigenvalues $\lambda'$ by the Lagrange interpolation polynomial
\begin{equation}
  p_k=\prod_{\lambda'\neq\lambda_k}\frac{D_k-\lambda'}{\lambda_k-\lambda'} ,
  \label{eq:b:proj}
\end{equation}
which equals one on the chosen component and zero on the others. For the octet path of the running example the seed
is $\ket{c_2}$, and
\begin{equation}
  p_2\ket{c_2}=\frac{D_2+C_F}{\Nc/2}\ket{c_2}=\ket{c_2}-\frac1\Nc\ket{c_1},
\end{equation}
the octet vector found in section~\ref{sec:cs:space}. Only the eigenvalues of $D_k$ enter, so no representation
matrices are ever needed, and the coefficients of the resulting colour tensors are rational functions of $\Nc$.

\subsection{Repeated multiplets}\label{sec:copies}

Coupling a gluon to a multiplet $X$ can return the same multiplet $X$ more than once, while any other outcome $Y\ne X$
occurs at most once. The simplest case is the pair of octets in
$\mathbf 8\otimes\mathbf 8=\mathbf 1\oplus\mathbf 8\oplus\mathbf 8\oplus\mathbf{10}\oplus\overline{\mathbf{10}}
\oplus\mathbf{27}$, which correspond to the antisymmetric colour tensor $f^{abc}$ and the symmetric tensor $d^{abc}$.
We call a gluon step $X\to X$ a \emph{copy step} and record which coupling is used alongside the path. For generic
$\Nc$ the number of copies equals the number of corners of the two diagrams of $X$, where a corner is a box that can
be removed while keeping a valid diagram. The octet $\irr{\yd1}{\yd1}$ and the $\mathbf{27}=\irr{\yd2}{\yd2}$ have two
corners and two copies; a multiplet such as $\irr{\yd1}{\yd{2,1}}$ has three.

The leg-by-leg projection of section~\ref{sec:casimirs} does not resolve these copies by itself. In the example of
eq.~\eqref{eq:b:88}, the octet is reached along three orthogonal leg routes, through the $\mathbf{15}$, the
$\bar{\mathbf 6}$ and the $\mathbf 3$. None of the three is a coupling of a gluon, because none of them is traceless in
the new quark--antiquark pair; the two gluon octets are two combinations of the three routes, orthogonal to the
trace. We therefore construct the copies differently. Starting from the vector with gluon $g$ omitted, we insert the
two legs of $g$ as a singlet pair (operation $e$), let the quark leg interact $r$ times with the earlier partons
through the colour correlation $\Tb_{g_q}\cdot\Tb_{<g}$, and finally remove the trace of the pair with the projector
$P_0$:
\begin{equation}
  \Phi_r\propto P_0\,(\Tb_{g_q}\cdot\Tb_{<g})^r\,e,\qquad r=1,2,\dots.
  \label{eq:b:copies}
\end{equation}
The singlet pair alone carries no colour; each application of the correlation couples it to the colour charge of the
earlier partons. One application attaches the new gluon through a single generator of the earlier partons, which for
an octet is the $f$ coupling. Two applications attach it through a symmetrised product of two generators, which after
subtraction of the $f$ part is the $d$ coupling, and higher powers give couplings through symmetrised products of more
generators, which we call $d_3,d_4,\dots$. Such polynomials in the generators were studied in ref.~\cite{Okubo:1977mx}. The candidates are made orthogonal by Gram--Schmidt orthogonalisation in the order $f,d,\dots$.

The step that makes this efficient is that the scalar products of the candidates do not depend on the rest of the
path. The candidates are polynomials in the total charge $\Tb_{<g}$ of the earlier partons, which act on a vector in
the multiplet $X$, and by Schur's lemma such an operator acts on all vectors of an irreducible multiplet in the same
way. The overlaps of the candidates, divided by the squared norm of the starting vector, are therefore numbers that
depend only on $X$ and $\Nc$. We compute them once per multiplet and use the same Gram--Schmidt coefficients for every
path through $X$ (appendix~\ref{app:proofs}). For the octet, for instance, the combination $4P_0(\Tb_{g_q}\cdot\Tb_{<g})^2e+\Nc\,\Phi_1$ is already
orthogonal to $\Phi_1$ and gives the $d$ coupling (appendix~\ref{app:p:norm}).

\subsection{Orthogonality and norms}\label{sec:norms}

The vectors constructed in this way are orthogonal for every process.

\begin{theorem}\label{thm:orth}
For any ordering of any number of quarks, antiquarks and gluons, two basis vectors with different paths or with
the same path and different copy names are orthogonal. The statement holds for symbolic $\Nc$ and for every
integer $\Nc$ at which the vectors are defined.
\end{theorem}

\noindent Different paths put the first $k$ partons in different irreducible representations for at least
one $k$. For different copies of the same path, the universal overlap factors ensure that the subtraction
at the copy step makes the full vectors orthogonal (appendix~\ref{app:proofs}).

For quarks and antiquarks, let $\ket v$ be the vector obtained by projecting the seed with coefficient
one. To compute its squared norm, first identify each leg $k$ that adds a box. It contributes the ratio
of multiplet dimensions after and before that leg. For each box $x$, draw an arc from the leg $a_x$
that adds it to the leg $r_x$ that next empties the same position. Two arcs in the same diagram cross
if $a_x<a_y<r_x<r_y$, with $x$ the earlier addition. Writing $c_x$ for the content of box $x$, we obtain
\begin{equation}
  \braket{v}{v}=\prod_{k\ \text{adds a box}}\frac{\dim X_k}{\dim X_{k-1}}
  \prod_{\{x,y\}\ \text{crossing}}\Bigl(1-\frac1{(c_x-c_y)^2}\Bigr).
  \label{eq:b:norm}
\end{equation}
Each crossing contributes once. The first crossings occur for three quark--antiquark pairs
(figure~\ref{fig:arcs} and appendix~\ref{app:p:norm}).

For $q_1\bar q_2q_3\bar q_4$ the singlet path adds boxes at legs 1 and 3. Its arcs join 1 to 2 and
3 to 4. The octet path adds boxes at legs 1 and 2. Its arcs join 1 to 4 and 2 to 3.
Neither path has a crossing, so the dimension ratios alone give
\begin{equation}
  \braket{c_1}{c_1}=\Nc\cdot\Nc=\Nc^2,\qquad
  \bigl\|\ket{c_2}-\ket{c_1}/\Nc\bigr\|^2=\Nc\cdot\frac{\Nc^2-1}{\Nc}=\Nc^2-1,
\end{equation}
in agreement with direct colour contraction. A complete worked example, the basis, norms and soft anomalous
dimension for $q\bar q\to gg$, is given in appendix~\ref{app:example}.

A gluon enters eq.~\eqref{eq:b:norm} through its two legs. Each copy step multiplies the squared norm by a universal
factor, so the norm of any basis vector is a product of $O(n)$ local factors (appendix~\ref{app:proofs}). That the
non-vanishing vectors also span $\mathcal V_n$ is checked by counting them (section~\ref{sec:val:basis}). The individual factors are known
from the representation theory of the symmetric group, which permutes like indices, and of the
walled Brauer algebra, which also allows contractions between a quark and an antiquark index
\cite{James:1979,Nazarov:1996,Mathas:2008,Rui:2008,Si:2023,Grinko:2023cdq}.
To our knowledge this is the first closed form for these explicit colour tensors with arbitrary
ordering of quarks and antiquarks and symbolic $\Nc$.

\subsection{Colour operators act locally}\label{sec:local}

A colour operator acting only on partons $i$ through $j$ can change only the part of the path between
them, provided it commutes with their total colour charge, as dipoles and exchanges of partons do. Exchanging neighbouring partons changes the order in
which we couple them. We use such exchanges to reorder the partons and to construct the hard amplitude
(section~\ref{sec:hard}).

For a dipole $\Tb_i\cdot\Tb_j$ with $i<j$, picture parton $i$ emitting a gluon that parton $j$ absorbs.
Between them we carry the exchanged gluon's adjoint index past each intervening parton. At each step
we express the same colour state in the two coupling orders. The overlaps between these orders are
\emph{recoupling coefficients}, the colour counterparts of angular-momentum $6j$ symbols \cite{Cvitanovic:2008zz}. We tabulate
them once for each set of neighbouring multiplets. For basis vectors $\ket a$ and $\ket b$, computing
$\bra b\Tb_i\cdot\Tb_j\ket a$ takes $O(|i-j|)$ table look-ups, independently of $\dim\mathcal V_n$.

Many paths can differ within the interval between $i$ and $j$. In our dipole calculation, the number
of basis vectors reached from one vector grows by about a factor of five per unit increase in $|i-j|$.
Computing a complete row of $\Gamma$ becomes impractical beyond about twelve gluons. We therefore
apply $\Gamma$ directly to vectors in section~\ref{sec:evolution}, without forming its matrix.

At $\Nc=3$ and $n=10^4$ gluons, one norm takes 1.3~s and one dipole element
$\bra b\Tb_1\cdot\Tb_{n-2}\ket a$ takes 9~s. All $n-1$ neighbouring exchanges on the timed basis
vector take 11~minutes. These calculations use at most 7.9~GB of memory.

We can also count paths and locate individual vectors without listing the basis. We build the
Bratteli diagram at $\Nc=3$ up to $n=1500$ gluons, where the number of singlet paths has 1343 decimal
digits. Given a path with its copy labels, we find its position in a fixed ordering of all such paths
(its rank), or recover it from its rank, in about 15~ms.

\subsection{Relation to earlier multiplet bases}\label{sec:ks}

We implement the construction of refs.~\cite{Keppeler:2012ih,Keppeler:2013yla} independently in explicit
tensors and compare the two bases vector by vector. For $q^m\bar q^m$ with $m\le5$, for three to five gluons and
for $q\bar qgg$, the two bases coincide up to phases, provided the same coupling order and the same convention for
repeated multiplets are used. With other orderings the bases differ by a recoupling, and with the opposite copy
convention by a rotation in the space of copies. For four and five gluons the matrices $\Gamma$ in the two bases
are related by the corresponding unitary transformation, and their spectra agree to a relative precision of $10^{-13}$. We did not test
more than two copies of a representation or more than five gluons.
\section{Exact evolution in full colour}\label{sec:evolution}

In this section we describe how $e^{-t\Gamma}$ is applied to colour states of up to $8.3\times10^8$ components.
Section~\ref{sec:size} quantifies the size of the problem, section~\ref{sec:matvec} explains how $\Gamma$ is
applied to a vector without being stored and how the exponential is computed from such products,
section~\ref{sec:hard} describes the hard colour states, and section~\ref{sec:scaling} gives the performance on
CPUs and on a GPU.

\subsection{Size of the problem}\label{sec:size}

Table~\ref{tab:dims} lists the dimension of $\mathcal V_n$ for $n$ gluons. At $\Nc=3$ it grows by a factor of about
six per gluon and reaches $8.3\times10^8$ at $n=15$. A single complex vector of this length occupies 13~GB, and
the evolution needs a few dozen of them. The number of single-trace colour flows, $(n-1)!$, grows faster still
and exceeds the dimension by a factor of about 100 at $n=15$. For $\Nc\ge n$ no basis vectors vanish, and the
dimension is larger than at $\Nc=3$. It is then the number of derangements of the $n$ gluons: the invariants are spanned by products of traces $\mathrm{Tr}(t^{a_i}t^{a_j}\cdots)$, one for each cycle of a permutation of the gluons, which are independent for $\Nc\ge n$, and a cycle of length one gives $\mathrm{Tr}\,t^a=0$.

\begin{table}[t]
  \centering
  \small
  \begin{tabular}{r rrr r}
    \toprule
    $n$ & $\dim\mathcal V_n$ ($\Nc=3$) & $\dim\mathcal V_n$ ($\Nc\ge n$) & $(n-1)!$ & one vector ($\Nc=3$) \\
    \midrule
     4 & 8               & 9      & 6                     & -- \\
     5 & 32              & 44     & 24                    & -- \\
     6 & 145             & 265    & 120                   & -- \\
     7 & 702             & 1\,854 & 720                   & -- \\
     8 & 3\,598          & 14\,833& 5\,040                & 58~kB \\
     9 & 19\,280         & 133\,496 & 40\,320              & 0.31~MB \\
    10 & 107\,160        & 1\,334\,961 & $3.6\times10^5$       & 1.7~MB \\
    11 & 614\,000        & 14\,684\,570 & $3.6\times10^6$       & 9.8~MB \\
    12 & 3\,609\,760     & 176\,214\,841 & $4.0\times10^7$       & 58~MB \\
    13 & 21\,695\,104    & $2.3\times10^9$ & $4.8\times10^8$       & 347~MB \\
    14 & 132\,904\,369   & $3.2\times10^{10}$ & $6.2\times10^9$       & 2.1~GB \\
    15 & 827\,888\,726   & $4.8\times10^{11}$ & $8.7\times10^{10}$    & 13~GB \\
    \bottomrule
  \end{tabular}
  \caption{Dimension of the colour space of $n$ gluons at $\Nc=3$ and for $\Nc\ge n$, the number of single-trace
  colour flows, and the memory of one complex double-precision vector at $\Nc=3$. The dimensions are the
  numbers of paths in the Bratteli diagram; for $\Nc\ge n$ they are the numbers of derangements of $n$ objects
  (permutations without fixed points).}
  \label{tab:dims}
\end{table}

A direct approach would form the matrix $\Gamma$ and exponentiate it. This is impossible already at moderate $n$:
as explained in section~\ref{sec:local}, $\Gamma$ is not sparse, and its dense form at $n=12$ would need
$2\times10^{14}$ bytes. Instead, we never store $\Gamma$ and compute only its products with vectors.

\subsection{Applying \texorpdfstring{$\Gamma$}{Gamma} without storing it}\label{sec:matvec}

\paragraph{Products $\Gamma x$.}
Section~\ref{sec:local} described $\Tb_i\cdot\Tb_j$ as the emission of a gluon at parton $i$, its transport past the
partons $i+1,\dots,j-1$, and its absorption at parton $j$. We now apply these steps to all components of a vector at
once. To do so, we cut each path after parton $k$ and group the components of $x$ by the representation $Y$ at the
cut. For each $Y$ we arrange the components in a matrix $x_Y$: a row specifies a path of partons $1,\dots,k$ ending in
$Y$, a column a continuation from $Y$ through partons $k+1,\dots,n$, and each entry is the coefficient of one complete
path. An operation on parton $k+1$ changes only the path labels next to the cut, so it acts on these matrices by
multiplication with a small table of section~\ref{sec:local}.

For a fixed emitter $i$, emission at parton $i$ adds an adjoint index, carried by the exchanged gluon, to the
representation at the cut. We then move the cut to the right one parton at a time. At each parton $j>i$ we absorb the
index at $j$ and add the result to the output with the weight $\omega_{ij}$; to reach the next partner we carry the
open index past $j$ (transport). Emission, transport and absorption are all products of small dense matrices. We call
this pass over all partners of $i$ a \emph{sweep}. One product $\Gamma x$ therefore costs $n-1$ sweeps, each of which
touches every component a number of times that depends only on the local tables. At $n=8$ this takes 1.6~ms, compared
to 4.5~s when $\Gamma$ is applied element by element.

The sequence of block operations depends only on the process and on which weights $\omega_{ij}$ are non-zero. It is
computed once, stored as flat index arrays, which we call the \emph{block plan}, and reused for every product. Within
each step all operations write to disjoint memory, so the result is identical, bit by bit, for any number of CPU
threads.

\paragraph{The exponential.}
To compute $x(t)=e^{-t\Gamma}x$ we use Krylov methods \cite{Park:1986wcd,Saad:1992,Hochbruck:1997,Sidje:1998}. For a
not too large step in $t$, $x(t)$ is well approximated by a linear combination of $x,\Gamma x,\dots,\Gamma^{m-1}x$
with $m$ of order 20; these vectors span the \emph{Krylov space}. We orthonormalise them and represent $\Gamma$ in this
$m$-dimensional space by an $m\times m$ matrix, whose exponential is computed exactly. This gives the coefficients of
$x(t)$ in the orthonormalised vectors. An estimate of the error comes for free from the next product. If it exceeds
the tolerance, the step in $t$ is divided into smaller steps, each starting from the result of the previous one.

Where memory allows, we keep all $m$ vectors and orthogonalise each new vector against all earlier ones (the Arnoldi method \cite{Arnoldi:1951}). Otherwise we use the symmetry of $\Gamma$: for real weights it is a real symmetric matrix, and for complex
weights complex symmetric (section~\ref{sec:cs:why}). The new vector then needs to be orthogonalised only against the
two preceding ones (the Lanczos method \cite{Lanczos:1950zz}). In the complex-symmetric case the orthogonality is defined with the bilinear form $u^Tv=\sum_au_av_a$, without complex conjugation \cite{Freund:1992}. A first pass computes the $m\times m$
matrix; a second pass regenerates the vectors and adds them up with the coefficients found in the first. Only four
vectors of full length are then kept in memory, at the price of computing the $m$ products twice. We use a relative
tolerance of $10^{-8}$, so that the results are accurate to seven to eight digits; the error estimates of the largest
runs are well below this tolerance. Further details are given in appendix~\ref{app:numerics}.

\subsection{Hard colour state}\label{sec:hard}

The evolution starts from the colour vector of a tree-level amplitude. For $n$ gluons we use the adjoint
decomposition of refs.~\cite{DelDuca:1999iql,DelDuca:1999rs},
\begin{equation}
  \begin{aligned}
  \ket{\mathcal M}&=\sum_{\sigma\in S_{n-2}}A(1,\sigma_2,\dots,\sigma_{n-1},n)\,\ket{F_\sigma},\\
  \ket{F_\sigma}&=\tilde f^{a_1a_{\sigma_2}b_1}\tilde f^{b_1a_{\sigma_3}b_2}\cdots\tilde f^{b_{n-3}a_{\sigma_{n-1}}a_n},
  \end{aligned}
  \label{eq:hard:ddm}
\end{equation}
where $\tilde f^{abc}=\mathrm{Tr}([t^a,t^b]t^c)$ and the colour-ordered amplitudes $A$ are computed with
Berends--Giele recursion \cite{Berends:1987me,Duhr:2006iq}. The colour factor of the identity ordering is a chain of
$f$ couplings in which every intermediate representation is an octet. This is, up to a constant, a single basis
vector of section~\ref{sec:bases}: the path $(\mathbf 8,\mathbf 8,\dots,\mathbf 8,\mathbf 1)$ with all copies of
type $f$. The colour factors of all other orderings follow from it by exchanges of neighbouring gluons, which act
locally on the labels. The sum over the $(n-2)!$ orderings is organised recursively, so that partial sums are
shared between orderings, much as subcurrents are shared in Berends--Giele recursion. At $\Nc=3$ the projection then
costs fewer than $2\times10^3$ multiply-adds per colour-ordered amplitude for $n\le12$, growing only slowly with $n$, which is less than the
computation of the amplitude itself. For $n\ge13$ a variant with bounded memory is used (appendix~\ref{app:trees}). For processes with one
quark line, $q\bar q+(n-2)$ gluons, the colour factors are the open strings
$(t^{a_{\sigma_1}}\cdots t^{a_{\sigma_{n-2}}})_{i\bar\jmath}$, summed over all $(n-2)!$ orderings, and the projection
proceeds in the same way. As a by-product, the norm of $\ket{\mathcal M}$ is the colour-summed squared
tree amplitude. The details of the projection and a comparison with other methods for colour-summed amplitudes are
given in appendix~\ref{app:trees}.

Other colour states can be specified directly in the basis. Colour flows, for instance, have exact coordinates and
are used for the leading-colour comparisons of section~\ref{sec:lcvsfc}.

\subsection{Implementation and scaling}\label{sec:scaling}

The sweep is implemented in C++ with OpenMP and vector instructions (AVX2 and AVX-512) and, for the GPU, in CUDA,
with the same block plan. Table~\ref{tab:matvec} shows the time for one product $\Gamma x$. On one CPU thread the
time grows by a factor of seven to nine per gluon, while the dimension grows by a factor of about six, and the gain from more threads
saturates at about 64 threads, where the sweep is limited by memory bandwidth. On one A100 GPU the product is
4--10 times faster than on 64 CPU threads and reaches about 60\% of the peak memory bandwidth. GPU and CPU results agree
to $10^{-15}$.

\begin{table}[t]
  \centering
  \small
  \begin{tabular}{r r rrrr r}
    \toprule
     & & \multicolumn{4}{c}{CPU threads} & \\
    $n$ & $\dim\mathcal V_n$ & 1 & 16 & 64 & 128 & A100 \\
    \midrule
    11 & 614\,000       & 0.98~s & 0.26~s & 0.19~s & 0.17~s & 0.020~s \\
    12 & 3\,609\,760    & 8.7~s  & 1.23~s & 0.84~s & 0.83~s & 0.12~s \\
    13 & 21\,695\,104   & 65~s   & 7.5~s  & 4.6~s  & 4.3~s  & 0.97~s \\
    14 & 132\,904\,369  & --     & 51~s   & 29~s   & --     & 7.2~s \\
    \bottomrule
  \end{tabular}
  \caption{Time for one product $\Gamma x$ with complex weights for $n$ gluons at $\Nc=3$. CPU: two AMD EPYC~9645
  processors (192 cores), best of two runs. GPU: one NVIDIA A100 (40~GB), double precision. The machines were shared
  with other jobs, so the numbers indicate throughput rather than idle-machine performance.}
  \label{tab:matvec}
\end{table}

A complete evolution needs a few dozen products per value of $t$. On the A100 an evolution of $n=13$ gluons to 13
values of $t$ took typically 4~minutes and one of $n=14$ gluons about one hour, with less than 8~GB of host memory.
This made it possible to evolve the hard states of 24 kinematic points at $n=13$ and 19 at $n=14$ with the
weights~\eqref{eq:cs:hardweights}, and, separately, the 21 and 20 events at $n=13$ and 14 used in
section~\ref{sec:exclusive}. At $n=15$, with $827\,888\,726$ colour states, the vectors no longer fit into the memory of the GPU, and we evolved
the hard states of four kinematic points exactly on CPU nodes with 64--80 threads. Building the hard state took
1.8--9.4~hours and each value of $t$ took 3--7 hours, with up to 857~GB of memory. To our knowledge, these are the
largest colour spaces in which soft evolution has been computed exactly.

The value $n=15$ is not a limit of the method. Nothing in the construction changes with $n$: the same block plan, sweep
and Krylov iteration are used at $n=4$ and at $n=15$, and the cost of one product grows by a factor of seven to nine per
gluon while the memory grows with the dimension, by a factor of about six. The practical limit is the memory of a
single node or GPU. Since each step of the sweep acts on independent blocks of the vector, the method lends itself
to distributing the vectors over several GPUs or nodes, which would extend the exact evolution to sixteen and more
partons; we have not implemented this.

We also use products with $\Gamma$ to find its lowest eigenvalues at $n=14$ (section~\ref{sec:spectrum}) and to act on compressed colour states for several hundred gluons (section~\ref{sec:beyond}). Before using the evolved amplitudes to test leading colour, we check the basis, the operators and the evolution independently.
\section{Validation}\label{sec:validation}

The results of this paper rest on three ingredients: the basis, the colour operators acting on it, and the
evolution. Each is checked against independent calculations and against published results.
Table~\ref{tab:checks} summarises the checks, and sections~\ref{sec:val:basis}--\ref{sec:val:evol} describe the
most important ones. The operator that adds a real gluon, which is needed only in section~\ref{sec:emission}, is
validated there.

\begin{table}[t]
  \centering
  \small
  \begin{tabular}{>{\raggedright\arraybackslash}p{0.43\textwidth} >{\raggedright\arraybackslash}p{0.34\textwidth} >{\raggedright\arraybackslash}p{0.14\textwidth}}
    \toprule
    check & range & agreement \\
    \midrule
    \multicolumn{3}{l}{\emph{Basis}} \\
    Gram matrix diagonal & every process with $n\le8$ (eight gluons: sampled pairs) & exact \\
    Gram matrix vs symbolic calculation with FORM & $g^2$--$g^5$, $q\bar qg$, $q\bar qgg$, $q\bar qq\bar q$ & exact in $\Nc$ \\
    Closed-form norms, eq.~\eqref{eq:b:norm} & 242\,904 paths, two values of $\Nc$ & exact \\
    Number of basis vectors at fixed $\Nc$ & $g^n$, $n\le9$; $q\bar qg^k$, $k\le6$; $\Nc=2$--5 & exact \\
    \midrule
    \multicolumn{3}{l}{\emph{Colour operators}} \\
    $\Tb_i\cdot\Tb_j$ and $\Gamma$ vs explicit $\mathrm{SU}(\Nc)$ tensors & up to five partons, $\Nc=2$--4 & $10^{-13}$ \\
    $\Gamma$ vs independent trace-basis code & $n\le5$, symbolic and $\Nc=5,6$ & exact \\
    Spectrum of $\Gamma$ vs ColorFull \cite{Sjodahl:2014opa} & $g^{4,5,6}$, $q\bar qg^{3,4}$, $\Nc=3,7$ & $2\times10^{-14}$ \\
    Eigenvalues of $\Gamma$ for $2\to2$ \cite{Kidonakis:1998nf,Dokshitzer:2005ig,Sjodahl:2008fz} & $q\bar q\to q\bar q$, $qq\to qq$, $gg\to gg$; $\Nc=3$--7 & $10^{-10}$ \\
    \midrule
    \multicolumn{3}{l}{\emph{Evolution}} \\
    Krylov vs dense exponential & $n\le7$, real and complex weights & $10^{-9}$ \\
    $R(t)$ vs independent tensor code & $n\le6$, $\Nc=3$--5 & $3\times10^{-14}$ \\
    $R(t)$ vs ColorFull & as for the spectrum & $7\times10^{-14}$ \\
    GPU vs CPU product $\Gamma x$ & $n=4$--13 & $7\times10^{-16}$ \\
    \bottomrule
  \end{tabular}
  \caption{Checks of the colour basis, the dipole operators and the evolved squared amplitude. The Gram matrix contains the scalar products of all pairs of basis vectors. ``Exact'' denotes equality in the arithmetic used for the check, either rational arithmetic or arithmetic modulo two large primes. Numerical entries give the largest relative deviation in the stated range.}
  \label{tab:checks}
\end{table}

\subsection{Basis}\label{sec:val:basis}
Orthogonality is proven (theorem~\ref{thm:orth}); in addition, every step of the proof is tested separately in exact
arithmetic, and the theorem is tested as a whole by computing Gram matrices. For every process with up to seven
partons and every process with eight partons other than eight gluons the Gram matrix is diagonal in exact arithmetic. For eight gluons, with
14\,833 basis vectors at generic $\Nc$, all pairs within blocks and 3000 random pairs between blocks were checked
in arithmetic modulo a prime. The Gram matrices agree with an independent symbolic calculation in FORM
\cite{Vermaseren:2000nd} as rational functions of $\Nc$, and for up to five gluons the decomposition into multiplets agrees with that of
ref.~\cite{Chargeishvili:2024pnq}. The closed-form norms of eq.~\eqref{eq:b:norm} were compared with direct
calculations for all 242\,904 paths of six quark--antiquark pairs. At fixed $\Nc$ some labels describe vanishing
vectors (section~\ref{sec:bratteli}). We checked that the number of non-vanishing vectors equals the dimension of
$\mathcal V_n$, computed independently from the characters of $\mathrm{SU}(\Nc)$, for $n\le9$ gluons and for
$q\bar q$ with up to six gluons at $\Nc=2$--5. For more gluons the number of paths at $\Nc=3$ agrees with the
character calculation up to $n=30$.

\subsection{Colour operators}\label{sec:val:ops}
The matrices of $\Tb_i\cdot\Tb_j$ and of $\Gamma$ were compared with explicit $\mathrm{SU}(\Nc)$ tensors and with an
independent implementation in a trace basis, in which the characteristic polynomials agree symbolically. A fully
independent comparison is provided by ColorFull \cite{Sjodahl:2014opa}, which shares neither code nor basis with our
implementation. For five processes with up to six partons, each at $\Nc=3$ and 7, and for random complex weights, the spectra
of $\Gamma$ agree to $2\times10^{-14}$ and $R(t)$ to $7\times10^{-14}$. For processes with quarks this agreement
required a change of sign of the quark generators in one routine of ref.~\cite{Sjodahl:2014opa}; the change is
fixed uniquely by colour conservation.

For four partons the soft anomalous dimension is known in closed form. We reproduce the eigenvalues of
ref.~\cite{Kidonakis:1998nf} for $q\bar q\to q\bar q$, $qq\to qq$ and $gg\to gg$, and of
refs.~\cite{Dokshitzer:2005ig,Sjodahl:2008fz} for $gg\to gg$, for $\Nc=3$--7, to $10^{-10}$ or better, after
accounting for conventions (normalisation of the generators, overall sign, and a shift proportional to the identity
from the imaginary parts). The characteristic polynomials agree symbolically in $\Nc$ and the kinematic variables for
$q\bar q\to q\bar q$, $qg\to qg$ and $q\bar q\to gg$. For $gg\to gg$ at $\Nc=3$, ref.~\cite{Kidonakis:1998nf} lists
nine eigenvalues, one of which belongs to a vector that vanishes at $\Nc=3$; our eight-dimensional basis correctly
omits it. The normalisation of $\Gamma$ and the relation between $t$ and $\alpha_s\ln(Q/Q_0)$ were checked against
ref.~\cite{Becher:2009cu}.

\subsection{Evolution}\label{sec:val:evol}
An independent code builds the colour generators, $\Gamma$ on the full tensor space $(\Nc^2-1)^n$, the colour-dressed
tree amplitudes and the exponential from scratch, without using any part of the basis construction. For $n=4$--6 and
$\Nc=3$--5, with and without the Coulomb phases, $R(t)$ agrees with our evolution to $3\times10^{-14}$. Beyond the reach
of dense calculations ($n\ge8$) we rely on internal consistency: the GPU and CPU products agree to $7\times10^{-16}$,
and at $n=14$ two different Krylov methods on the GPU and a CPU run agree to seven digits. The four points at $n=15$ were each computed once, with error estimates well below the tolerance; they demonstrate
that the exact evolution is feasible at this size and do not enter the comparisons of section~\ref{sec:lcvsfc}, for which
no leading-colour counterparts were computed. Finally, the
difference between full and leading colour scales as $1/\Nc^2$ for $\Nc=3$, 5 and 7 (section~\ref{sec:exclusive}).

Full-colour soft evolution has been published for up to seven partons \cite{Gerwick:2014gya}, for observables
different from ours; to our knowledge there is no published evolved colour-summed amplitude against which to test our
results for $n\ge8$. The independent checks through $n=7$ test the same local colour operations that we use at higher multiplicity. The separate comparison with the bases of ref.~\cite{Keppeler:2012ih} extends to five gluons (section~\ref{sec:ks}). With these checks in place, we use the exact evolution as the reference for leading colour.
\section{Full colour versus leading colour}\label{sec:lcvsfc}

We first compare full and leading colour for the evolution of the hard amplitude (section~\ref{sec:window}), then for exclusive jet rates and rapidity gaps (sections~\ref{sec:exclusive} and~\ref{sec:gaps}). We use the spectrum of $\Gamma$ to interpret the late-time behaviour and examine the probability that hard jets form a colour singlet (sections~\ref{sec:spectrum} and~\ref{sec:singlet}). These calculations evolve a fixed set of hard partons. In section~\ref{sec:emission} we allow real gluons outside the gap to radiate into it and examine the effect of Coulomb phases.

We use $\Nc=3$ except when checking the large-$\Nc$ limit. We generate events uniformly in phase space \cite{Kleiss:1985gy} for two incoming partons and $n-2$ outgoing jets,
in the partonic centre-of-mass frame with energy $\sqrt s$. All jets are separated by $\Delta R>0.4$ in rapidity and
azimuth and have $p_T>f\sqrt s$ with $f=0.05/(n-2)$. The purpose of the cut is only to ensure that every jet is a hard,
resolved jet; it should not otherwise shape the sample. With $n-2$ jets sharing the energy $\sqrt s$, the average jet
transverse momentum in flat phase space is about $0.8\sqrt s/(n-2)$, so the cut lies at about 6\% of the average at every
multiplicity and removes only configurations with an unusually soft jet; the acceptance of all cuts stays above 20\% up to
$n=13$. Since the weights of the veto observables depend only on the jet directions, the cut affects the results only
through the angular distribution of the jets, and its precise value is unimportant: with a twice harder cut,
$f=0.1/(n-2)$, the medians of $R_{\rm LC}/R_{\rm FC}$ change by at most 1.4\% at $n=10$ (200 events each) and by less
than one standard deviation at $n=11$ (60 events each), for both observables and all three values of $t$. For the
exclusive-jet samples of section~\ref{sec:exclusive} the jets have $|y|<2.5$ and all events have maximally
helicity-violating (MHV) helicities. For the gap samples of
section~\ref{sec:gaps} the jets lie in $0.5<|y|<3$; the helicities are random for the
main samples up to $n=12$ and MHV for the independent sample at $n=12$ and at $n=13$, 14. We give medians and 16--84\% ranges over events with equal weight. Each event contributes its own ratio of evolution factors.

We compute $R_{\rm FC}(t)$ with the exact evolution of section~\ref{sec:evolution}. For gluon trees the sum for $R_{\rm LC}(t)$ in eq.~\eqref{eq:cs:RLC} runs over $(n-1)!$ cyclic orderings. With one quark line it runs over the $(n-2)!$ orderings of the gluons along the line. We evaluate this sum directly for the exclusive-jet samples up to $n=10$ and for the main gap samples up to $n=12$.
For the larger samples, where it has millions of terms per event, we sample the flows with a Markov chain instead, with
a statistical error below 1\% per event; in the figures the two methods are not distinguished. The sampling requires
the squared amplitude for arbitrary orderings in closed form, which is why the sampled events have MHV helicities
(appendix~\ref{app:numerics}). LC+ cannot be sampled in this way: because it keeps the overlaps between the evolved flows, eq.~\eqref{eq:cs:RLCp},
every flow has to be projected onto the basis. We computed it for $n\le9$ in the exclusive-jet samples of section~\ref{sec:exclusive}, which contain
100 events per $n$, and up to $n=13$ for the smaller samples of section~\ref{sec:window}. For gluon trees, colour-diagonal shifts of the weights, including the jet-radius logarithms, cancel in $R_{\rm LC}/R_{\rm FC}$ and $R_{\rm LC+}/R_{\rm FC}$ by colour conservation (section~\ref{sec:cs:gamma}).

\subsection{Validity of leading colour as a function of multiplicity}\label{sec:window}

\begin{figure}[t]
  \centering
  \includegraphics[width=\textwidth]{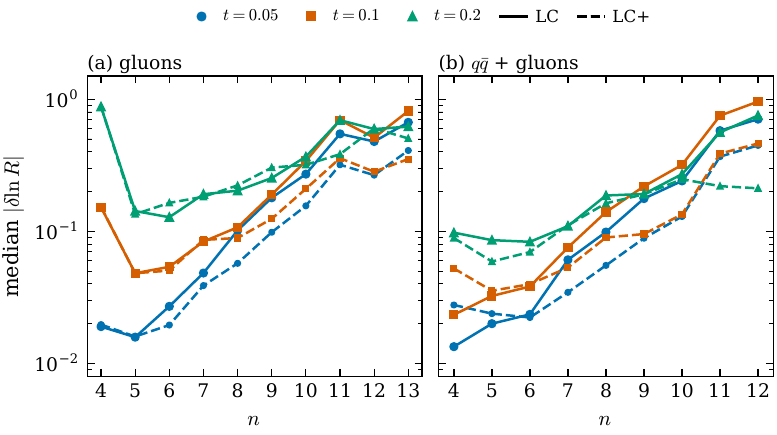}
  \caption{Median deviation $|\delta\ln R|=|\ln(R_{\rm approx}/R_{\rm FC})|$ of leading colour (solid lines, large
  markers) and LC+ (dashed lines, small markers) from full colour, for the anomalous dimension of the hard process with
  Coulomb phases (mean-subtracted hard weights), at $t=0.05$, 0.1 and 0.2, for (a) $n$-gluon and (b) $q\bar q+(n-2)$-gluon
  tree amplitudes. The medians are over 60 states (kinematic points times helicity configurations, MHV and alternating)
  for $n\le9$ and 6--48 states for larger $n$. At $t=0.05$ the error of leading colour grows from a few per cent at
  $n\le6$ to 0.5--0.7 in $\ln R$ at $n\ge11$. LC+ shows the same growth; for $n\ge8$ and $t\le0.1$ its error is smaller by a factor of 1.2--2.4.}
  \label{fig:window}
\end{figure}

We first compare the colour treatments for the anomalous dimension of the hard process, that is, for the
weights~\eqref{eq:cs:hardweights}, with and without the Coulomb phases. We subtract the mean of the weights, which only
multiplies $R$ by a constant (section~\ref{sec:cs:gamma}); we refer to these as the mean-subtracted hard weights. Because of this
constant, $R$ itself is not meaningful here, only the ratio of $R$ between two colour treatments. We measure the
deviation of leading colour by $|\delta\ln R|=|\ln(R_{\rm LC}/R_{\rm FC})|$, whose median is shown in
figure~\ref{fig:window} as a function of $n$. For four to six partons it is 0.02--0.03 at $t=0.05$. It then grows steadily: it is
0.10 at $n=8$, 0.27 at $n=10$ 0.48 at $n=12$ and 0.67 at $n=13$. At
$t=0.1$ the values are similar or larger, e.g.\ 0.51 at $n=12$. The one large value at $n=4$ and $t=0.2$ is due to the Coulomb phases, which
leading colour drops entirely (section~\ref{sec:cs:lc}): without them the deviation is 0.11 instead of 0.88. Since the
Coulomb part of $\Gamma$, eq.~\eqref{eq:cs:coulomb}, does not grow with $n$, while the real part does, its relative
importance is largest for four partons. LC+ shows the same growth. For $n\ge8$ and $t\le0.1$ its error is smaller than that of leading colour by a factor of 1.2--2.4; at $t=0.2$ the gain varies between none and a factor of 3.6. For processes with a quark line the errors are
similar up to $n=10$ and somewhat larger at $n=11$ and 12, 0.58 and 0.71 at $t=0.05$ (figure~\ref{fig:window}(b)). The
growth with $n$ is expected: the number of colour-connected pairs grows like $n$, while the number of pairs that are not
connected in a given flow, whose exchanges leading colour neglects, grows like $n^2$.

\subsection{Exclusive jet rates}\label{sec:exclusive}

An exclusive $(n-2)$-jet event has two incoming partons and a veto on additional resolved radiation. We compare the colour evolution for two definitions of the veto. In the first, the soft veto, we veto gluons with transverse momentum above $Q_0$ in $|y|<2.5$ outside cones of radius 0.4 around the jets. We integrate the veto weights of eq.~\eqref{eq:cs:vetoweights} over this region.

In the second definition we use $N$-jettiness $\tau_N$ \cite{Stewart:2010tn}, which measures radiation relative to the $N=n-2$ jet directions and the two beam directions. At next-to-leading logarithmic accuracy its colour evolution uses the weights
\begin{equation}
  \omega_{ij}=-\ln\frac{n_i\cdot n_j}{2},
  \label{eq:lc:tauweights}
\end{equation}
where $n_i$ are the light-like jet and beam directions. The remaining energy and scale dependence is colour-diagonal. For small cones and a large rapidity range, the non-diagonal parts of the two sets of weights coincide. We check this numerically to $3\times10^{-5}$ for cones of radius 0.002.

\begin{figure}[t]
  \centering
  \includegraphics[width=\textwidth]{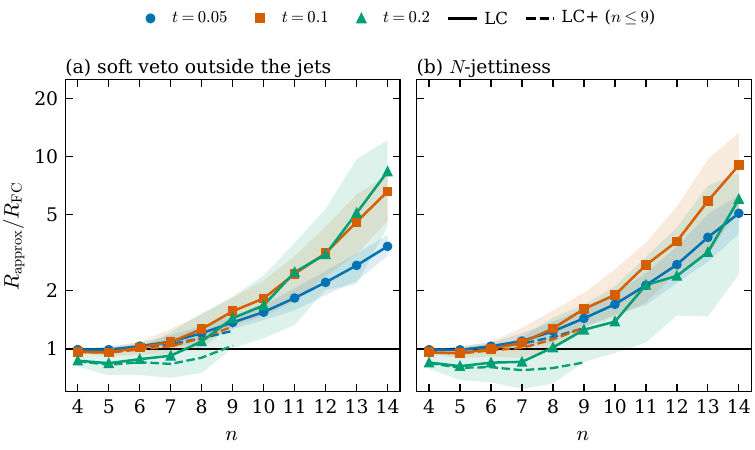}
  \caption{Ratio of the approximate (LC or LC+) and full-colour evolution factors for exclusive $(n-2)$-jet
  rates, (a) with a soft veto outside the jet cones and (b) for $N$-jettiness, as a function of $n$, at $t=0.05$, 0.1 and
  0.2. Solid lines: median of $R_{\rm LC}/R_{\rm FC}$ over events, with the 16--84\% range as bands; dashed lines: median
  of $R_{\rm LC+}/R_{\rm FC}$, shown for $n\le9$ only, because LC+ requires the overlaps of all evolved colour flows
  (section~\ref{sec:lcvsfc}). For $n\le10$ leading colour is the exact sum over all $(n-1)!$ flows; for $n\ge11$ it is
  obtained by Monte Carlo sampling of the flows, with a statistical error below 1\% per event, smaller than the
  symbols. The samples contain 100 events per $n$ for $n=4$--9, 30 for $n=10$--12,
  21 at $n=13$ and 20 at $n=14$. Leading colour overestimates the exclusive rate increasingly with $n$, by a factor of 6.6
  at $n=14$ and $t=0.1$.}
  \label{fig:exclusive}
\end{figure}

Figure~\ref{fig:exclusive} shows the ratio $R_{\rm LC}/R_{\rm FC}$ for both observables. For $n=4$--6 leading colour
agrees with full colour to within 5\% at $t\le0.1$. With more jets it overestimates the rate, increasingly so. For the soft veto at
$t=0.1$ the median ratio is $1.27$ at $n=8$, $1.83$ at $n=10$, $3.14$ at $n=12$, $4.56$ at $n=13$ and $6.56$ at
$n=14$. At $n=13$ and 14 leading colour exceeds full colour in every one of the 41 events. LC+ is closer to full colour
but shows the same trend. At $t=0.2$ the ratio falls below one for $n\le7$: leading colour then underestimates the rate. This reflects the asymptotic behaviour discussed in section~\ref{sec:spectrum}, where full colour always
decays more slowly than leading colour. As a check of the leading-colour limit, we varied $\Nc$ over 3, 5 and 7 at fixed
$\Nc t$ for a five-gluon event with the $N$-jettiness weights: the quantity $\Nc^2\ln(R_{\rm FC}/R_{\rm LC})$ is then constant to within 4\%
(0.81--0.84), so full colour approaches leading colour as $1/\Nc^2$, as it should.

\subsection{Gap fractions and quark lines}\label{sec:gaps}

\begin{figure}[t]
  \centering
  \includegraphics[width=\textwidth]{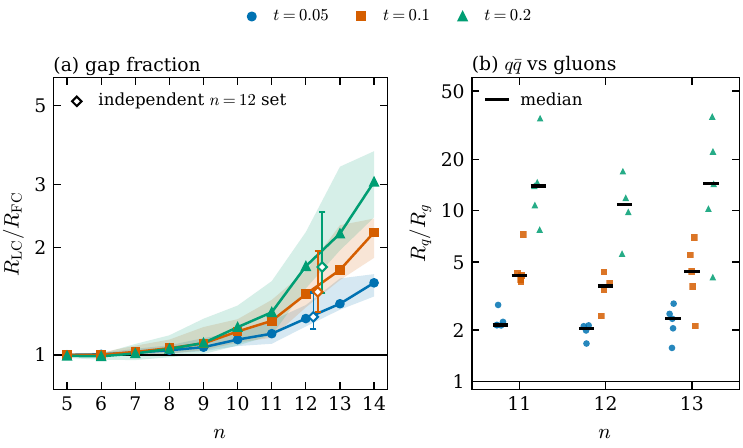}
  \caption{(a) Ratio of leading-colour and full-colour gap fractions for a rapidity gap $|y|<0.5$ as a function of
  $n$: medians (lines) and 16--84\% ranges (bands) over 200 events per $n$ for $n\le8$, 100 for $n=9$, 10, 50 at $n=11$,
  20 at $n=12$, 6 at $n=13$ and 5 at $n=14$. For the independent $n=12$ sample and at $n=13$, 14, leading colour is obtained by Monte Carlo sampling of the colour
flows (error below 1\% per event). Diamonds: an
  independent sample of 10 events at $n=12$ (for $t=0.05$, 0.1, 0.2 from left to right), with its 16--84\% range as
  bars. (b) Ratio $R_q/R_g$ of the evolution factors of $q\bar q+(n-2)$ gluons and of $n$ gluons at the same momenta, with
  the hard weights at a common reference scale $\mu$, for $n=11$--13 (14 kinematic points), at $t=0.05$, 0.1 and 0.2; bars mark the medians.
  Leading colour overestimates the gap fraction by a factor of 2.2 at $n=14$ and $t=0.1$, and the hard amplitude of the
  quark-line process decreases 2--13 times less than that of the pure-gluon process.}
  \label{fig:gapquark}
\end{figure}

The same pattern appears for the gap fraction, in which radiation is vetoed only in a rapidity slice
$|y|<0.5$ between jets that all lie outside it. Here $R_{\rm FC}(t)$ is itself the colour factor of the probability of
finding no emission above $Q_0$ in the gap; at $t=0.1$ it is about $0.02$--$0.04$ for $n=12$--14. The median of
$\ln(R_{\rm LC}/R_{\rm FC})$ at $t=0.1$ is $0.04$ at $n=8$, $0.15$ at $n=10$, $0.39$ at $n=12$, $0.55$ at $n=13$ and
$0.79$ at $n=14$, so that leading colour overestimates the gap fraction by factors of 1.04, 1.16, 1.48, 1.73 and 2.21
(figure~\ref{fig:gapquark}(a)). At $n=12$--14, leading colour lies above full colour in every event and at every value of $t$ we computed, up to $t=0.3$. Two
independent event samples at $n=12$ give $0.39$ and $0.41$.

Processes with a quark line, $q\bar q+(n-2)$ gluons, evolve differently from pure gluon processes. We evolve both with the hard weights~\eqref{eq:cs:hardweights}, including the Coulomb phases, for the same momenta
and the same fixed reference scale $\mu$. The ratio $R_q/R_g$ of the quark and gluon evolution
factors has a median of 2.1, 4.1 and 12.9 at $t=0.05$, 0.1 and 0.2 for $n=11$--13 (figure~\ref{fig:gapquark}(b)): the
hard amplitude of the quark process decreases 2 to 13 times less than that of the gluon process. Much of this difference
comes from the smaller total colour charge, $2C_F$ instead of $2C_A$ for the two partons of the quark line, which enters
through the colour-diagonal part of $\Gamma$; since this part depends on the choice of $\mu$
(section~\ref{sec:cs:gamma}), the size of $R_q/R_g$, unlike the ratios of leading to full colour, depends on this
choice. The error of leading colour grows with $n$ for quark processes as for gluon processes (section~\ref{sec:window}).

\subsection{Spectrum of the soft anomalous dimension}\label{sec:spectrum}

\begin{figure}[t]
  \centering
  \includegraphics[width=0.6\textwidth]{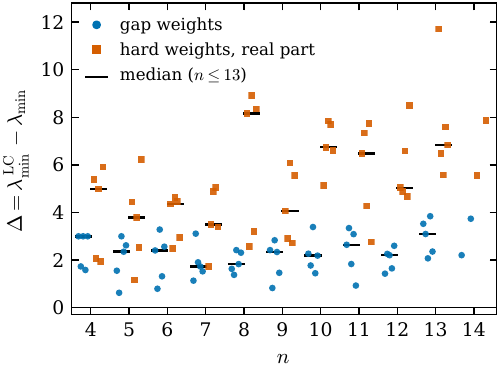}
  \caption{Difference $\Delta$ between the smallest leading-colour flow eigenvalue and the smallest eigenvalue of
  $\Gamma$, in the units of the weights $\omega_{ij}$, for gap weights (circles) and for the real part of the mean-subtracted
  hard weights (squares), for five kinematic points per $n$ and two at $n=14$. Bars mark the medians for $n\le13$. $\Delta$ is
  positive at every point and does not decrease with $n$.}
  \label{fig:spectrum}
\end{figure}

For real weights $\Gamma$ is a real symmetric matrix, and at large $t$ the evolution is dominated by its smallest eigenvalue $\lambda_{\min}$, for any state with a
non-zero component along the corresponding eigenvector: $R_{\rm FC}(t)\propto e^{-2t\lambda_{\min}}$. At leading colour the corresponding quantity is
the smallest flow eigenvalue $\lambda^{\rm LC}_{\min}=\min_\sigma\lambda^{\rm LC}_\sigma$, eq.~\eqref{eq:cs:lambdaLC}. Finding it
means minimising a sum of weights over all cyclic orderings of the partons, a problem of the travelling-salesman type,
which we solve exactly by dynamic programming over subsets of partons \cite{Bellman:1962,Held:1962} (appendix~\ref{app:numerics}). We compute $\lambda_{\min}$ with the Lanczos
method on the full space. Figure~\ref{fig:spectrum} shows the difference $\Delta=\lambda^{\rm LC}_{\min}-\lambda_{\min}$
for the gap weights of section~\ref{sec:gaps} and for the real part of the mean-subtracted hard weights. At every kinematic point and
every $n$ up to 14, $\Delta$ is positive, between 0.6 and 12, with medians of 2--3 for the gap weights and 3--8 for the
hard weights: the lowest eigenvalue of $\Gamma$ lies below the leading-colour eigenvalue of every flow. Mixing
between flows lowers the ground state by an amount that is not suppressed by $1/\Nc$ and does not decrease with $n$.
 Consequently, at large $t$ the ratio
$R_{\rm LC}/R_{\rm FC}$ falls like $e^{-2t\Delta}$, and leading colour underestimates the rate.

This is compatible with the overestimate found in section~\ref{sec:exclusive}. At small $t$, $R(t)$ is determined by the
expectation values of $\Gamma$ and its powers in the hard state (the coefficients of section~\ref{sec:moments}), not by the
lowest eigenvalue, and these give a stronger suppression in full colour than in leading colour when $n$ is large. The
lowest eigenvalue dominates only once $t$ is comparable to the inverse spacing of the low-lying eigenvalues. In
figure~\ref{fig:exclusive} the ratio has crossed one by $t=0.2$ only for $n\le7$.

\subsection{Colour-singlet probabilities}\label{sec:singlet}

The probability that a subset $S$ of partons is in a colour singlet, i.e.\ carries no net colour, is
$P_S=\|\Pi_S\mathcal M\|^2/\|\mathcal M\|^2$, where $\Pi_S$ projects onto the zero eigenspace of the Casimir
$(\sum_{i\in S}\Tb_i)^2$. It is relevant, for example, for colour reconnection \cite{Christiansen:2015yqa} and for the radiation pattern of
colour-singlet subsystems. In a single colour flow every proper subset of partons is joined to the rest by at least one
colour line, so the leading-colour value of $P_S$ is exactly zero. A first estimate beyond leading colour keeps the flows
separate: in each flow in which $S$ is a group of neighbours, it takes the exact singlet probability of that group, and it
assigns zero to the other flows. This estimate is of order $1/\Nc^2$ and contains no interference between flows. We
computed $P_S$ in full colour for five events at each $n=5$--10, generated as in section~\ref{sec:gaps}. For pairs of the
three hardest jets the median of $P_S$ at $t=0$ lies between 4\% and 8\%, depending on $n$, with individual events
between 3\% and 9\%. This is about twice the estimate from separate flows, which we attribute to interference between
flows. At $t=0.2$ the medians are 3--6\%. For three hard states with $n=13$, evolved with the hard weights, the singlet
probability of the group of partons in the forward hemisphere grows from 0.5\% at $t=0$ to 6\% at $t=0.2$, while the
probabilities of the pairs decrease.

\subsection{Real emission}\label{sec:emission}

So far only virtual corrections were included, which is exact for the global part of the observables. Two further
effects require real emissions. A soft gluon emitted outside the vetoed region can itself radiate a softer gluon into it;
these correlations give \emph{non-global logarithms} \cite{Dasgupta:2001sh}. And once real emissions collinear to an
incoming parton are included, the Coulomb phases no longer cancel between real and virtual corrections; this gives
\emph{super-leading logarithms}, terms with more powers of $\ln(Q/Q_0)$ than of $\alpha_s$. They were found in ref.~\cite{Forshaw:2006fk}, are related
to the breakdown of collinear factorisation \cite{Catani:2011st,Forshaw:2012bi}, and have since been studied at fixed
order \cite{Forshaw:2008cq,Keates:2009dn,Angeles-Martinez:2015rna}, resummed to all orders
\cite{Becher:2021zkk,Becher:2023mtx,Boer:2024hzh,Becher:2024kmk} and computed for dijet, top-pair and
vector-boson-fusion processes \cite{Becher:2024nqc,Banerjee:2025kkq,Becher:2026uih}; related enhanced logarithms and
a first parton-shower implementation are discussed in refs.~\cite{Banfi:2025mra,Schalch:2026xxk}. Both
require the operator that adds a real gluon to the colour state. In our basis it acts locally, like the dipoles, and
maps a vector of $\mathcal V_n$ to one of $\mathcal V_{n+1}$.

For a rapidity gap of width $Y=2$, the gap fraction $F$ is the fraction of events without emission above the veto scale
$Q_0$ in the gap. We expand it in the number $m\le3$ of out-of-gap gluons with transverse momentum above $Q_0$, each of which is either
a real gluon, emitted outside the gap, or the corresponding virtual correction at the same point, as in
ref.~\cite{Forshaw:2006fk}. We write $F=F_0(1+r_1+r_2+r_3)$, where $F_0$ is the purely virtual (global) part and $r_m$
the relative contribution of $m$ out-of-gap gluons. Each term is an overlap of two colour states, the amplitude and its
conjugate, obtained by alternating the in-gap virtual evolution $e^{-\delta t\,\Gamma}$ with real-emission or virtual
insertions; when all insertions are real emissions on both sides, it is a squared norm. The terms are integrated over
the directions of the out-of-gap gluons and over their emission scales, ordered in transverse momentum. A real gluon
becomes a further parton of the colour state and takes part in the subsequent evolution. Real and virtual insertions
cancel point by point when no evolution follows, and the $r_m$ capture the non-global and, through the Coulomb phases, the
super-leading logarithms order by order in $m$.

We validate this calculation on two known results. For a rapidity gap in $qq\to qq$, we reproduce the coefficient of the
first super-leading logarithm of ref.~\cite{Forshaw:2006fk} to $4\times10^{-6}$. The hemisphere mass distribution in
$e^+e^-$ annihilation, in which radiation into one hemisphere is restricted, has non-global logarithms with coefficients
$S_2$ and $S_3$ at second and third order \cite{Dasgupta:2001sh,Schwartz:2014wha}. We reproduce $S_2$ to $10^{-5}$, and
our Monte Carlo result $S_3=2.419\pm0.054$ lies within $0.3$ standard deviations of $C_FC_A^2\zeta_3/6=2.404$ at
$\Nc=3$.

\begin{figure}[t]
  \centering
  \includegraphics[width=\textwidth]{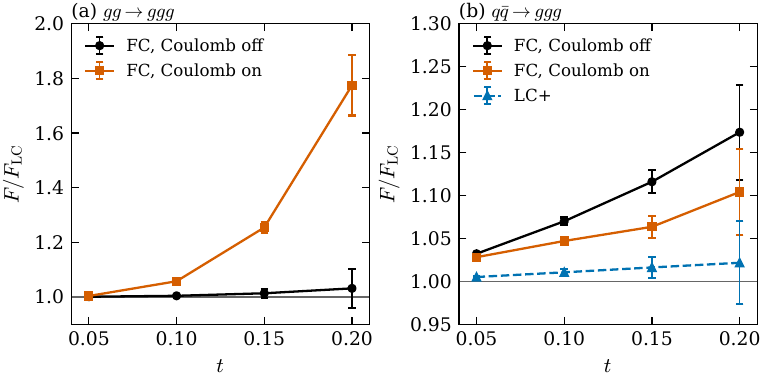}
  \caption{Gap fraction $F$ with up to three out-of-gap gluons (real or virtual) in full colour, divided by its
  leading-colour value $F_{\rm LC}$ (note: full over leading colour, the inverse of figures~\ref{fig:exclusive}
  and~\ref{fig:gapquark}), for (a) $gg\to ggg$ and (b) $q\bar q\to ggg$, averaged over 100 events, as a function of $t$:
  full colour without (circles) and with (squares) Coulomb phases, and LC+ for the quark channel (triangles). Here LC+ is used in its flow-diagonal form, with the
  eigenvalues~\eqref{eq:cs:lambdaLCp} but without the overlaps between flows; for gluons it then coincides with leading
  colour. Only $t\le0.2$ is shown, where the expansion in the number of emissions converges.
  The error bars combine the Monte Carlo errors of numerator and denominator as if independent; since both are computed
  from the same samples, this is expected to overestimate the error of the ratios. Coulomb phases dominate the
  full-colour effect for $gg\to ggg$; for $q\bar q\to ggg$ the non-Coulomb part dominates.}
  \label{fig:emission}
\end{figure}

We then compare full colour with and without Coulomb phases and leading colour for $gg\to ggg$ and $q\bar q\to ggg$, with
100 events each (figure~\ref{fig:emission}). The expansion in $m$ converges for $t\le0.2$, and we restrict the
discussion to this range; beyond it the terms $r_2$ and $r_3$ become comparable to $r_1$. For $gg\to ggg$, full colour without Coulomb phases agrees with leading colour to 0.4\% at
$t=0.1$ and to 3\% at $t=0.2$. The Coulomb phases, in contrast, increase the gap fraction by 5\%, 24\% and 72\% at
$t=0.1$, 0.15 and 0.2. Most of this increase does not come from the super-leading terms in the real emissions but from
the virtual evolution itself, where the Coulomb phases contribute from order $t^2$ onwards, through the commutator of
the real and imaginary parts of $\Gamma$: at $t=0.2$ the emission terms raise $F$ by 7\% and the virtual evolution by
61\%. For $q\bar q\to ggg$ already the non-Coulomb part differs from leading colour: full colour exceeds
leading colour by 7\% at $t=0.1$ and by 17\% at $t=0.2$. Here the Coulomb phases lower the gap fraction, by 6\% at $t=0.2$. The super-leading terms in the emissions again raise $F$, by 6\%, but the Coulomb effect in the virtual
evolution has the opposite sign to the gluon case and lowers $F$ by 11\%. The sign of this effect depends on the
colour structure of the process, since the Coulomb operator~\eqref{eq:cs:coulomb} is the Casimir of a $q\bar q$ pair in
one case and of a $gg$ pair in the other. At five partons the full-colour effects are therefore dominated by the Coulomb phases in $gg\to ggg$ and by the
non-Coulomb colour mixing in $q\bar q\to ggg$. The present
implementation of the emission operator limits this study to five partons in the hard process.

\section{Beyond exact evolution}\label{sec:beyond}

The exact evolution of section~\ref{sec:evolution} stores complete vectors, of 13~GB each at $n=15$; distributing them
over several devices would extend it by a few partons, but not to the several hundred partons of a parton shower. In
this section we ask what can be computed for such multiplicities without storing complete vectors. Section~\ref{sec:mps} describes an approximate evolution with matrix
product states and determines where it can be trusted. Section~\ref{sec:moments} shows that the first terms of the
expansion of $R(t)$ in $t$ can be computed exactly for several hundred gluons.

\subsection{Matrix product states}\label{sec:mps}

A basis vector is a path $(X_1,\dots,X_n)$ through the Bratteli diagram, and a colour state assigns one coefficient
to every path. Because the paths are built parton by parton, the coefficient can be written as a product of matrices,
one per parton,
\begin{equation}
  x_{(X_1,\dots,X_n)}=A_1[X_1]\,A_2[X_1,X_2]\cdots A_n[X_{n-1},X_n],
  \label{eq:mps}
\end{equation}
where we suppress the copy names and the row and column indices of the matrices. For a fixed path,
$A_k[X_{k-1},X_k]$ is a $\chi_{k-1}\times\chi_k$ matrix, and the product is a number. This is a \emph{matrix product state}
(MPS), a representation widely used for one-dimensional quantum systems
\cite{Fannes:1990ur,White:1992zz,Ostlund:1995zz,Vidal:2003lvx,Schollwoeck:2010uqf,Orus:2013kga}. The matrix size $\chi_k$
at the cut between partons $k$ and $k+1$ is called the \emph{bond dimension}. It counts how many independent
combinations of the first halves of the paths the second half has to distinguish, that is, how much colour information
has to be passed from the first $k$ partons to the rest. For $\chi_k=1$ at every cut the coefficient factorises into local
factors; larger $\chi_k$ allow correlations between distant partons. Since our labels are representations rather than
individual colour states, the bonds count multiplets, as in the MPS that exploit the $\mathrm{SU}(\Nc)$ symmetry in
refs.~\cite{McCulloch:2007azj,Weichselbaum:2012fox}.

Every state can be written in this form if $\chi_k$ may grow to the number of path halves at the cut, which grows
exponentially with $n$. An MPS is useful when a small maximal bond dimension $\chi$ suffices. To compress a state, we write
its coefficients at each cut as a sum of products of orthonormal combinations of left and right path halves, ordered
by their weights (the singular values), and keep the $\chi$ largest. Because our basis is orthonormal, this truncation \cite{Eckart:1936}
gives the smallest possible error in the colour-summed norm at that cut.

Some colour states have a very small bond dimension. The colour factor of the adjoint decomposition in the coupling
order (section~\ref{sec:hard}), the all-octet $f$ chain, is a single basis vector and has $\chi=1$; an open string
$(t^{a_3}\cdots t^{a_n})_{i\bar\jmath}$ in the coupling order also has $\chi=1$, and a single trace has $\chi=2$. A trace in a
random ordering, however, has a bond dimension that grows like $1.7^n$, and a generic tree amplitude, which is a sum
over all orderings, is not a low-bond state. The operator $\Gamma$ itself can be written in the same product form. Its
bond at a cut is set by the number of independent exchanges across the cut, the rank of the matrix of weights
$\omega_{ij}$ with $i\le k<j$, which is about $n/2$ for generic kinematics. Applying $\Gamma$ increases the bond
dimension, which is then reduced again by compression. We compute $e^{-t\Gamma}x$ with the Krylov method of
section~\ref{sec:matvec}, with all vectors in MPS form \cite{Paeckel:2019yjf}.

\begin{figure}[t]
  \centering
  \includegraphics[width=\textwidth]{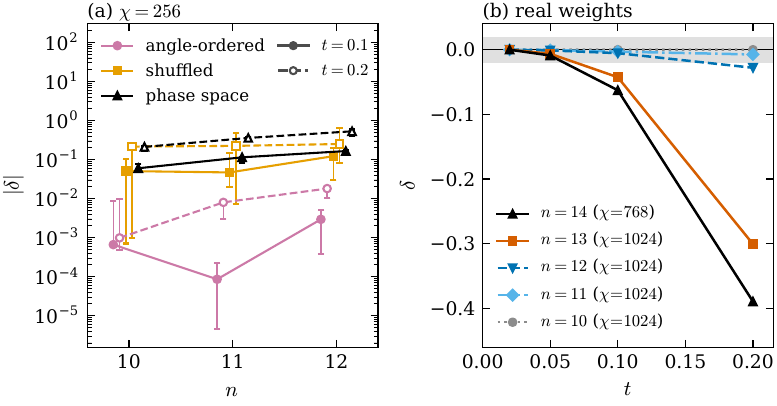}
  \caption{Accuracy of the MPS evolution, measured by $\delta=\|e^{-t\Gamma}x\|^2_{\rm MPS}/\|e^{-t\Gamma}x\|^2_{\rm
  exact}-1$. (a) $|\delta|$ at bond dimension $\chi=256$ for angle-ordered weights, the same weights with randomly
  relabelled partons (shuffled), and the weights of random phase-space points, at $t=0.1$ (solid lines, filled markers)
  and $t=0.2$ (dashed lines, open markers); median and range of three kinematic points. (b) $\delta$ as a function of $t$
  for $n=10$--14 at the largest bond dimension available ($\chi=1024$, and 768 at $n=14$), real weights. The grey band marks
  $\pm2\%$. Ordering the partons as in an angular-ordered shower reduces the error by a factor of 30 to more than 1000,
  while for random kinematics the MPS misses 30--40\% of the norm at $t=0.2$.}
  \label{fig:mps}
\end{figure}

We compared this approximate evolution with the exact results of section~\ref{sec:evolution} for $n=8$--14, starting
from the all-octet $f$ chain and with bond dimensions up to $\chi=1024$. Figure~\ref{fig:mps}(b) shows the relative error of
$\|e^{-t\Gamma}x\|^2$ at the largest bond dimension for each $n$. For $t\lesssim0.05$ the error stays within 1--2\% up to
$n=14$. At larger $t$ it grows quickly: at $n=13$ and 14 the MPS misses 4--6\% of the norm at $t=0.1$ and 30--40\% at
$t=0.2$, and more for smaller $\chi$ or complex weights. The error is
negative wherever it exceeds the numerical precision ($|\delta|<10^{-6}$ at $t=0.02$), because compression removes norm, and it decreases only slowly with $\chi$. The physical hard states of section~\ref{sec:lcvsfc} have a large bond dimension already at $t=0$, and we expect the
errors for them to be larger. A truncated MPS is therefore not a substitute for the exact evolution at the values of $t$
relevant for the observables of section~\ref{sec:lcvsfc}.

The situation is different when the weights have a hierarchical structure. In an angular-ordered parton shower \cite{Marchesini:1983bm,Marchesini:1987cf}, emissions are
ordered in angle, so that a parton is mostly correlated with the partons emitted close to it. We generated weights with this
structure: a random binary tree over the partons with opening angles that decrease strictly along every branch, with the
partons numbered along the tree. Figure~\ref{fig:mps}(a) shows that the MPS evolution is then 30 to more than 1000 times more
accurate at the same bond dimension than for the weights of a random phase-space point. If the same ordered weights are
assigned to randomly relabelled partons, the advantage disappears completely. The accuracy is therefore a consequence of
the ordering along the chain, not of the distribution of the weights. Generic events generated flat in phase space are
not angle-ordered in this sense, and the advantage decreases with $n$ at fixed $\chi$. Nevertheless, the result suggests
that colour states produced by an angular-ordered evolution are compressible, which may be relevant for full-colour
parton showers.

\subsection{Exact Taylor coefficients for hundreds of gluons}\label{sec:moments}

The expansion of the evolved norm in $t$,
\begin{equation}
  \|e^{-t\Gamma}\mathcal M\|^2=\sum_{k\ge0}\frac{(-t)^k}{k!}\,c_k,
  \label{eq:taylor}
\end{equation}
has the coefficients
\begin{equation}
  \begin{aligned}
  c_1&=2\,\mathrm{Re}\bra{\mathcal M}\Gamma\ket{\mathcal M},\qquad
  c_2=2\,\mathrm{Re}\bra{\mathcal M}\Gamma^2\ket{\mathcal M}+2\|\Gamma\mathcal M\|^2,\\
  c_3&=2\,\mathrm{Re}\bra{\mathcal M}\Gamma^3\ket{\mathcal M}+6\,\mathrm{Re}\braket{\Gamma\mathcal M}{\Gamma^2\mathcal M}.
  \end{aligned}
  \label{eq:ck}
\end{equation}
Divided by $\|\mathcal M\|^2$, they give the first three terms of the fixed-order expansion of $R(t)$. For a state with a
small bond dimension they can be computed exactly with the MPS representation: for the all-octet $f$ chain ($\chi=1$),
$\Gamma\mathcal M$ has a bond dimension of only $3n-4$. We first
remove the part of the weights of the form $a_i+a_j$, which only shifts $\Gamma$ by a constant
(section~\ref{sec:cs:gamma}), compute $\Gamma\mathcal M$ exactly, and obtain the terms with $\Gamma^2$ and $\Gamma^3$ by
contracting $\Gamma$ between $\Gamma\mathcal M$ and $\mathcal M$ or $\Gamma\mathcal M$, without ever forming
$\Gamma^2\mathcal M$.

Table~\ref{tab:moments} shows the cost. For 100 gluons the three coefficients take 15--55~s on one core, and
for 500 gluons with real weights they take 7~hours and 12~GB. The results agree with the products of
section~\ref{sec:evolution} to $4\times10^{-15}$ for $n=6$--12. The next coefficient, $c_4$, requires
$\Gamma^2\mathcal M$, whose bond dimension grows like $n^2$; at 100 gluons this would need about 600~GB.

\begin{table}[t]
  \centering
  \small
  \begin{tabular}{r r rr r}
    \toprule
    $n$ & bond of $\Gamma\mathcal M$ & time, real weights & time, complex weights & memory \\
    \midrule
     10 &   26 & 0.2~s  & 0.2~s  & 0.08~GB \\
     40 &  116 & 0.6~s  & 1.2~s  & 0.11~GB \\
    100 &  296 & 15~s   & 55~s   & 0.56~GB \\
    500 & 1496 & 7.1~h  & --     & 11.7~GB \\
    \bottomrule
  \end{tabular}
  \caption{Cost of the exact coefficients $c_1$, $c_2$, $c_3$ for $n$ gluons at $\Nc=3$, starting from the all-octet $f$ chain, on one core of a shared machine.}
  \label{tab:moments}
\end{table}

We test the short-time expansion by retaining terms through $t^3$ in $\ln R(t)$, exponentiating, and comparing with the exact $R(t)$ for $n\le14$. We obtain 1\% accuracy up to $t\approx0.08$ for six gluons and up to $t\approx0.05$ for thirteen and fourteen. The tested range decreases slowly with multiplicity, suggesting that the coefficients may also describe evolution at $t\lesssim0.05$ for larger $n$. The exact comparison at thirteen and fourteen gluons already places $t\approx0.1$--0.2, where section~\ref{sec:lcvsfc} finds large colour corrections, outside the 1\% range.
\section{Conclusions}\label{sec:conclusions}

We have presented a construction of orthogonal multiplet bases of colour space for any process with quarks, antiquarks
and gluons, and used it to compute the one-loop soft evolution of colour states exactly for up to fifteen gluons. The
basis vectors are labelled by paths of representations and fixed by a chain of Casimir operators, applied separately
to the quark and antiquark index of each gluon; their norms are given
by a closed formula, and the colour operators act locally on the labels. This locality allows us to apply the soft
anomalous dimension to vectors with $8.3\times10^8$ components without storing it, on CPUs and on a GPU. Fifteen partons
is not a limit of the method: the algorithm is the same at every $n$, and the largest multiplicity is set by the
available memory.

Our main result concerns the accuracy of leading colour for many partons. For four to six partons leading colour is
accurate to a few per cent at the scales relevant for the LHC. With more partons its error grows steadily, and for
exclusive jet rates with a soft veto it overestimates the full-colour result by a factor of about 1.8 at ten partons
and of about six and a half at fourteen, at $t=0.1$. The direction of the deviation is the same in every event at the
largest multiplicities. At late times, reached within $t\le0.2$ only for up to seven partons, the sign reverses, because the lowest eigenvalue of the soft anomalous dimension lies
below the leading-colour eigenvalue of every colour flow by an amount of order one. These results show that the formal $1/\Nc^2$ suppression of the
subleading terms does not guarantee their smallness when the number of partons is large.

These findings bear directly on the physics programme of the LHC and of its high-luminosity phase. With an
integrated luminosity of about $3~\text{ab}^{-1}$, final states with many hard jets will be measured with small
statistical uncertainties, and many analyses select events by vetoing additional jets or by sorting them into jet
bins, from Higgs boson production in gluon fusion and vector-boson fusion to top-quark final states and searches for
new physics with many jets \cite{Azzi:2019yne,Cepeda:2019klc,CidVidal:2018eel,Stewart:2011cf}. The theoretical
description of such selections relies on soft-gluon resummation and on parton showers, whose colour evolution is
almost always that of leading colour. Our results show that, for the colour factors of exclusive rates with eight or more partons,
leading colour can be wrong by a large factor, and that the error grows
with the number of jets. The colour treatment therefore becomes more important, not less, as measurements extend to
higher jet multiplicities. The exact results provide full-colour reference values against which improved colour
treatments in parton showers and resummations \cite{Platzer:2018pmd,Nagy:2019pjp,Hamilton:2020rcu,vanBeekveld:2024wws}
can be tested.

Several statements may be useful beyond this analysis. As is known \cite{Catani:1998bh,Forshaw:2006fk}, the Coulomb part
of the one-loop soft anomalous dimension is the Casimir of the incoming partons; in our bases this makes it diagonal
when the incoming partons are coupled first. Colour-singlet probabilities
of subsets of jets are of the order of several per cent, although they vanish at leading colour. And the compressed (matrix-product-state) evolution is 30 to more than 1000 times more accurate for angle-ordered weights
than for generic kinematics, which may be of use for full-colour parton showers. The compression was tested, however, on a single
basis vector as the initial state, and the exact Taylor coefficients for hundreds of gluons are available only for such
simple states.

Some questions remain open. Our events are generated flat in phase space, mostly with maximally helicity-violating
helicities, and are not weighted by the cross section; the full-colour corrections to physical distributions require the
convolution with realistic event samples. We have included real emissions only for up to five hard partons, and we have
not studied the interplay of the colour evolution with recoil and with running-coupling effects beyond one loop.

For improvements we see three directions: the extension of the exact evolution to sixteen and more partons by
distributing the vectors over several GPUs; a compiled implementation of the emission operator, which would make the
real-emission study possible for many-jet final states; and the use of the basis and of the compression of ordered
states in a full-colour parton shower.

The basis construction, the colour operators and the evolution are available in the public code
\orthobase{}~2.0 (appendix~\ref{app:code}).

\acknowledgments
The work of B.C.\ was supported by the Deutsche Forschungsgemeinschaft (DFG) under grant 396021762 --- TRR~257
``Particle Physics Phenomenology after the Higgs Discovery''.

\appendix
\section{Conventions}\label{app:conventions}

\paragraph{Kinematics.}
All partons are treated as outgoing, $\sum_ip_i=0$, with incoming partons carrying negative energy; the two incoming
partons are numbered 1 and 2. We use the metric $(+,-,-,-)$, massless partons, $s_{ij}=2p_i\cdot p_j$, and no averaging
over colours or helicities.

\paragraph{Generators.}
In the text the generators are normalised as $\mathrm{Tr}\,t^at^b=\frac12\delta^{ab}$, so that $C_F=(\Nc^2-1)/(2\Nc)$ and
$C_A=\Nc$. A quark carries $t^a$, an antiquark $-(t^a)^T$ and a gluon $(T^a)_{bc}=-if^{abc}$. Our code uses
$\mathrm{Tr}\,t^at^b=\delta^{ab}$, in which $\Tb_i\cdot\Tb_j$ and all Casimirs are twice as large. The code writes the soft
anomalous dimension as $\sum_{i<j}w_{ij}\Tb_i\cdot\Tb_j$ with $w_{ij}=\omega_{ij}/2$ (in the paper we use $\omega_{ij}$
throughout), so that $\Gamma$ and all its
eigenvalues are the same in both conventions, and the evolution variable $t$ of eq.~\eqref{eq:cs:evolution} has the same
meaning. The unnormalised basis vectors are built with the code's generators; their norms, e.g.\ the norm
$(2\Nc)^{n-2}(\Nc^2-1)$ of the all-octet $f$ chain, refer to this normalisation. All results are quoted in the orthonormal
frame $u_a=v_a/\sqrt{\braket{v_a}{v_a}}$, in which they do not depend on it. In this frame $\Gamma$ is complex symmetric,
and real symmetric for real weights.

\paragraph{Relation to the literature.}
For $2\to2$ processes with incoming partons $A$, $B$ and outgoing partons 1, 2, the matrix $\Gamma_{S'}$ of
ref.~\cite{Kidonakis:1998nf}, in units of $\alpha_s/\pi$, equals eq.~\eqref{eq:cs:gamma} with
$\omega_{A1}=\omega_{B2}=-\mathcal T$, $\omega_{A2}=\omega_{B1}=-\mathcal U$ and $\omega_{AB}=\omega_{12}=0$, where $\mathcal T=\ln(-\hat t/\hat s)+i\pi$
and $\mathcal U=\ln(-\hat u/\hat s)+i\pi$ in terms of the Mandelstam variables $\hat s,\hat t,\hat u$, up to a multiple of the identity. The anomalous dimension of ref.~\cite{Becher:2009cu} satisfies
$\Gamma_{\rm BN}/\gamma_{\rm cusp}=\Gamma$ with the weights~\eqref{eq:cs:hardweights}, up to colour-diagonal terms. At fixed
coupling the evolution from $Q$ to $Q_0$ with $\Gamma_{\rm BN}$ is therefore $e^{-t\Gamma}$ with
$t=(\alpha_s/2\pi)\ln(Q^2/Q_0^2)$, and with one-loop running $t=(2/\beta_0)\ln[\alpha_s(Q_0)/\alpha_s(Q)]$.
\section{Proofs}\label{app:proofs}

We prove theorem~\ref{thm:orth} and the norm formula~\eqref{eq:b:norm}. We write $N=\Nc$ for the number of colours. The first question is how to distinguish two ways of coupling the partons to a singlet. As in section~\ref{sec:bratteli}, we split each gluon into a quark leg followed by an antiquark leg. This retains information that the Casimir after a complete gluon can miss. For example, the $\mathbf{10}$ and $\overline{\mathbf{10}}$ in $\mathbf 8\otimes\mathbf 8$ have equal Casimirs, but the representations after the quark leg differ.

Coupling one leg adds or removes one box in a Young diagram. If the box has content $c$ (column minus row), the leg operator $L_r$ at leg $r$ has eigenvalue $c$ for addition and $-(N+c)$ for removal. When the allowed eigenvalues are distinct, a polynomial in $L_r$ selects the required representation. A separate choice arises when adding a gluon returns the preceding representation $X$. The scalar products of the candidate couplings then depend only on $X$ and $N$. We can therefore orthogonalise the couplings once and use them on every path through $X$.

Appendices~\ref{app:p:setting} and~\ref{app:p:vectors} give the operators and construct the vectors of section~\ref{sec:bases}. The three lemmas in appendix~\ref{app:p:lemmas} give the orthogonality proof in appendix~\ref{app:p:orth}. Appendix~\ref{app:p:norm} gives the norm as a product of dimension ratios, factors from crossings of the kind shown in figure~\ref{fig:arcs}, and one factor for each repeated coupling. Only its proof, in appendix~\ref{app:p:normproof}, uses symmetric-group representation theory. Appendix~\ref{app:p:scope} gives the assumptions, the numerical checks and the relation to the literature. Table~\ref{tab:p:glossary} collects the notation.

\begin{table}[t]
  \centering
  \small
  \begin{tabular}{@{}l>{\raggedright\arraybackslash}p{0.56\textwidth}l@{}}
    \toprule
    Term & Meaning & Details\\
    \midrule
    q-leg, b-leg & a fundamental (quark) or antifundamental (antiquark) index, respectively &
      app.~\ref{app:p:setting}\\
    chain & the ordered legs of the process, with a q-leg followed by a b-leg for each gluon &
      app.~\ref{app:p:setting}\\
    leg operator $L_r$ & the sum of swaps with earlier legs of the same type minus the sum of contractions with earlier legs of the opposite type, eq.~\eqref{eq:p:L} &
      app.~\ref{app:p:setting}\\
    content & column minus row for a box in a Young diagram & app.~\ref{app:p:setting}\\
    isotypic component & the subspace containing all copies of a specified representation on a given set of legs &
      app.~\ref{app:p:setting}\\
    leg path $\mathcal P$ & the sequence of representations after each individual leg &
      app.~\ref{app:p:vectors}\\
    line $x$ & the lifetime of one box, from its addition at leg $a_x$ to its removal at leg $r_x$ &
      app.~\ref{app:p:vectors}\\
    seed & the product of Kronecker deltas joining the two ends of each line &
      app.~\ref{app:p:vectors}\\
    copy step & a gluon that returns the preceding representation and carries a copy name $f,d,\dots$ &
      app.~\ref{app:p:vectors}\\
    reduced path, vector & the path after deleting its first copy-step gluon, and a vector for that path &
      app.~\ref{app:p:vectors}\\
    candidate $\Phi_r$ & the map that inserts a singlet pair, applies $r$ powers of the colour correlation on its q-leg and removes the trace, eq.~\eqref{eq:p:cand} &
      app.~\ref{app:p:vectors}\\
    Gram--Schmidt remainder & the part of a candidate orthogonal to all earlier candidates &
      app.~\ref{app:p:vectors}\\
    Markov property & tracing a projector over the added index gives the identity on the parent multiplet times a ratio of dimensions &
      app.~\ref{app:p:lemmas}\\
    crossing & two lines in the same diagram whose ends alternate, $a_x<a_y<r_x<r_y$ (figure~\ref{fig:arcs}) &
      app.~\ref{app:p:norm}\\
    \bottomrule
  \end{tabular}
  \caption{Terms used in appendix~\ref{app:proofs}.}
  \label{tab:p:glossary}
\end{table}

\subsection{Setting}\label{app:p:setting}

\paragraph{Legs and colour space.}
We specify a process by an ordered list of $n$ outgoing partons, with equally many quarks and antiquarks. Its \emph{chain} lists the individual colour legs. A quark gives one \emph{q-leg}, an antiquark gives one \emph{b-leg} (b for ``bar''), and each gluon gives a q-leg $g_q$ followed by a b-leg $g_b$. We use $k$ to count partons and $\ell,r$ to count legs. The chain has $m$ legs of each type, where $m$ is the number of quarks plus the number of gluons. A q-leg carries the fundamental space $V=\mathbb C^N$ and a b-leg its dual $V^*$. We write $V_\ell$ for the space on leg $\ell$ and $W=\bigotimes_\ell V_\ell$ for their product in chain order.

Throughout this appendix we use the code convention $\mathrm{Tr}(t^at^b)=\delta^{ab}$ (appendix~\ref{app:conventions}). Relative to section~\ref{sec:colourspace}, the colour charges acquire a factor $\sqrt2$ and their products a factor $2$. Thus $C_F=(N^2-1)/N$ and $C_A=2N$. A gluon is a traceless matrix $Z_{i\bar\jmath}$ on its pair of legs. Its components $Z^a=\mathrm{Tr}(t^aZ)$ have the same scalar product as the matrices because the $t^a$ form an orthonormal basis of traceless matrices. The Casimir relation in eq.~\eqref{eq:b:Deig} scales on both sides by the same factor. The norm in eq.~\eqref{eq:b:norm} involves only dimensions and box contents.

The colour space $\mathcal V_n$ of eq.~\eqref{eq:cs:space} consists of the $\mathrm{SU}(N)$ invariants in $W$ that are traceless in every gluon pair $(g_q,g_b)$. In $W$, which has equally many legs of each type, products of Kronecker deltas span the invariants by the first fundamental theorem of invariant theory \cite{Goodman:2009}. Each product pairs every q-leg with a different b-leg. Writing $S_m$ for the group of permutations of $m$ objects, we call the resulting tensors \emph{permutation states},
\begin{equation}
  \ket\sigma=\prod_{a=1}^m\delta_{i_a\bar\jmath_{\sigma(a)}},\qquad \sigma\in S_m ,
  \label{eq:p:perm}
\end{equation}
where $i_a$ and $\bar\jmath_b$ label the $a$-th q-leg and the $b$-th b-leg in chain order. Contracting all indices as in section~\ref{sec:cs:space} gives one free colour sum for each cycle of $\sigma^{-1}\tau$. Hence $\braket\sigma\tau=N^{\mathrm{cyc}(\sigma^{-1}\tau)}$, where $\mathrm{cyc}$ counts cycles. The states $\ket\sigma$ are linearly independent for $N\ge m$. We construct the vectors below as finite sums of these states with coefficients in $\mathbb Q(N)$, the field of rational functions of $N$ with rational coefficients.

\paragraph{Operators.}
On leg $\ell$ the generator $t^a_{(\ell)}$ acts as $t^a$ on a q-leg and as $-(t^a)^T$ on a b-leg. For a set $S$ of legs, $T^a_S=\sum_{\ell\in S}t^a_{(\ell)}$ is its total charge. The dot denotes a sum over the generator index $a$. The Fierz identity $\sum_at^a_{ij}t^a_{kl}=\delta_{il}\delta_{kj}-\delta_{ij}\delta_{kl}/N$ gives, for $\ell\ne r$,
\begin{equation}
  t_{(\ell)}\cdot t_{(r)}=(\ell\,r)-\frac1N\quad\text{(same type)},\qquad
  t_{(\ell)}\cdot t_{(r)}=-E_{\ell r}+\frac1N\quad\text{(opposite type)},
  \label{eq:p:fierz}
\end{equation}
where $(\ell\,r)$ exchanges two indices of the same type. On a q-leg and a b-leg, $E_{\ell r}$ takes the trace over the two indices and inserts a Kronecker delta in their place.

Let $n_{\rm s}(r)$ and $n_{\rm o}(r)$ count the earlier legs of the same and of the opposite type as leg $r$. We remove the constant terms in eq.~\eqref{eq:p:fierz} to define the \emph{leg operator}
\begin{equation}
  L_r=\sum_{\ell<r}t_{(r)}\cdot t_{(\ell)}+\frac{n_{\rm s}(r)-n_{\rm o}(r)}{N}
  =\sum_{\substack{\ell<r\\\text{same type}}}(\ell\,r)-\sum_{\substack{\ell<r\\\text{opposite type}}}E_{\ell r}.
  \label{eq:p:L}
\end{equation}
The $L_r$ are Hermitian; $L_r$ is the operator $\ell_k$ of eq.~\eqref{eq:b:content} in section~\ref{sec:casimirs}. Each $L_r$ couples to the earlier legs only through their total charge, so it commutes with every colour-invariant operator on those legs. In particular, it commutes with all earlier leg operators. For q-legs alone, $L_r=\sum_{\ell<r}(\ell\,r)$ is the Jucys--Murphy element of the symmetric group \cite{Jucys:1974,Murphy:1981,Okounkov:1996}. For any interleaving of q- and b-legs, the corresponding swaps and contractions generate the walled Brauer algebra on these tensors. The $L_r$ are its Jucys--Murphy elements \cite{Brundan:2012,Jung:2020,Grinko:2023cdq}.

To compare with the parton operator $D_k$ in eq.~\eqref{eq:b:D}, we keep $D_k$ in the units of section~\ref{sec:colourspace}. Let $s_{<r}$ be the number of q-legs minus the number of b-legs before leg $r$. For a quark or antiquark $k$ on that leg, $2D_k=L_r\mp s_{<r}/N$, with the upper sign for a quark. For a gluon $k$, the two leg operators give $2D_k=L_{g_q}+L_{g_b}$ on tensors traceless in that gluon.

\paragraph{Representations.}
We label an irreducible representation by two Young diagrams, $X=\irr\alpha\beta$, and write $V_X$ for its representation space \cite{King:1971rs,Stembridge:1987}. A Young diagram consists of left-aligned rows of boxes with non-increasing lengths. The diagram $\beta$ describes the quark-like indices and $\alpha$ the antiquark-like indices. For either diagram $\lambda$, let $\lambda_i$ be the length of row $i$, $\ell(\lambda)$ the number of rows and $|\lambda|$ the total number of boxes. Put $p=\ell(\beta)$ and $q=\ell(\alpha)$. The representation exists at $N$ if $p+q\le N$. Its $\mathrm U(N)$ highest weight is $(\beta_1,\dots,\beta_p,0,\dots,0,-\alpha_q,\dots,-\alpha_1)$. On a fixed set of legs the $\mathrm U(1)$ charge $|\beta|-|\alpha|$ is fixed, so the $\mathrm{SU}(N)$ and $\mathrm U(N)$ decompositions coincide.

The \emph{content} of the box in row $i$ and column $j$ is $c=j-i$. The quadratic Casimir is
\begin{equation}
  C(X)=\sum_{i}\beta_i(\beta_i+N+1-2i)+\sum_{i}\alpha_i(\alpha_i+N+1-2i)-\frac{(|\beta|-|\alpha|)^2}{N},
  \label{eq:p:casimir}
\end{equation}
which gives $C_F=N-1/N$ and $C_A=2N$. To couple a q-leg, we add a box to $\beta$ or remove a box from $\alpha$, keeping the allowed diagrams. Each resulting representation occurs exactly once in $V_X\otimes V$ (Pieri's rule \cite{Goodman:2009}). For $V_X\otimes V^*$ we exchange $\alpha$ and $\beta$.

For a set $S$ of legs, the \emph{$Y$-isotypic component} is the subspace of $W$ containing all copies in which those legs transform as $Y$. We write $\Pi^S_Y$ for the orthogonal projector onto this subspace. Projectors for different $Y$ on the same $S$ are mutually orthogonal. Each $\Pi^S_Y$ commutes with the generators $T^a_S$, with every operator invariant under colour rotations on $S$, and with operators acting outside $S$.

\subsection{The basis vectors}\label{app:p:vectors}

A basis vector carries the parton path $(X_1,\dots,X_n)$ of section~\ref{sec:bratteli}. Here $X_k$ is the representation after $k$ complete partons, $X_0$ denotes the initial singlet and $X_n$ is the final singlet. A gluon $k$ with $X_k=X_{k-1}$ is a \emph{copy step}. The vector carries a copy name at each such step. A \emph{leg path} instead records the representation after each individual leg,
$\mathcal P=(X^{(0)}=\varnothing,X^{(1)},\dots,X^{(2m)}=\varnothing)$. Here $\varnothing$ denotes the singlet with both diagrams empty, and $X^{(r)}$ labels legs $1,\dots,r$. Each leg changes one box as prescribed by its type.

\begin{figure}[t]
  \centering
\begin{tikzpicture}[
    leg/.style={circle,fill=black!70,inner sep=1.3pt},
    arc/.style={thick,draw=MidnightBlue!80},
    xarc/.style={line width=1.6pt,draw=BurntOrange!80!black},
    st/.style={font=\footnotesize,text=black!85},
    ar/.style={font=\footnotesize},
    cn/.style={font=\footnotesize,fill=white,inner sep=1pt}]
  \newcommand{\linearc}[6]{%
    \pgfmathsetmacro{\h}{0.42*(#3-#2)+0.25}
    \draw[#1] (#2,0) .. controls (#2,\h) and (#3,\h) .. (#3,0) node[cn,pos=#5,anchor=#6] {$c=#4$};}
  \begin{scope}
    \draw[black!30] (0.6,0) -- (6.4,0);
    \linearc{arc}{1}{6}{0}{0.5}{center}
    \linearc{xarc}{2}{4}{-1}{0.2}{east}
    \linearc{xarc}{3}{5}{1}{0.8}{west}
    \foreach \x/\l in {1/q_1,2/q_2,3/q_3,4/\bar q_4,5/\bar q_5,6/\bar q_6}
      {\node[leg] at (\x,0) {}; \node[st] at (\x,-0.35) {$\l$};}
    \foreach \x/\l in {2/a_x,3/a_y,4/r_x,5/r_y} \node[ar,text=BurntOrange!60!black] at (\x,-0.75) {$\l$};
    \foreach \x in {0.5,6.5} \node[font=\footnotesize] at (\x,-1.25) {$\varnothing$};
    \foreach \x/\d in {1.5/{1},2.5/{1,1},3.5/{2,1},4.5/{2},5.5/{1}}
      \node at (\x,-1.25) {\expandafter\yd\expandafter{\d}};
    \node[font=\small] at (3.5,-1.85) {(a) crossing: $a_x<a_y<r_x<r_y$};
  \end{scope}
  \begin{scope}[xshift=7.4cm]
    \draw[black!30] (0.6,0) -- (4.4,0);
    \linearc{arc}{1}{4}{0}{0.5}{center}
    \linearc{arc}{2}{3}{1}{0.5}{south}
    \foreach \x/\l in {1/q_1,2/q_2,3/\bar q_3,4/\bar q_4}
      {\node[leg] at (\x,0) {}; \node[st] at (\x,-0.35) {$\l$};}
    \foreach \x/\l in {1/a_x,2/a_y,3/r_y,4/r_x} \node[ar,text=MidnightBlue] at (\x,-0.75) {$\l$};
    \foreach \x in {0.5,4.5} \node[font=\footnotesize] at (\x,-1.25) {$\varnothing$};
    \foreach \x/\d in {1.5/{1},2.5/{2},3.5/{1}}
      \node at (\x,-1.25) {\expandafter\yd\expandafter{\d}};
    \node[font=\small] at (2.5,-1.85) {(b) nested: $a_x<a_y<r_y<r_x$};
  \end{scope}
\end{tikzpicture}
  \caption{Lines drawn as arcs over the legs. Each arc runs from the leg $a_x$ that adds a box to the leg $r_x$
  that removes the box at the same position; $c$ is its content, and the row below shows the Young diagram
  $\beta$ after each leg. (a) The leg path $\varnothing\to\yd1\to\yd{1,1}\to\yd{2,1}\to\yd2\to\yd1\to\varnothing$
  of $q_1q_2q_3\bar q_4\bar q_5\bar q_6$: the lines of $q_2$ and $q_3$ cross, $a_x<a_y<r_x<r_y$, and contribute the factor $1-(c_x-c_y)^{-2}=3/4$ to the squared norm~\eqref{eq:p:norm}. (b) The leg path
  $\varnothing\to\yd1\to\yd2\to\yd1\to\varnothing$ of $q_1q_2\bar q_3\bar q_4$: the two lines are nested,
  $a_x<a_y<r_y<r_x$, and contribute no factor. Nested and disjoint arcs are free; only crossing arcs in the same
  diagram change the norm.}
  \label{fig:arcs}
\end{figure}
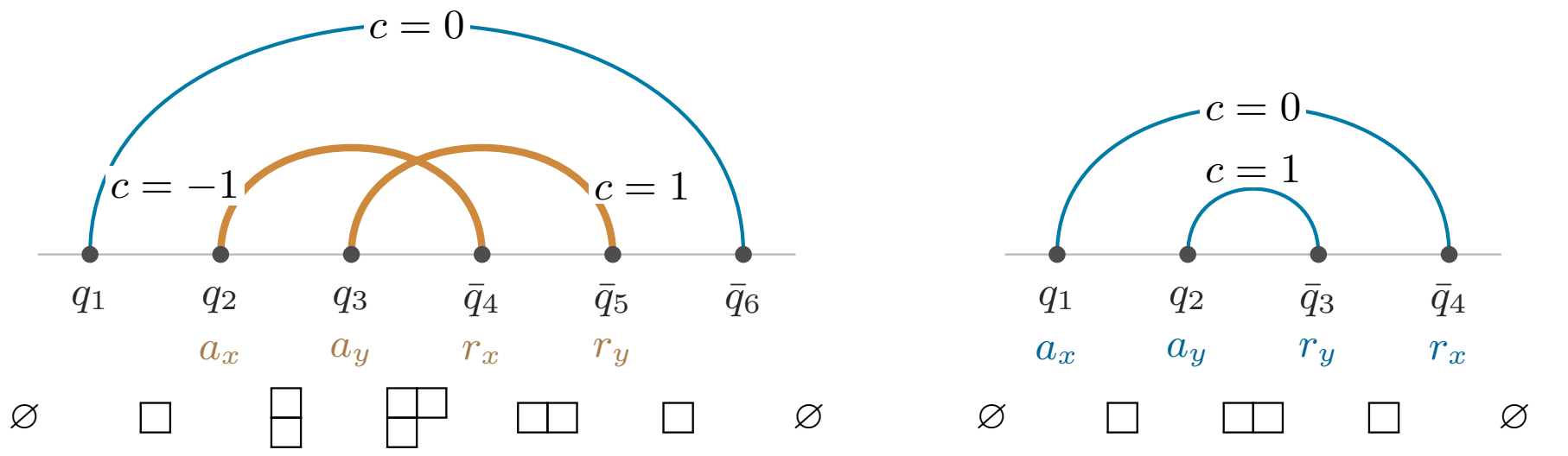

\paragraph{Paths without copy steps.}
If $X_k\ne X_{k-1}$ at a gluon, the two labels fix the intermediate representation after its q-leg. A parton path without copy steps therefore fixes a unique leg path. Along this path a box position can be filled, emptied and used again. We call each interval of occupation a \emph{line} $x$. It starts at the adding leg $a_x$, ends at the next leg $r_x$ that removes that box, and has the content $c_x$ of the box. We write $V_x=V_{a_x}$ for the space carried by its starting leg. Figure~\ref{fig:arcs} draws each line as an arc. Two lines in the same diagram cross when their ends alternate, $a_x<a_y<r_x<r_y$.

Every leg is the end point of exactly one line. We join the two ends of each line by a Kronecker delta. Their product is the \emph{seed}, the permutation state $\ket{\sigma_{\mathcal P}}$ to which we apply the projectors. The resulting vector is
\begin{equation}
  v_{\mathcal P}=p_{2m}\cdots p_1\ket{\sigma_{\mathcal P}},\qquad
  p_r=\prod_{\lambda'\ne\lambda_r}\frac{L_r-\lambda'}{\lambda_r-\lambda'},
  \label{eq:p:base}
\end{equation}
where $\lambda_r$ is the eigenvalue for the required transition $X^{(r-1)}\to X^{(r)}$. The product runs over the eigenvalues $\lambda'$ on the other representations in $X^{(r-1)}\otimes V_r$ (lemma~\ref{lem:p:spec}). This is the projection of eq.~\eqref{eq:b:proj} applied one leg at a time. It also enforces tracelessness at every gluon on this path. The trace of the gluon pair is a singlet, so its contribution after $g_b$ still transforms as $X_{k-1}$. The projector selects $X_k\ne X_{k-1}$.

\paragraph{Copy steps.}
Let $k$ be the first copy step and $X=X_k=X_{k-1}$. Deleting gluon $k$ and the entry $X_k$ gives the \emph{reduced process} and \emph{reduced path}. This path has one fewer copy step. Its basis vectors $w$, the \emph{reduced vectors}, are therefore defined recursively, starting from eq.~\eqref{eq:p:base}.

To restore the gluon, $e(w)=w\otimes\delta_{i_{g_q}\bar\jmath_{g_b}}$ inserts a singlet pair at position $k$. Let $T_{<k}$ be the total charge of all legs belonging to partons before $k$. The operator $G=t_{(g_q)}\cdot T_{<k}$ couples the q-leg of the inserted pair to that charge. The projector $P_0=1-E_{g_qg_b}/N$ removes the trace of the pair. With $\nu(X)$ the total number of removable boxes in $\alpha$ and $\beta$, our \emph{candidates} are
\begin{equation}
  \Phi_rw=P_0\,G^r\,e(w),\qquad r=1,\dots,\nu(X).
  \label{eq:p:cand}
\end{equation}
Here $r$ counts powers of $G$. In comparing with eq.~\eqref{eq:b:copies}, the leg on which the charge acts matters. The full gluon charge $\Tb_g$ annihilates the inserted singlet $e(w)$.

We orthogonalise the candidates in order of increasing $r$. From $\Phi_rw$ we subtract its projections onto the preceding non-zero orthogonal vectors. The result is the \emph{Gram--Schmidt remainder}. We drop remainders of norm zero and name the surviving copies $f,d,d_3,\dots$. The code uses fixed linear combinations of the candidates, for example $\Phi_d=(2P_0G^2+NG)e$. Adding lower candidates leaves the orthogonal remainder unchanged, while multiplying the leading candidate by $\kappa_r$ multiplies its remainder by $\kappa_r$. The code uses $\kappa=(1,2,1,1,\dots)$.

\subsection{Three lemmas}\label{app:p:lemmas}

\begin{lemma}[Spectrum]\label{lem:p:spec}
Let $X=\irr\alpha\beta$. On vectors in the $X$-isotypic component of the legs $<r$ and in the $Y$-isotypic
component of the legs $\le r$, the operator $L_r$ acts as the scalar $\lambda=c$ if leg $r$ adds a box of content $c$ to its own
diagram ($\beta$ for a q-leg, $\alpha$ for a b-leg), and as $\lambda=-(N+c)$ if it removes a box of content $c$
from the other diagram.
\end{lemma}

\begin{proof}
By eq.~\eqref{eq:p:fierz}, $L_r=\frac12(C_{\le r}-C_{<r}-C_F)+(n_{\rm s}-n_{\rm o})/N$, where $C_S$ is the
Casimir of the legs in $S$. Take a q-leg and let $s=|\beta|-|\alpha|=n_{\rm s}-n_{\rm o}$. Adding a box in row $i$
of $\beta$, with content $c=\beta_i+1-i$, changes the first sum in eq.~\eqref{eq:p:casimir} by $2c+N$ and the last
term by $-(2s+1)/N$, so that $\frac12(C(Y)-C(X)-C_F)=c-s/N$ and $\lambda=c$. Removing the box of content
$c=\alpha_i-i$ from row $i$ of $\alpha$ changes the second sum by $-(2c+N)$, so that $\lambda=-(N+c)$. The
b-leg follows by exchanging $\alpha$ and $\beta$.
\end{proof}

\noindent The contents of the addable boxes of a diagram are pairwise distinct, and so are those of its removable
boxes. The eigenvalues of lemma~\ref{lem:p:spec} on the constituents of $X\otimes V_r$ are therefore distinct
functions of $N$, and they are distinct numbers for all but finitely many $N$, in particular for
$N>2\max|c|$. On a vector in the $X$-isotypic component, the Lagrange polynomial $p_r$ is then the isotypic
projector $\Pi^{\le r}_Y$. For the gluon operator $G$ the same argument gives $G=\eta_Y$ on the constituent $Y$ of
$X\otimes V_{g_q}$, with $\eta_Y=\lambda_Y-s/N$.

\begin{lemma}[Isotypy]\label{lem:p:iso}
Let $v$ be a basis vector with path $(X_1,\dots,X_n)$ and let $S_j$ be the legs of the partons $1,\dots,j$. Then
$\Pi^{S_j}_{X_j}v=v$ for every $j$, i.e.\ the first $j$ partons of $v$ transform as $X_j$.
\end{lemma}

\begin{proof}
\emph{Paths without copy steps.} We follow the legs in order. After projection through leg $r-1$, the earlier legs transform as $X^{(r-1)}$. On that subspace, lemma~\ref{lem:p:spec} identifies $p_r$ with $\Pi^{\le r}_{X^{(r)}}$, the projector for legs $1,\dots,r$. Induction therefore puts the image of $p_r\cdots p_1$ in the required component at leg $r$.

Subsequent projections preserve this component. For $r'>r$ we have $L_{r'}=t_{(r')}\cdot(T_{\le r}+T_{(r,r')})+\text{const}$, where $T_{(r,r')}$ sums the charges of legs strictly between $r$ and $r'$. The projector $\Pi^{\le r}_{X^{(r)}}$ commutes with the total charge $T_{\le r}$ and with all operators on the later legs. It therefore commutes with $L_{r'}$ and its polynomial $p_{r'}$. Applying this result at the last leg of each complete parton gives the assertion.

\emph{Copy steps.} We expand a matrix $Z$ on the inserted pair in the basis $\{1/\sqrt N,t^a\}$. The trace projector removes its identity component. By eq.~\eqref{eq:p:fierz}, every candidate in eq.~\eqref{eq:p:cand}, and every linear combination of candidates, then has the form
\begin{equation}
  \Phi w=\sum_at^a_{(g)}\otimes A^aw,\qquad
  A^a=\sum_{r\ge1}\pi_r\,\mathrm{Tr}\bigl(t^at^{b_r}\cdots t^{b_1}\bigr)\,T^{b_r}_{<k}\cdots T^{b_1}_{<k},
  \label{eq:p:Aform}
\end{equation}
where $t^a_{(g)}$ is the matrix $t^a$ on the gluon pair, the repeated generator indices $b_i$ are summed, and the $\pi_r\in\mathbb Q(N)$ are the coefficients of the chosen combination. The traces are invariant tensors. Thus the operators $A^a$ transform together as an adjoint on the earlier legs, and contracting their index with $t^a_{(g)}$ gives an invariant coupling.

For $j<k$, $\Pi^{S_j}_Y$ commutes with $T^a_{<k}$ and with $t^a_{(g)}$, hence with $\Phi$. For $j\ge k$, consider a colour rotation of the first $j$ partons. Rotating the reduced tensor and then inserting the gluon gives the same tensor as inserting the gluon first and rotating all of $S_j$. Consequently $\Pi^{S_j}_Y\Phi=\Phi\,\Pi^{S_j\setminus k}_Y$, where $S_j\setminus k$ omits both legs of gluon $k$. Deleting the repeated entry $X_k=X_{k-1}$ preserves the representation at each corresponding boundary. Induction on the number of copy steps completes the proof.
\end{proof}

\noindent The same commutation shows that $\Phi$ preserves tracelessness in every other gluon $j$:
$E_{q_jb_j}$ is $N$ times the projector onto the singlet of the pair $(q_j,b_j)$ and commutes with $T^a_{<k}$.

\begin{lemma}[Universal copy overlaps]\label{lem:p:overlap}
Let $\Phi,\Phi'$ be of the form~\eqref{eq:p:Aform} at gluon $k$, let $w'$ be $X$-isotypic on the legs $<k$, and
let $w$ be arbitrary. Then $\braket{\Phi w}{\Phi'w'}=O_{\Phi\Phi'}(X)\braket{w}{w'}$, where $O_{\Phi\Phi'}(X)$
depends only on $X$, $N$ and the coefficients $\pi_r,\pi'_r$. For the candidates of eq.~\eqref{eq:p:cand},
\begin{equation}
  \begin{aligned}
  \braket{\Phi_rw}{\Phi_{r'}w'}&=O_{rr'}(X)\braket w{w'},\\
  O_{rr'}(X)&=\sum_Y\eta_Y^{r+r'}d_Y-\frac1N\Bigl(\sum_Y\eta_Y^rd_Y\Bigr)\Bigl(\sum_Y\eta_Y^{r'}d_Y\Bigr),
  \end{aligned}
  \label{eq:p:S}
\end{equation}
where $Y$ runs over the constituents of $X\otimes V$, $\eta_Y$ is the eigenvalue of $G$ and
$d_Y=\dim Y/\dim X$.
\end{lemma}

\begin{proof}
Taking the scalar product $\mathrm{Tr}(Z^\dagger Z')$ on the gluon pair contracts the two adjoint indices in eq.~\eqref{eq:p:Aform}. This gives $\braket{\Phi w}{\Phi'w'}=\bra wK\ket{w'}$ with $K=\sum_aA^{a\dagger}A'^a$. The operator $K$ is a polynomial in the total generators $T^b_{<k}$ and commutes with them all. By Schur's lemma it is a scalar on each irreducible copy of $X$. The same polynomial in the generators gives the same scalar on every copy. For fixed $\Phi,\Phi'$, this leaves only the dependence on $X$ and $N$.

To determine the scalar in eq.~\eqref{eq:p:S}, let $\Pi_Y$ select $Y$ on the earlier legs together with $g_q$. Decomposing the inserted pair gives $e(w')=\sum_Y\Pi_Ye(w')$ and hence $G^re(w')=\sum_Y\eta_Y^r\Pi_Ye(w')$. The singlet pair is $\delta=\sum_i\ket i_{g_q}\ket i_{g_b}$, with the second ket in the dual basis. Contracting the pair takes a partial trace, a sum over the $g_q$ index alone. In particular,
$\bra{e(w)}\Pi_Y\ket{e(w')}=\bra w\mathrm{Tr}_{g_q}\Pi_Y\ket{w'}$ and
$\bra{\delta}\Pi_Ye(w')=(\mathrm{Tr}_{g_q}\Pi_Y)w'$.

On the $X$-isotypic component, this partial trace commutes both with colour rotations and with every invariant operator on the earlier legs. It is therefore a multiple of the identity on all copies of $X$. On one copy its full trace is $\dim Y$, so the multiplier is $d_Y=\dim Y/\dim X$. This partial-trace identity is the \emph{Markov property}.

We now subtract the singlet with $P_0=1-\ket\delta\bra\delta/N$. The preceding identity gives
$\braket{P_0\Pi_Ye(w)}{P_0\Pi_{Y'}e(w')}=(d_Y\delta_{YY'}-d_Yd_{Y'}/N)\braket w{w'}$.
Weighting by $\eta_Y^r\eta_{Y'}^{r'}$ and summing gives eq.~\eqref{eq:p:S}. At an integer $N$ where a constituent $Y$ of the given parent is absent, its dimension polynomial vanishes and the formula still applies. We write $O(X)$ for the matrix with entries $O_{rr'}(X)$. For $w'=w$, the matrix of pairwise scalar products, called the Gram matrix, is $O(X)\braket ww$.
\end{proof}

\subsection{Proof of theorem~\ref{thm:orth}}\label{app:p:orth}

\emph{Different paths.} If two paths differ at parton $j$, $X_j\ne X'_j$, lemma~\ref{lem:p:iso} gives
$\braket v{v'}=\braket{\Pi^{S_j}_{X_j}v}{\Pi^{S_j}_{X'_j}v'}=\bra v\Pi^{S_j}_{X_j}\Pi^{S_j}_{X'_j}\ket{v'}=0$.

\emph{Same path, different copies.} We use induction on the number of copy steps; a path without copy steps
carries a single vector. Otherwise let $k$ be the first copy step and $\{w_i\}$ the reduced vectors,
which are pairwise orthogonal by induction and $X$-isotypic on the legs $<k$ by lemma~\ref{lem:p:iso}. A full
vector is $\Phi_cw_i$, with $\Phi_c$ a Gram--Schmidt remainder of the form~\eqref{eq:p:Aform}. Two vectors
with different reduced vectors, $i\ne i'$, are orthogonal by lemma~\ref{lem:p:overlap}:
$\braket{\Phi_cw_i}{\Phi_{c'}w_{i'}}=O_{cc'}(X)\braket{w_i}{w_{i'}}=0$, whatever the copies $c,c'$. Two vectors
with the same reduced vector and different copies are different remainders of one Gram--Schmidt sequence and are
orthogonal by construction. By lemma~\ref{lem:p:overlap} the Gram matrix of the candidates is $O(X)\braket ww$,
so the Gram--Schmidt coefficients are rational functions of the entries of $O(X)$. They are the same for every reduced vector and
every surrounding path and can be tabulated once, which is what the code does. Two copy-name tuples differ
either at a later copy step (different $w_i$) or only at step $k$ (same $w_i$), so all cases are covered.

\emph{Symbolic $N$.} For all integers $N$ above a bound that depends on the process, the hypotheses hold and the
vectors are the values of fixed vectors over $\mathbb Q(N)$. Their scalar products are rational functions that
vanish at infinitely many integers, and hence vanish identically. \hfill$\square$

\subsection{Norms}\label{app:p:norm}

\begin{theorem}\label{thm:p:norm}
Let $\mathcal P$ be a leg path for an arbitrary interleaving of $m$ q-legs and $m$ b-legs, and let $N$ be an
integer at which every $X^{(r)}$ exists and the local spectra of lemma~\ref{lem:p:spec} are non-degenerate. Then
\begin{equation}
  \braket{v_{\mathcal P}}{v_{\mathcal P}}=\prod_{r:\ \text{leg $r$ adds a box}}\frac{\dim X^{(r)}}{\dim X^{(r-1)}}
  \prod_{\{x,y\}\ \text{crossing}}\Bigl(1-\frac1{(c_x-c_y)^2}\Bigr),
  \label{eq:p:norm}
\end{equation}
where two lines $x,y$ \emph{cross} if their boxes lie in the same diagram and $a_x<a_y<r_x<r_y$
(figure~\ref{fig:arcs}). Both sides are rational functions of $N$, so eq.~\eqref{eq:p:norm} also holds in
$\mathbb Q(N)$. A copy step at $X$ with copy $c$ multiplies the norm of the reduced vector by
\begin{equation}
  \tilde s_c(X)=\kappa_c^2\,\frac{\det O_{[1..c]}(X)}{\det O_{[1..c-1]}(X)},
  \label{eq:p:copyfactor}
\end{equation}
with $O_{[1..c]}$ the leading $c\times c$ block of eq.~\eqref{eq:p:S}.
\end{theorem}

\noindent Written for individual legs, eq.~\eqref{eq:b:norm} gives eq.~\eqref{eq:p:norm}. For a path without copy steps, both legs of each gluon enter the products. In figure~\ref{fig:arcs}, nested and disjoint arcs contribute a factor of unity. Each crossing in the same diagram contributes the factor shown in eq.~\eqref{eq:p:norm}.

For the copy factor, $c$ denotes the position of the copy in the ordered sequence $f$, $d$, $d_3$, \dots. By lemma~\ref{lem:p:overlap}, the Gram matrix of $\Phi_1w,\dots,\Phi_cw$ is $O_{[1..c]}(X)\braket ww$. Orthogonalising these vectors preserves their Gram determinant and makes it the product of their squared norms. Dividing two consecutive determinants therefore gives the squared norm of the last remainder, with the empty determinant equal to one. Including the leading coefficient $\kappa_c$ gives eq.~\eqref{eq:p:copyfactor}. For a full vector we use eq.~\eqref{eq:p:norm} on the path with all copy steps deleted and multiply by one factor for each deleted step.

\paragraph{Examples.}
For the octet $X=\irr{\yd1}{\yd1}$ the constituents of $X\otimes V$ are
$\irr{\yd1}{\yd2}$, $\irr{\yd1}{\yd{1,1}}$ and $\irr{\varnothing}{\yd1}$, with
\begin{equation}
  \eta_Y=1,\,-1,\,-N,\qquad d_Y=\frac{N(N+2)}{2(N+1)},\,\frac{N(N-2)}{2(N-1)},\,\frac{N}{N^2-1},
\end{equation}
and eq.~\eqref{eq:p:S} gives
\begin{equation}
  O(\mathbf 8)=\begin{pmatrix}2N&-N^2\\-N^2&N(N^2-2)\end{pmatrix}.
  \label{eq:p:S8}
\end{equation}
Since $2O_{12}+NO_{11}=0$, the candidate $\Phi_d=(2P_0G^2+NG)e$ is already orthogonal to $\Phi_f$ for the octet;
the same holds for the $\mathbf{27}$, which is also self-conjugate. In general it does not hold, and the
Gram--Schmidt step is needed: at $X=\irr{\yd1}{\yd2}$ the overlap of the two candidates is
$(N-2)(N+1)(N^2+7N-2)/N^2$ in units of $\braket ww$.
The copy factors~\eqref{eq:p:copyfactor} for the octet are then $\tilde s_f=2N$ and $\tilde s_d=4\cdot N^2(N^2-4)/(2N)=2N(N^2-4)$. For the $\mathbf{27}$, $X=\irr{\yd2}{\yd2}$, the factors are $\tilde s_f=4(N+1)$ and $\tilde s_d=4(N-2)(N+1)(N+2)(N+4)/N$.

Let $A$ be the all-octet $f$-chain of $n$ gluons. Its path is $(\mathbf 8,\mathbf 8,\dots,\mathbf 8,\mathbf 1)$, with copy $f$ at gluons $2,\dots,n-1$. Deleting these copy steps leaves the two-gluon path. Its two adding legs give $N\cdot(N^2-1)/N$, and restoring each of the $n-2$ copy steps supplies a factor $2N$. Thus
\begin{equation}
  \braket AA=(2N)^{n-2}(N^2-1).
  \label{eq:p:fchain}
\end{equation}
In the normalisation of section~\ref{sec:colourspace}, $A$ equals $2^{(n-2)/2}$ times the chain of structure constants $\mathrm{i}f^{a_1a_2c_1}\,\mathrm{i}f^{c_1a_3c_2}\cdots$, up to a phase. The squared norm of the structure-constant chain is the standard $N^{n-2}(N^2-1)$.

The first crossing occurs with three quark--antiquark pairs. Figure~\ref{fig:arcs}(a) shows the path $\varnothing\to\yd1\to\yd{1,1}\to\yd{2,1}\to\yd2\to\yd1\to\varnothing$ for $q_1q_2q_3\bar q_4\bar q_5\bar q_6$. The box of $q_2$, of content $-1$, closes while the later box of $q_3$, of content $1$, is still open. Their crossing gives $\dim\yd{2,1}\cdot(1-\frac14)=N(N^2-1)/4$ in eq.~\eqref{eq:p:norm}, in agreement with explicit construction at $N=3,4,5$.

\subsection{Proof of the norm formula}\label{app:p:normproof}

Only this proof uses symmetric-group representation theory. At an adding leg $r$, we project onto $X^{(r)}$ and trace over the added index. The Markov property then gives a factor $\dim X^{(r)}/\dim X^{(r-1)}$. At a removing leg we join the two ends of a line $x$. To do so, we move its index past each later index in the same diagram whose line $y$ is still open. Each such exchange contributes $1-(c_x-c_y)^{-2}$ to the squared norm. The removal itself contributes no dimension factor.

\paragraph{Ingredients.}
We use four facts below. A linear map is \emph{equivariant} if rotating its input and then applying the map gives the same result as rotating its output. For a map $F$ we use the squared Hilbert--Schmidt norm $\|F\|^2_{\rm HS}=\mathrm{Tr}\,F^\dagger F$, the sum of the squared absolute values of its matrix elements in orthonormal bases.

(A) \emph{Markov property.} If $U\cong V_X$ is irreducible and $\Pi_Y$ the isotypic projector of $U\otimes V$ (or
$U\otimes V^*$) onto $Y$, then $\mathrm{Tr}_V\Pi_Y=(\dim Y/\dim X)\,\mathds{1}_U$.

(B) \emph{Top component.} Let $B=V^{\otimes b}\otimes V^{*\otimes a}$ have individually labelled factors. Every constituent $\irr{\alpha'}{\beta'}$ has $|\beta'|\le b$ and $|\alpha'|\le a$, as follows by counting the quark and antiquark indices in a tensor of the given weight. Take $X=\irr\alpha\beta$ with $|\beta|=b$ and $|\alpha|=a$. This is a top component, reached with all boxes present and no contraction. Its isotypic subspace is $V_X\otimes M_X$. Colour rotations act on $V_X$, while the \emph{multiplicity space} $M_X$ labels the different copies of $X$.

Permuting the quark indices and the antiquark indices separately acts on $M_X$ as $S_b\times S_a$. Under these permutations $M_X$ is isomorphic to $S^\beta\otimes S^\alpha$, where $S^\lambda$ is the irreducible representation of the permutation group labelled by diagram $\lambda$ (its Specht module \cite{Specht:1935,James:1978}). To see this, consider the highest-weight vectors of weight $X$. Their quark indices lie in $\{1,\dots,\ell(\beta)\}$ and their antiquark indices in $\{N-\ell(\alpha)+1,\dots,N\}$. Schur--Weyl duality \cite{Goodman:2009} pairs each colour representation occurring in a block of fundamental tensors with the permutation representation of the same diagram. Applying this to the two blocks gives $S^\beta\otimes S^\alpha$, the top component of the mixed-tensor decomposition \cite{Koike:1989}.

An equivariant map $F:B\to U\cong V_X$ selects a linear combination of the copies. Write $\pi_X$ for the orthogonal projector onto $V_X\otimes M_X$ and choose a scalar-product-preserving identification $\iota$ of $V_X$ with $U$. Then $F=(\iota\otimes\bra g)\pi_X$ for a unique $g\in M_X$, and $\|F\|^2_{\rm HS}=\dim X\,\|g\|^2$.

(C) \emph{Young's orthogonal form.} For a diagram $\lambda$ with $n$ boxes, restricting $S^\lambda$ successively to $S_{n-1},S_{n-2},\dots$ resolves it into one-dimensional pieces. Each piece corresponds to a \emph{standard tableau} $\theta$, a filling of $\lambda$ by $1,\dots,n$ in increasing order along each row and down each column. Write $\theta(i)$ for the box containing $i$ and $c(\theta(i))$ for its content. The corresponding unit vectors $\psi_\theta$ form the Gelfand--Tsetlin basis of $S^\lambda$, the permutation analogue of the path basis in section~\ref{sec:bratteli}.

The transposition $s_i=(i\;i{+}1)$ exchanges two consecutive labels. Its action involves only the original tableau and the tableau $\theta\circ s_i$ with these labels exchanged \cite{Okounkov:1996}. With $r=c(\theta(i{+}1))-c(\theta(i))$, it reads
$s_i\psi_\theta=r^{-1}\psi_\theta+\sqrt{1-r^{-2}}\,\psi_{\theta\circ s_i}$,
where $\psi_{\theta\circ s_i}=0$ if the exchanged tableau is not standard.

(D) \emph{Moving one label to the end.} We write $\psi^o_\theta$ for the Gelfand--Tsetlin vectors defined with the
labels ordered as $o=(o_1,\dots,o_n)$ instead of $(1,\dots,n)$, so that the successive restrictions remove $o_n$ first,
then $o_{n-1}$, and so on. Let $y=\theta(\ell)$ be a
removable box and let $o'$ be the order $o$ with $\ell=o_j$ moved to the end. The projection of $\psi^o_\theta$
onto the span of the $\psi^{o'}_{\theta'}$ with $\theta'(\ell)=y$ is
$\pm\prod_{i>j}\sqrt{1-(c(\theta(o_i))-c(y))^{-2}}\;\psi^{o'}_\theta$. This follows from (C): moving $\ell$ to the
end is the product $s_{n-1}\cdots s_j$, and the condition that $\ell$ ends on $y$ selects the term in which every
factor moves $\ell$ on.

\begin{proof}[Proof of eq.~\eqref{eq:p:norm}]
Let $\Lambda_r$ be the set of lines open after leg $r$; they are in one-to-one correspondence with the boxes of
$X^{(r)}$. Let $B_r=\bigotimes_{x\in \Lambda_r}V_x$, and let $\sigma_r:B_r\to V_1\otimes\cdots\otimes V_r$ place
every open line on its starting leg and contract the end points of every closed line. Then
$F_r=p_r\cdots p_1\sigma_r$ maps $B_r$ equivariantly onto a copy of $V_{X^{(r)}}$, with $F_0=1$ and
$\|F_{2m}\|^2_{\rm HS}=\braket{v_{\mathcal P}}{v_{\mathcal P}}$. By (B), $F_r=(\iota\otimes\bra{g_r})
\pi_{X^{(r)}}$. We show by induction on $r$ that $g_r$ is a multiple of the product of the Gelfand--Tsetlin
vectors of the two diagrams for the tableaux ``line $\mapsto$ box'', with the lines of each diagram ordered by
their starting legs, and that $\|F_r\|^2_{\rm HS}$ is the right-hand side of eq.~\eqref{eq:p:norm} restricted to
adding legs $\le r$ and to the crossing pairs whose earlier line is closed by leg $r$. The tableaux are standard, because a box to
the left of or above an open box cannot be removed before it.

\emph{Adding leg.} Opening line $\ell$ gives $F_{r+1}=p_{r+1}(F_r\otimes\mathds{1})$. Tracing over the new leg by (A) gives $\|F_{r+1}\|^2_{\rm HS}=(\dim X^{(r+1)}/\dim X^{(r)})\|F_r\|^2_{\rm HS}$. It remains to identify the tableau after this step. Write $\alpha^{(r)},\beta^{(r)}$ for the two diagrams of $X^{(r)}$. The map $g_r\mapsto g_{r+1}$ commutes with permutations of the old lines. Under those permutations $M_{X^{(r+1)}}$ contains $S^{\beta^{(r)}}\otimes S^{\alpha^{(r)}}$ exactly once. By Schur's lemma, $g_{r+1}$ is therefore proportional to the Gelfand--Tsetlin vector obtained by putting the new label $\ell$ in the added box.

\emph{Removing leg.} A line $\ell$ at box $y$ closes, say in $\beta$. Then $F_{r+1}=p_{r+1}(F_r\otimes
\mathds{1})(\mathds{1}\otimes\delta)$, where $\mathds{1}\otimes\delta$ inserts a Kronecker delta that joins leg $r{+}1$ to the starting leg of $\ell$. By (B)
applied to $B_{r+1}$ and Young's branching rule (the restriction of $S^\lambda$ from $S_n$ to $S_{n-1}$ is the sum
of $S^{\lambda'}$ over the diagrams $\lambda'$ obtained by removing one box), $M_{X^{(r)}}=\bigoplus_{y'}M''_{y'}$
over the removable boxes $y'$ of $\beta^{(r)}$, and $p_{r+1}$ keeps only $y'=y$. On that summand the map
$J=(\Pi\otimes\mathds{1})(\mathds{1}\otimes\delta)$, which inserts the delta and then applies the projector $\Pi$ onto the copy of
$V_{X^{(r+1)}}$ in $V_{X^{(r)}}\otimes V_{r+1}$, satisfies
$J^\dagger J=\dim X^{(r)}/\dim X^{(r+1)}$ by (A) and Schur's lemma, so that
$\|F_{r+1}\|^2_{\rm HS}=\|F_r\|^2_{\rm HS}\,\|g_y\|^2/\|g_r\|^2$, with $g_y$ the component of $g_r$ in $M''_y$. No
dimension factor arises. By (D), $\|g_y\|^2/\|g_r\|^2=\prod_z(1-(c_z-c_y)^{-2})$ over the open lines $z$ of $\beta$
that started after $\ell$, which are exactly the lines that cross $\ell$ with $\ell$ the earlier one. The lines
of $\alpha$ are spectators. Every crossing pair is counted once, when its earlier line closes. Finally, $g_y$ is
proportional to $\psi^{o'}_\theta$ with $\ell$ in the last position, so after the restriction that removes $\ell$,
$g_{r+1}$ is proportional to the Gelfand--Tsetlin vector of the tableau with $\ell$ deleted, which continues the
induction.
\end{proof}

\subsection{Assumptions and scope}\label{app:p:scope}

The proofs require every representation along the path to exist at $N$ and the local eigenvalues in lemma~\ref{lem:p:spec} to be distinct. Both conditions hold for all $N$ above a bound of order $m$. The construction uses $N\ge m$ so that the permutation states are linearly independent. Under the stated hypotheses, orthogonality and the norm formula hold for every ordering of quarks, antiquarks and gluons. They also hold as rational identities in symbolic $N$. Three further issues concern the implementation and its use at fixed $N$.

First, the code reconstructs rational functions of $N$ from evaluations at integers or from calculations modulo a prime $p$ \cite{Wang:1982}. We assume the reconstruction returns the true functions. In modular arithmetic, a non-zero norm can vanish modulo $p$, causing the code to drop a vector with probability of order $1/p$. This can leave the basis incomplete while preserving the orthogonality of the retained vectors.

Second, at fixed small $N$, such as $N=3$, we specialise the symbolic vectors. After we clear poles, identically vanishing scalar products remain zero. Completeness also requires enough vectors to survive. At generic $N$, $X$ occurs $\nu(X)$ times in the product of $X$ with the adjoint. The candidates $\Phi_1w,\dots,\Phi_{\nu(X)}w$ are independent by a Vandermonde argument. At fixed $N$ we compare the number of non-zero vectors with the dimension of the invariant space, obtained independently by integrating the character over the group. They agree for $N=2,\dots,5$ with up to nine gluons and for $q\bar qg^k$ with $k\le6$. The argument leaves open the case in which a lower Gram--Schmidt remainder vanishes at a special $N$ and a higher one survives. We find no such case below ten gluons at $N\ge2$.

Third, the identification of the code operations with the tensor operators above rests on tests. We also verify eq.~\eqref{eq:p:norm} as an identity in $\mathbb Q(N)$ for up to six quark--antiquark pairs in several interleavings. In modular arithmetic we verify it for all orderings with $m\le5$ and all 924 orderings with $m=6$.

\paragraph{Relation to the literature.}
The two kinds of factor in eq.~\eqref{eq:p:norm} are known individually. The dimension ratios are the weights obtained by tracing over an added index in the orthogonal form of the walled Brauer algebra \cite{Nazarov:1996,Grinko:2023cdq,Si:2023}. The factors $1-(c_x-c_y)^{-2}$ occur when restricting a permutation representation to one fewer label. They enter both Gram determinants and the norms of seminormal basis vectors, for which the Jucys--Murphy elements are diagonal \cite{James:1979,Murphy:1981,Mathas:2008,Rui:2008,Si:2023}.

For the standard order $q^m\bar q^m$, the Gelfand--Tsetlin basis of ref.~\cite{Grinko:2023cdq} also gives eq.~\eqref{eq:p:norm}. Inserting $\delta$ between the middle legs and moving the b-legs by a shortest sequence of neighbouring exchanges reproduces both products, using eqs.~(59)--(62) of that reference. The seminormal vectors of refs.~\cite{Rui:2008,Si:2023} use signed sums of diagrams. Their norms differ by path-dependent factors from the norm for our single-diagram seed $\sigma_{\mathcal P}$. We have found neither this closed form for that seed nor a treatment of interleaved q- and b-legs in the literature. Interleaving changes the sequence of Jucys--Murphy operators \cite{Brundan:2017}. Our proof uses the $\mathrm U(N)$ representation on each successive set of legs and covers all interleavings.
\section{A worked example}\label{app:example}

In this appendix we construct the basis for $q\bar q\to gg$ explicitly, compute its norms from
eq.~\eqref{eq:b:norm} and directly, and give the soft anomalous dimension in this basis. All partons are outgoing
and ordered as in the code, $q_1\bar q_2g_3g_4$; the physical process $q\bar q\to gg$ differs only by the
labels of the two incoming partons. We write $i$ and $\bar\jmath$ for the colour indices of partons 1 and 2 and
$a,b$ for those of gluons 3 and 4. Vectors and norms are given with $\mathrm{Tr}\,t^at^b=\delta^{ab}$ as in
appendix~\ref{app:proofs}, and $\Gamma$ in the normalisation of section~\ref{sec:colourspace}. All results below
were computed with \orthobase{} at symbolic $\Nc$ and checked independently with explicit generators of
$\mathrm{SU}(\Nc)$ for $\Nc=3,4,5$.

\paragraph{Paths.}
After parton 1 the colour state is the fundamental $\irr{\varnothing}{\yd1}$. Parton 2 couples it to a singlet or
an octet, and the first gluon maps either of these to an octet; the second gluon must return to the singlet. At
generic $\Nc$ there are therefore two paths,
\begin{equation}
  (\mathbf 3,\mathbf 1,\mathbf 8,\mathbf 1),\qquad (\mathbf 3,\mathbf 8,\mathbf 8,\mathbf 1),
\end{equation}
where we label the representations by their dimensions at $\Nc=3$. The second path contains the copy step
$\mathbf 8\to\mathbf 8$ at gluon 3, and the octet has two removable boxes, so it carries the two copies $f$ and
$d$. The space has three dimensions for every $\Nc\ge3$. At $\Nc=2$ the $d$ copy vanishes and two vectors remain.

\paragraph{Vectors.}
In terms of colour structures the three basis vectors are
\begin{equation}
  \begin{gathered}
  u_1=\delta_{i\bar\jmath}\,\delta^{ab},\qquad
  u_f=\bigl(t^at^b-t^bt^a\bigr)_{i\bar\jmath},\\
  u_d=\Nc\bigl(t^at^b+t^bt^a\bigr)_{i\bar\jmath}-2\,\delta_{i\bar\jmath}\,\delta^{ab}.
  \end{gathered}
  \label{eq:ex:vectors}
\end{equation}
The singlet $u_1$ is the vector of the first path. In terms of the legs of section~\ref{sec:bratteli}, its seed
(appendix~\ref{app:p:vectors}) connects the quark with the antiquark and the quark leg of each gluon with the antiquark leg of the other, and the
projectors remove the traces in the gluon pairs; by the Fierz identity the result is
$\delta_{i\bar\jmath}\delta^{ab}$. The two other vectors are obtained from the reduced vector
$w=(t^b)_{i\bar\jmath}$ of $q_1\bar q_2g_4$ by inserting gluon 3 with eq.~\eqref{eq:p:cand}. For $r=1$ the
colour charge of the preceding partons acts on $w$ as a commutator, which gives $u_f$. For the octet the second
candidate is already orthogonal to the first (appendix~\ref{app:proofs}), and it gives $u_d$. In the
normalisation of section~\ref{sec:colourspace}, $u_f=2\,\mathrm i f^{abc}(t^c)_{i\bar\jmath}$ and
$u_d=2\Nc\,d^{abc}(t^c)_{i\bar\jmath}$, so that the basis is the $s$-channel basis of singlet,
antisymmetric and symmetric octet exchange.

\paragraph{Norms.}
For the singlet path the legs that add a box are the quark ($\dim\mathbf 3/\dim\mathbf 1=\Nc$), the quark leg of
gluon 3 (again $\Nc$) and its antiquark leg ($(\Nc^2-1)/\Nc$). No two lines cross, so eq.~\eqref{eq:b:norm} gives
$\Nc(\Nc^2-1)$. For the copies we delete gluon 3. The reduced path $(\mathbf 3,\mathbf 8,\mathbf 1)$ has norm
$\Nc\cdot(\Nc^2-1)/\Nc=\Nc^2-1$, and the copy factors $\tilde s_f=2\Nc$ and $\tilde s_d=2\Nc(\Nc^2-4)$ of
appendix~\ref{app:p:norm} multiply it. Together,
\begin{equation}
  \braket{u_1}{u_1}=\Nc(\Nc^2-1),\quad
  \braket{u_f}{u_f}=2\Nc(\Nc^2-1),\quad
  \braket{u_d}{u_d}=2\Nc(\Nc^2-1)(\Nc^2-4).
  \label{eq:ex:norms}
\end{equation}
These agree with a direct calculation. With $\sum_bt^bt^b=(\Nc^2-1)/\Nc$ and $\sum_bt^bt^at^b=-t^a/\Nc$ in the normalisation of
appendix~\ref{app:proofs},
$\braket{u_f}{u_f}=2\,\mathrm{Tr}(t^at^bt^bt^a)-2\,\mathrm{Tr}(t^at^bt^at^b)=2(\Nc^2-1)^2/\Nc+2(\Nc^2-1)/\Nc=2\Nc(\Nc^2-1)$.
Similarly $\|\{t^a,t^b\}\|^2=2(\Nc^2-1)(\Nc^2-2)/\Nc$ and
$\braket{\delta\delta}{\{t^a,t^b\}}=2(\Nc^2-1)$, which give the norm of $u_d$. The three vectors are pairwise
orthogonal: $u_f$ is antisymmetric in $a\leftrightarrow b$ and the other two are symmetric, and
$\braket{u_1}{u_d}=2\Nc(\Nc^2-1)-2\Nc(\Nc^2-1)=0$. At $\Nc=3$ the norms are 24, 48 and 240, the values found
by the code.

\paragraph{Soft anomalous dimension.}
In the orthonormal basis $(\hat u_1,\hat u_f,\hat u_d)$, $\hat u=u/\|u\|$, the normalisation of the generators
used for the vectors drops out, and the operator $\Gamma$ of eq.~\eqref{eq:cs:gamma} reads
\begin{equation}
  \Gamma=\begin{pmatrix}
  -C_F\,\omega_{12}-\Nc\,\omega_{34} & -\delta\omega/\sqrt2 & 0\\[2pt]
  -\delta\omega/\sqrt2 & A & -\tfrac14\sqrt{\Nc^2-4}\,\delta\omega\\[2pt]
  0 & -\tfrac14\sqrt{\Nc^2-4}\,\delta\omega & A
  \end{pmatrix},
  \label{eq:ex:gamma}
\end{equation}
with
\begin{equation}
  \delta\omega=\omega_{13}-\omega_{14}-\omega_{23}+\omega_{24},\qquad
  A=\frac{\omega_{12}}{2\Nc}-\frac{\Nc}{2}\,\omega_{34}-\frac{\Nc}{4}\bigl(\omega_{13}+\omega_{14}+\omega_{23}+\omega_{24}\bigr),
\end{equation}
and $C_F=(\Nc^2-1)/(2\Nc)$. The matrix is complex symmetric, as it must be in an orthonormal basis. For equal
weights it is $-(C_F+C_A)$ times the identity, which is colour conservation. Only the combination $\delta\omega$ mixes
the basis vectors. The reason is that the remaining part of the weights between the pairs $(12)$ and $(34)$ has
the form $\omega_{ij}=x_i+y_j$, and by colour conservation it reduces to the Casimirs of the pairs, which are
diagonal in this basis (section~\ref{sec:cs:gamma}). The singlet does not mix with the symmetric octet at this
order. The entries of eq.~\eqref{eq:ex:gamma} were computed with \orthobase{} as rational functions of $\Nc$ and
the weights, and agree to machine precision with an explicit evaluation of
$\bra{\hat u_a}\Gamma\ket{\hat u_b}$ with $\mathrm{SU}(\Nc)$ generators at $\Nc=3,4,5$ and random weights.

\paragraph{Comparison with the literature.}
Reference~\cite{Kidonakis:1998nf} gives the soft anomalous dimension of $q\bar q\to gg$ in the $s$-channel basis
$c_1=\delta_{r_Ar_B}\delta_{r_1r_2}$, $c_2=d^{r_1r_2c}(t^c)_{r_Br_A}$, $c_3=\mathrm if^{r_1r_2c}(t^c)_{r_Br_A}$
(their eqs.~(62)--(63)), as a matrix $\Gamma_{IJ}$ defined by $\Gamma c_J=\sum_Ic_I\Gamma_{IJ}$ and in units of
$\alpha_s/\pi$. We identify their partons $(A,B,1,2)$ with our $(1,2,3,4)$ and set
\begin{equation}
  \omega_{13}=\omega_{24}=-\mathcal T,\qquad \omega_{14}=\omega_{23}=-\mathcal U,\qquad \omega_{12}=\omega_{34}=0,
\end{equation}
with $\mathcal T=\ln(-\hat t/\hat s)+\mathrm i\pi$ and $\mathcal U=\ln(-\hat u/\hat s)+\mathrm i\pi$ as in
their eq.~(45), where $\hat s,\hat t,\hat u$ are the Mandelstam variables; the normalisation conventions are
collected in appendix~\ref{app:conventions}. Then $\delta\omega=2(\mathcal U-\mathcal T)$ and
$A=\Nc(\mathcal T+\mathcal U)/2$. Transforming eq.~\eqref{eq:ex:gamma} to the basis $c_1=u_1$,
$c_2=u_d/(2\Nc)$, $c_3=-u_f/2$, with the norms of eq.~\eqref{eq:ex:norms}, gives
\begin{equation}
  \begin{pmatrix}
  0 & 0 & \mathcal U-\mathcal T\\
  0 & \frac{\Nc}{2}(\mathcal T+\mathcal U) & \frac{\Nc}{2}(\mathcal U-\mathcal T)\\
  2(\mathcal U-\mathcal T) & \frac{\Nc^2-4}{2\Nc}(\mathcal U-\mathcal T) & \frac{\Nc}{2}(\mathcal T+\mathcal U)
  \end{pmatrix},
\end{equation}
which is eq.~(63) of ref.~\cite{Kidonakis:1998nf} without the factor $\alpha_s/\pi$, entry by entry and for
symbolic $\Nc$. The sign of $c_3$ relative to $u_f$ is a phase convention. The asymmetry of this matrix reflects
only the different norms of $c_1,c_2,c_3$; in the orthonormal basis the same operator is the symmetric
matrix~\eqref{eq:ex:gamma}.
\section{Numerical details}\label{app:numerics}

\subsection{Evolution}

The exponential $e^{-t\Gamma}x$ is computed with a Krylov space of dimension $m=24$ and a relative tolerance of $10^{-8}$
($10^{-10}$ for the comparisons of section~\ref{sec:window}). The error of a step is estimated a posteriori from the next
Krylov vector \cite{Saad:1992,Hochbruck:1997}; if it exceeds the tolerance, the step is split into substeps of adaptive
length. On the CPU and, where memory allows, on the GPU, all $m$ Krylov vectors are kept (Arnoldi). At $n=14$ on the GPU we use
instead a symmetric Lanczos recursion with the bilinear form $x^Ty$, which is valid for the complex symmetric $\Gamma$. It
keeps four vectors and runs twice, once to build the tridiagonal matrix and once to assemble the result, at the cost of
$2m-1$ products. All products are in double precision. For each value of $t$ the evolution starts again from $t=0$, so that
the results at different $t$ are independent.

The block plan of the sweep, section~\ref{sec:matvec}, is computed once per process and set of non-zero weights and stored as
index arrays. Within each step all operations write to disjoint memory, so that the result does not depend on the number of
threads and needs no atomic operations on the GPU. The CPU code is written in C++ with OpenMP and has kernels for AVX2 and
AVX-512, selected at run time; the GPU code uses CUDA. The timings of section~\ref{sec:scaling} were measured on nodes with two
AMD EPYC processors and on an NVIDIA A100 with 40~GB of memory; the largest runs used nodes with about 1~TB of memory.

\subsection{Spectrum}

The lowest eigenvalues of $\Gamma$ for real weights are computed with a thick-restart Lanczos method \cite{Wu:2000} with full
reorthogonalisation and a tolerance of $10^{-9}$, checked against dense diagonalisation for $n\le6$. For one of the two points
at $n=14$ the excited levels did not converge within the limit of 5000 products, while the lowest eigenvalue did. The smallest
leading-colour eigenvalue, the minimum of eq.~\eqref{eq:cs:lambdaLC} over all cyclic orderings, is a shortest-tour problem; we
solve it exactly by dynamic programming over subsets \cite{Bellman:1962,Held:1962}, which we checked against complete enumeration for $n\le10$.

\subsection{Leading colour and LC+}

The leading-colour factor~\eqref{eq:cs:RLC} is summed over all $(n-1)!$ flows for the exclusive-jet samples up to $n=10$
and for the main gap samples up to $n=12$. For the larger samples we sample the cyclic
orderings with a Metropolis algorithm \cite{Metropolis:1953am} with weight $|A_\sigma|^2$ and average $e^{-2t\,\mathrm{Re}\,\lambda^{\rm LC}_\sigma}$. This
requires $|A_\sigma|^2$ for arbitrary orderings, which is available in closed form for MHV amplitudes \cite{Parke:1986gb};
for this reason the sampled events of sections~\ref{sec:exclusive} and~\ref{sec:gaps} have MHV
helicities. The statistical error is below 1\%
per event, and the method reproduces the exact sum where both are available. The overlaps of
the evolved flows needed for LC+ are obtained by projecting each flow onto the orthonormal basis, which has the cost of one
projection of $(n-1)!$ coefficients.

\subsection{Matrix product states}

The MPS evolution uses a Krylov space of dimension 16 in MPS form; intermediate states are compressed to at most four times
the target bond dimension $\chi$ during the products and to $\chi$ afterwards, with a tolerance of $10^{-12}$. The Taylor coefficients of section~\ref{sec:moments} are computed without truncation.
\section{Code and data availability}\label{app:code}

The construction of the basis, the colour operators, the compiled core for the evolution, the CUDA implementation and the
matrix product states are available in version 2.0 of the code \orthobase{} at
\url{https://github.com/rakab/OrthoBase} under the BSD 3-clause licence. Version 2.0 replaces the implementation of
ref.~\cite{Chargeishvili:2024pnq}, which is kept in the repository for reference. The old implementation also illustrates how a general-purpose constraint solver can take over combinatorial tasks in colour
algebra: the decomposition of products of irreducible representations is stated as integer constraints on the
entries of Young diagrams, and the satisfiability-modulo-theories solver Z3 \cite{deMoura:2008} enumerates all
solutions. The rules of the combinatorics are thus stated declaratively and the search is left to the solver, an
approach that may be useful for other enumeration problems in particle physics, such as the construction of
admissible colour or helicity structures. The code is written in Python with a C++
core; its documentation describes the installation and contains short examples for each of the steps of
sections~\ref{sec:bases} and~\ref{sec:evolution}.

The projection of tree amplitudes onto the basis (section~\ref{sec:hard} and appendix~\ref{app:trees}) and the scripts used
to produce the results of sections~\ref{sec:lcvsfc} and~\ref{sec:beyond} are not part of the public code. The numerical
results are available on request.
\section{Tree-level colour sums}\label{app:trees}

\subsection{Projection of the adjoint decomposition}

In the adjoint decomposition~\eqref{eq:hard:ddm} the colour factor of the identity ordering is a single basis vector. Every
intermediate representation of the chain $\tilde f^{a_1a_2b_1}\tilde f^{b_1a_3b_2}\cdots$ is an octet, and every coupling
$\mathbf 8\otimes\mathbf 8\to\mathbf 8$ is antisymmetric, which is the copy of type $f$. Hence
\begin{equation}
  \ket{F_{\rm id}}=K_n\ket{u_f},\qquad K_n=\sqrt{(2\Nc)^{n-2}(\Nc^2-1)},
\end{equation}
in the normalisation of the code, where $\ket{u_f}$ is the normalised all-octet $f$ chain. The colour factor of any other
ordering is obtained by neighbour exchanges, $\ket{F_\sigma}=K_n\,P(\sigma)\ket{u_f}$, where $P(\sigma)$ is a product of the
orthogonal exchange matrices of section~\ref{sec:local}. Thus
\begin{equation}
  \ket{\mathcal M}=K_n\sum_{\sigma\in S_{n-2}}A_\sigma\,P(\sigma)\ket{u_f}\equiv K_n\,c,\qquad
  \sum_{\rm colours}|\mathcal M|^2=K_n^2\,\|c\|^2 .
\end{equation}
The sum over orderings is organised by the orderings of the last legs: the orderings that agree in their last legs share the corresponding product of
exchanges, so the sum is computed level by level, from short sub-orderings to the full one, in the same way as subcurrents are
shared in Berends--Giele recursion \cite{Berends:1987me}. At $\Nc=3$ the number of basis vectors reached at intermediate levels
grows more slowly than the number of orderings, and the projection costs 1117, 1350 and 1527 complex multiply-adds per
colour-ordered amplitude at $n=10$, 11 and 12, approaching about $2\times10^3$. The colour-ordered amplitudes are computed with
a Berends--Giele recursion in which the four-gluon vertex is evaluated at the cost of a three-gluon vertex
\cite{Duhr:2006iq}; the recursion reproduces the Parke--Taylor amplitudes \cite{Parke:1986gb} exactly.

For $n\ge13$ storing all $(n-2)!$ amplitudes needs tens of GB (22.1~GiB at $n=13$ for table~\ref{tab:trees}). A
variant with bounded memory fixes the last $L$ legs, computes the projection
of each of the resulting subproblems with a Berends--Giele recursion for that fixed tail, and walks over the tails depth
first. On four threads this takes 117~s and 6.0~GB at $n=13$ and 29~minutes and 18.6~GB at $n=14$. The hard states at
$n=15$ of section~\ref{sec:scaling} were built in this way on 64--80 threads.

\subsection{Comparison with other methods}

Colour-summed tree amplitudes for up to eleven gluons were recently computed with three methods, a multiplet recursion and
Fourier transforms over the symmetric group in the trace and in the adjoint basis \cite{Bolinder:2025gbj,Frederix:2026cay}.
We ran the public implementation of these methods (version \texttt{345f8de} of the code of
ref.~\cite{Frederix:2026cay}) and our code on idle nodes with two AMD EPYC~9654 processors, both codes on the same node
for each $n$, one process per core, for twelve phase-space points per $n$. The colour-summed $|\mathcal M|^2$ of the two codes agree at every point to $2.7\times10^{-12}$.
Table~\ref{tab:trees} gives the time per evaluation for a fixed helicity configuration with $\lfloor n/2\rfloor$ positive helicities, for
which both codes use their general recursion.

\begin{table}[t]
  \centering
  \small
  \begin{tabular}{r r rrr}
    \toprule
    $n$ & this work & adjoint Fourier~\cite{Frederix:2026cay} & multiplet~\cite{Bolinder:2025gbj} & trace Fourier~\cite{Frederix:2026cay} \\
    \midrule
     4 & 2.0~\textmu s & 0.36$\times$ & 0.51$\times$ & 0.58$\times$ \\
     5 & 3.4~\textmu s & 0.71$\times$ & 1.62$\times$ & 1.36$\times$ \\
     6 & 11~\textmu s  & 1.05$\times$ & 4.3$\times$ & 2.3$\times$ \\
     7 & 64~\textmu s  & 1.17$\times$ & 8.1$\times$ & 2.8$\times$ \\
     8 & 0.45~ms       & 1.21$\times$ & 13.2$\times$ & 3.2$\times$ \\
     9 & 3.9~ms        & 1.21$\times$ & 18.9$\times$ & 3.6$\times$ \\
    10 & 38~ms         & 1.25$\times$ & 29.2$\times$ & 4.8$\times$ \\
    11 & 0.49~s        & 1.78$\times$ & 31.4$\times$ & 6.6$\times$ \\
    12 & 6.6~s         & -- & 28$\times$ & 49$\times$ \\
    13 & 103~s         & -- & -- & 51$\times$ \\
    \bottomrule
  \end{tabular}
  \caption{Time per colour-summed squared tree amplitude for $n$ gluons at $\Nc=3$ and a helicity configuration with $\lfloor n/2\rfloor$
  positive helicities, on one core: our implementation, and the ratio of the time of the three methods of
  refs.~\cite{Bolinder:2025gbj,Frederix:2026cay} to ours; ratios above one mean that our implementation is faster. Median
  over twelve phase-space points (three for the trace method at $n=13$); for the central 68\% of the points the ratio lies
  within 4\% of the median, except for the adjoint method at $n=8$ (12\%) and the trace method at $n=12$ (6\%).}
  \label{tab:trees}
\end{table}

For these helicities our implementation is 1.2 times faster than the adjoint-Fourier method for $n=7$--10 and 1.8 times
faster at $n=11$, and slower for $n\le5$. The published range of the adjoint-Fourier method ends at $n=11$. At $n\ge12$ the
public implementation stops at an internal consistency check, which reports that the transformed adjoint colour kernel is
not symmetric; we did not investigate the cause.

The multiplet and trace methods run at $n=12$. Their reach is set by memory rather than by the colour algebra. The
multiplet recursion of ref.~\cite{Bolinder:2025gbj} uses tabulated SU(3) recoupling coefficients, which are distributed
for up to 13 gluons and can be generated for more, but it keeps a current for every subset of external legs, each expanded
in the multiplet states of that subset; its memory grows from 4.0~GiB at $n=11$ to 43~GiB at $n=12$, and at $n=13$ the
estimate exceeds the memory of our node, so we did not run it. The trace method of ref.~\cite{Frederix:2026cay} completed
at $n=13$ for three of twelve points, at 51 times our time and 58~GiB; the remaining points were terminated close to
the limit of 60~GiB per process that we set. Our own table ends at $n=13$ for the same reason: beyond it the $(n-2)!$ colour-ordered amplitudes no longer
fit in memory, and our bounded-memory variant described above, which can reach further, runs on several threads, so it is
not included in a single-core comparison.

The multiplet basis of ref.~\cite{Bolinder:2025gbj} serves as an intermediate representation inside the recursion, with
numerical coefficients for $\Nc=3$, and the output is the colour-summed $|\mathcal M|^2$. In our construction the
orthonormal basis and the dipole operators acting on it are explicit and depend on $\Nc$ symbolically, which is what the
evolution of the main text requires; the tree-level colour sum is a by-product.

\clearpage
\bibliographystyle{JHEP}
\bibliography{refs_v2}
\end{document}